\documentclass[aps, pre, 10pt, twocolumn, a4paper, notitlepage, longbibliography]{revtex4-2}

\usepackage[T1]{fontenc}
\usepackage{amsmath, amssymb}
\usepackage{bm}
\usepackage{graphicx}
\usepackage{color}
\usepackage{multirow}

\newcommand{\kB}{k_\mathrm{B}}                     
\newcommand{\iu}{\mathrm{i}}                       
\newcommand{\ee}{\mathrm{e}}                       
\newcommand{\Hamiltonian}{\mathcal{H}}             
\newcommand{\dd}{\mathrm{d}}                       
\newcommand{\corrfunc}{\mathfrak{C}}               
\newcommand{\hdim}{\bar{h}}                        
\newcommand{\phidim}{\bar{\phi}}                   
\newcommand{\lsc}{\zeta}                           
\newcommand{\bfrho}{\bm{\rho}}                     
\newcommand{\BesselJ}{\mathcal{J}}                 
\newcommand{\BesselK}{\mathcal{K}}                 
\newcommand{\BesselKmod}{\mathcal{M}}              
\newcommand{\prob}{\mathfrak{p}}                   
\newcommand{\bfq}{{\bm{q}}}                        
\newcommand{\bfz}{{\bm{0}}}                        
\newcommand{\bfr}{{\bm{r}}}                        
\newcommand{\qcc}{{\left|\bfq\right|<\Lambda}}     
\newcommand{\const}{\operatorname{const}}          
\newcommand{\im}{\operatorname{Im}}                
\newcommand{\re}{\operatorname{Re}}                
\newcommand{\partf}{\mathcal{Q}}                   
\newcommand{\pathint}{\mathcal{D}}                 
\newcommand{\Adim}{\bar{A}}                        
\newcommand{\exads}{\Gamma_\text{ex}}              
\newcommand{\exvol}{V_\text{ex}}                   
\newcommand{\Tc}{T_\mathrm{c}}                     
\newcommand{\kei}{\operatorname{kei}}              
\newcommand{\transpose}{{\operatorname{T}}}
\newcommand{\rad}{\mathfrak{R}}

\allowdisplaybreaks

\begin{document}

\title{Protein--lipid domains in heterogeneous membranes\\ beyond spontaneous curvature effects}
\author{Piotr Nowakowski}
\affiliation{Max-Planck-Institut f{\"u}r Intelligente Systeme Stuttgart, Heisenbergstr.~3, 70569 Stuttgart, Germany}
\affiliation{Institut f{\"u}r Theoretische Physik IV, Universit{\"a}t Stuttgart, Pfaffenwaldring 57, 70569 Stuttgart}
\affiliation{Group for Computational Life Sciences, Division of Physical Chemistry, Ru\dj{}er Bo\v{s}kovi\'c Institute, Bijeni\v{c}ka 54, 10000, Zagreb, Croatia}

\author{Bernd Henning Stumpf}
\affiliation{PULS Group, Institut f{\"u}r Theoretische Physik, IZNF, Friedrich-Alexander-Universit{\"a}t Erlangen-N{\"u}rnberg, Cauerstra\ss{}e 3, 91058 Erlangen, Germany}

\author{Ana-Sun\v{c}ana Smith}

\affiliation{PULS Group, Institut f{\"u}r Theoretische Physik, IZNF, Friedrich-Alexander-Universit{\"a}t Erlangen-N{\"u}rnberg, Cauerstra\ss{}e 3, 91058 Erlangen, Germany}
\affiliation{Group for Computational Life Sciences, Division of Physical Chemistry, Ru\dj{}er Bo\v{s}kovi\'c Institute, Bijeni\v{c}ka 54, 10000, Zagreb, Croatia}

\author{Anna Macio\l{}ek}
\affiliation{Max-Planck-Institut f{\"u}r Intelligente Systeme Stuttgart, Heisenbergstr.~3, 70569 Stuttgart, Germany}
\affiliation{Institute of Physical Chemistry, Polish Academy of Sciences, Kasprzaka 44/52, 01-224 Warsaw, Poland}

\begin{abstract}
We study a model of a lipid bilayer membrane described by two order parameters: the chemical composition described using the Gaussian model and the spatial configuration described with the elastic deformation model of a membrane with a finite thickness, or equivalently, for an adherent membrane. We assume and explain on physical grounds the linear coupling between the two order parameters. Using the exact solution, we calculate the correlation functions and order parameters profiles. We also study the domains that form around inclusions on the membrane. We propose and compare six distinct ways to quantify the size of such domains. Despite of its simplicity, the model has many interesting features like Fisher--Widom line or two distinct critical regions.  
\end{abstract}
\maketitle

\section{Introduction}\label{secA}

Continuous research effort is put into understanding the relationship between physical properties, functionality and the mutual influence of lipids and  proteins in cell membranes. The central issues concern membrane domains. 
 It is well established that model lipid bilayer membranes containing cholesterol can support two coexisting liquid phases, called liquid--ordered (Lo) and liquid--disordered (Ld)~\cite{Dietrich2001, Veatch2007, HonerkampSmith2009}.
Near the critical point of miscibility  domains of different phases form, giving rise to a lateral heterogeneity of a lipid membrane. Recent studies suggest that this impacts the organization and function of plasma membrane proteins, which in turn may affect, e.g., membrane signaling~\cite{Stone2017a}. 

Formation of domains and functional lateral heterogeneity is observed also in  living cell membranes~\cite{Stone2017b, Voci2018, Roobala2018}.
These domains  are  nanoscopic and dynamical \cite{Meder2006,Pralle2000,Levental2020}, therefore the physical  basis of   heterogeneity in cell membranes is supposed to be more complex than a miscibility transition alone. 

Diverse equilibrium mechanisms that can lead to membrane segregation on a smaller length scale have been identified theoretically and experimentally~\cite{B822956B,destainville2018rationale}.
Several  of them involve nanometer--sized membrane inclusions, such as proteins. In general,  both direct and indirect interactions between  inclusion and  membrane lipids can generate domains.  Direct specific interactions cause chemically favored lipids to be attracted to the protein, creating an adsorption domain with a concentration of preferred lipids greater than that of the bulk membrane. The extent of such a domain is of the order of  the  composition correlation length $\xi $, and thus, near the critical point of demixing, where $\xi $ grows significantly, it can be as large as several microns~\cite{Hanke1999,honerkamp2008line}.

Protein inclusions are causing membrane disturbances which result in indirect interactions.
Among such perturbations are changes due to hydrophobic mismatch
between membrane lipids and inclusion, i.e., when the hydrophobic part of the inclusion has a thickness slightly different from the hydrophobic part of the membrane~\cite{Venturoli2005,Bitbol2012}. 
The energetic cost of hydrophobic mismatch deformation can be reduced by attracting lipids of a suitable characteristic. 
One possibility is that lipids that match the curvature of the membrane caused by the protein will be effectively attracted~\cite{Leibler1987,Sens2000}.
Thus inclusions with hydrophobic core  larger/smaller than that of  the  membrane  would tend  to attract lipids  of positive/negative  spontaneous curvature, thereby building a concave/convex shape  to fill in the height
mismatch. For example, cholesterol and saturated lipids exhibit a negative spontaneous curvature, whereas unsaturated lipids, with smaller acyl chain area to polar head group area ratio, exhibit positive spontaneous curvature. In this mechanism, the inclusion--induced deformation is strongly  dependent  upon  the  spontaneous  curvature, which in turn is coupled to the composition of the membrane.
A different possibility   is that in order to accommodate hydrophobic mismatch, lipids of matching length of acyl chain are effectively attracted to the protein inclusion. In model membranes, such as those considered experimentally  in Refs.~\cite{Dietrich2001, Veatch2007, HonerkampSmith2009}, the hydrophobic thickness of the membrane is nonuniform. The Lo phase, rich in saturated lipids, shows higher extension in the lipid acyl chains than the Ld phase, rich in unsaturated lipids~\cite{Brown1998}.  Depending on the sign of the hydrophobic mismatch,  the lipid composition around the protein will preferentially be in one of the two phases (Lo or Ld). In this complementary mechanism, the composition of the membrane is coupled to its thickness~\cite{PhysRevE.102.060401,Stumpf2021}, and not necessarily to the overall spontaneous curvature of the membrane.

Another possible mechanism for creating lipid domains around proteins would be by deforming the membranes through ligand--receptor interactions with the structures in the extracellular space. In this case either curvature effect could play a role, but also the composition of the membrane could be affected by the change of the average separation from the opposing, adherent surface.  Namely, the displacement of the membrane may result in the expulsion of the proteins of the glycocalyx  from the region of the contact. Likewise,  a significant redistribution of charged moieties may take place.  In both cases,  the nonspecific interactions of the adherent membrane will be affected, which can drive  modification of the membrane composition.

From the theoretical point of view, the curvature effects were conceptually formalized already two decades ago~\cite{Sens2000}. Further detailed analysis was both performed analytically and in simulations by several groups~\cite{Honigmann2014,Rautu2015,Ayton2005,Veksler2007,Sadeghi2014,Simunovic11226,Prevost2015}. The thickness mechanism was suggested in a numerical study~\cite{PhysRevE.102.060401}, while the analytic approach was then put froward only last year by us~\cite{Stumpf2021}. In Ref.~\cite{Stumpf2021}, we proposed a model which couples the elasticity theory of lipid bilayer thickness deformations \cite{dan1993membrane,refId0,PhysRevE.102.060401,Bitbol2012}
with the Landau--Ginzburg theory of critical demixing transition and with the inclusions.  Our  model was able to reproduce experimental observations of the formation of lipid domains around lipids linked to a reconstituted actin cortex filament in a  model membrane~\cite{Honigmann2014,Stumpf2021}. In these experiments the lipid bilayer was supported, which strongly suppresses 
spontaneous curvature effects.

Interestingly, we note that, following the work of Bibtol et al.~\cite{Bitbol2012}, it is possible to draw a direct analogy between the thickness and the shape deformation of the membrane in the lowest order of theory. That means it is possible, on a different scale, to address theoretically both problems within the same framework. The role of this paper is to provide  this analogy and to describe our model in more detail, which we could not do in the letter format. Furthermore, we want to understand the adsorption of lipids onto the protein inclusion given that no direct attractive interaction is imposed. 

We first explore the general features of the model for both, the shape and thickness  deformation fields by computing and analysing correlation functions. In the absence of the protein inclusions, we calculate two--point correlation functions and discuss different forms of their asymptotic decay in connection with the poles appearing in their  integral representation. 
Due to the presence of two order parameters and higher order derivatives  in the Hamiltonian, the behavior of these functions is rich and interesting. For example, the correlation length that governs 
 the asymptotic decay of the correlation functions, shows  a curious nonmonotonic and nonanalytic behavior as  function of the temperature deviation from the critical temperature $\tau$.
 This is far from a typical behavior of a binary mixture upon approaching the critical point.
From our analytical results we derive the asymptotic behavior of the correlation functions and the correlation length in the several  limiting cases of the three relevant parameters of our model. 

Furthermore, we investigate adsorption phenomena around  membrane--embedded protein in the context of a domain formation upon approaching the critical temperature. Given that there is no direct attraction between proteins and lipids that would be responsible  for the classical critical adsorption, it is interesting to understand  what kind of universal scaling law is obeyed for the adsorption phenomena that take place entirely due to the coupling of the two order parameters. 
 For this purpose we calculate the order parameter profiles around protein inclusion. Based on the integral and local properties of these profiles, we propose several definitions of the size of the domain. We determine their asymptotic behavior in several limits, including $\tau \to 0$.

The paper is organized as follows: In Sec.~\ref{secB} we define the model, explain the physical mechanisms behind our assumptions and briefly explain the method of calculation. Sec.~\ref{secD} is devoted to the correlation functions. We define them and use their properties to distinguish three zones in the space of parameters. In Sec.~\ref{secE}, we discuss the order parameter profiles and study the formation of domains. We introduce and compare six different ways to identify the size of induced domains. Our research is summarized and discussed in Sec.~\ref{secG}. Finally, we have included three appendixes to give more details of our calculations: in Appendix~\ref{appA}, we present the method used to calculate the order parameters profiles and correlation functions; in Appendix~\ref{appB}, we discuss the behavior or the roots of a certain polynomial, which determine the properties of correlation functions; and in Appendix~\ref{appC} we study the correlation functions in various limiting cases.

\section{Model}\label{secB}


We start from introducing the model discussed in this paper. In order to describe the system we use two order parameters.

\subsection{Configuration order parameter}

The spatial configuration of the membrane is described by the order parameter $\hdim\left(\bfr\right)$, the height of the membrane above the reference plane in a point given by the two--dimensional vector $\bfr$. For simplicity we assume that the membrane is infinite, such that $\bfr$ can be any vector from the plane. 
The energy associated with the configuration of a membrane can be approximated by 
~\cite{Nelson2004,lipowsky1995structure,Helfrich1973, Bruinsma1994}
\begin{equation}\label{secB:HelfrichModel}
    \beta \Hamiltonian_{\text{MD}}=\int \dd \bfr\left[ \kappa \big(\varkappa \left(\bfr\right)\big)^2+V_{\text{ext}}\big(\hdim\left(\bfr\right)\big)\right],
\end{equation}
which we call the membrane deformation  model. In this Hamiltonian, the bending stiffness $\kappa$ is a dimensionless parameter describing the energy cost of bending the membrane, $\varkappa$ is a mean curvature of the membrane, and an external potential $V_\text{ext}$ keeps the membrane above the reference plane. For simplicity we assume that there is no term proportional to the area of the membrane, i.e., we neglect the surface tension.  In \eqref{secB:HelfrichModel} we have introduced $\beta=\left(\kB T\right)^{-1}$ (where $T$ is a temperature and $k_B$ is the Boltzmann constant) in order to keep the formula dimensionless, and we have assumed that the membrane has no spontaneous curvature.

\begin{figure*}[t]
\begin{center}
\begin{tabular}{llll}
(a)& & (b) &\\
&\includegraphics[width=0.45\textwidth]{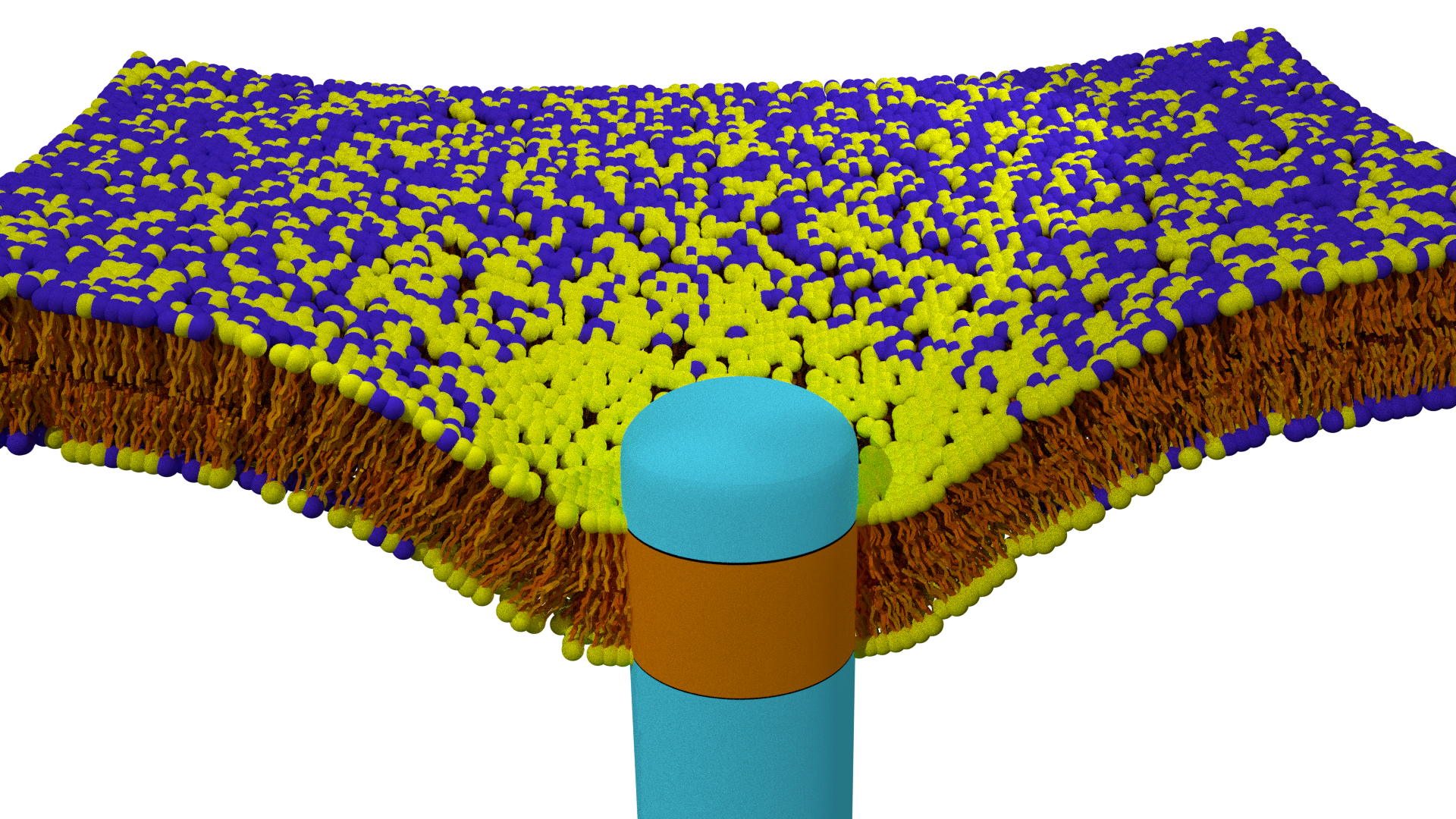} && \includegraphics[width=0.45\textwidth]{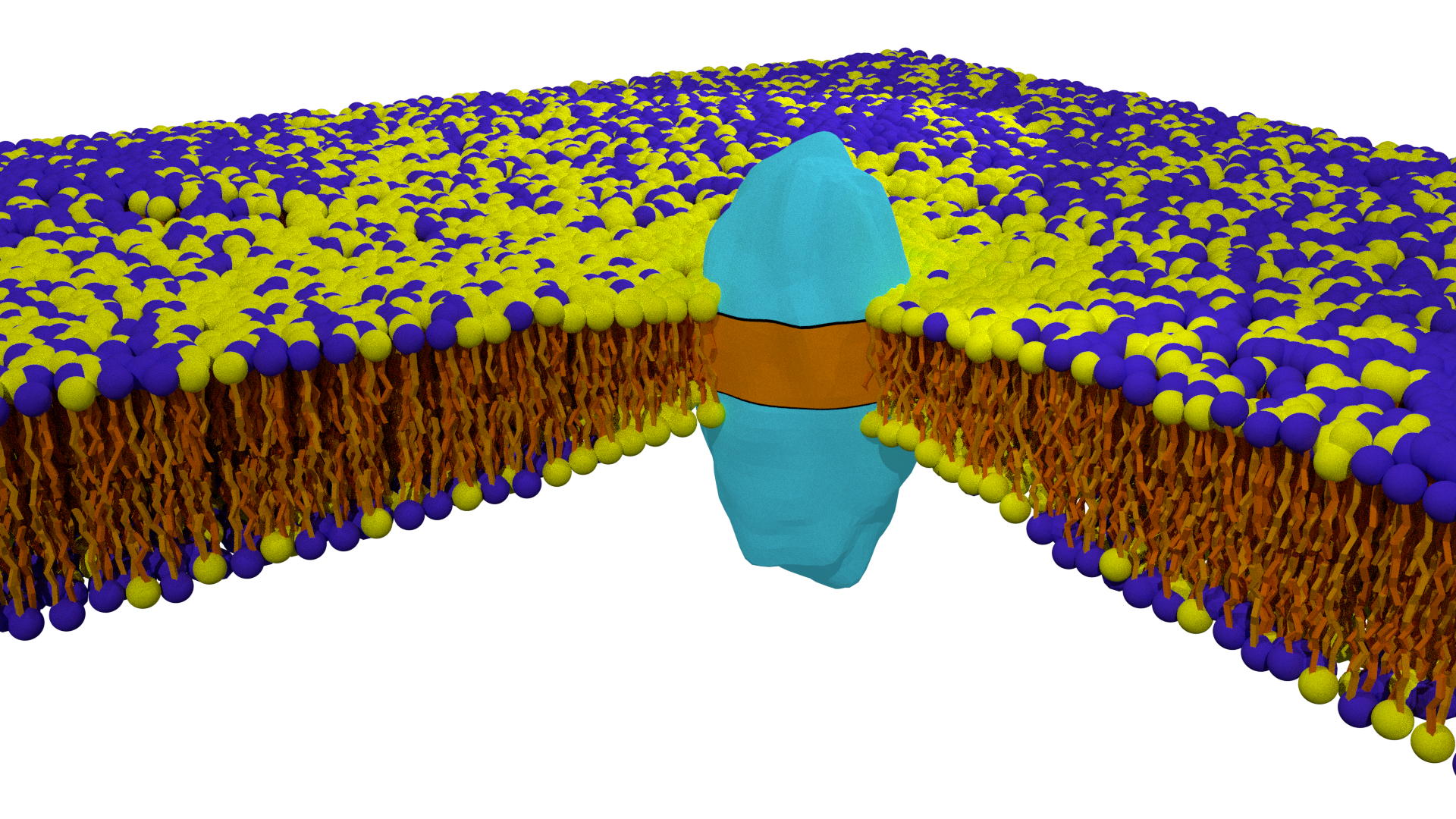} \\[0.0cm]
&\includegraphics[width=0.45\textwidth]{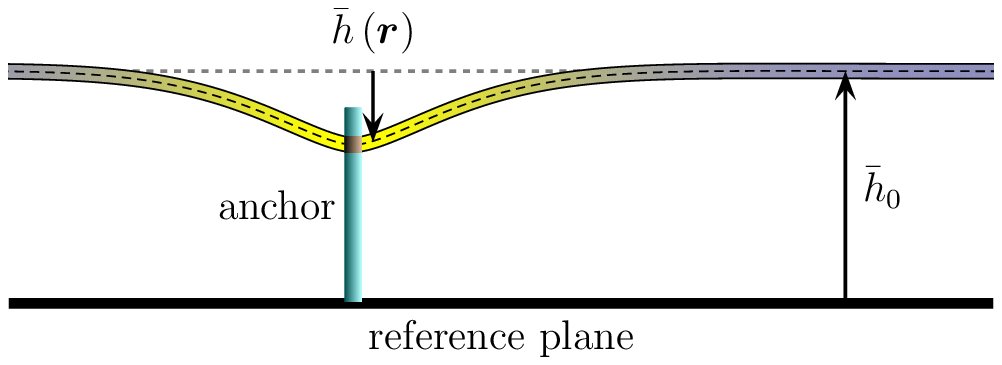} && \includegraphics[width=0.45\textwidth]{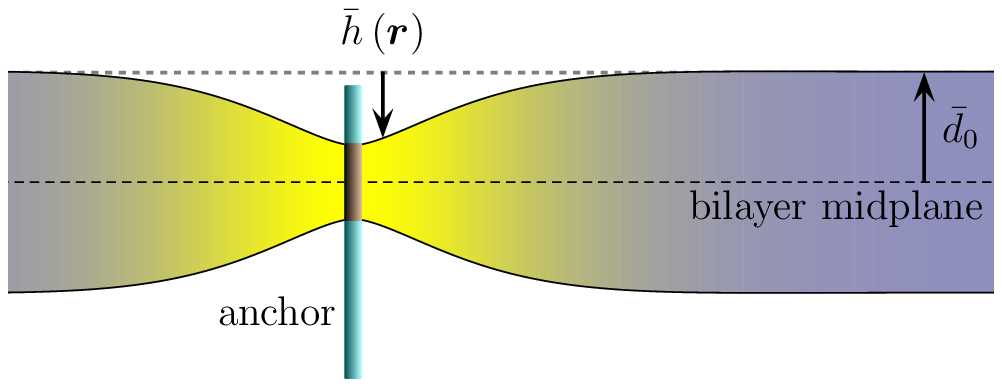}
\end{tabular}
\end{center}
\caption{\label{secB:schem}Schematic plot of the two possible interpretations of the model. Different types of lipids are denoted with blue and yellow color. (a) The membrane is not flat and its height over the reference plane is described by the order parameter $\hdim\left(\bfr\right)$. In this case the anchor (illustrated in the middle of the picture) is locally fixing the height of the membrane and, as a result, the membrane is  bent. The external potential that acts differently on different types of lipids changes the composition of the membrane around the anchor. (b) The membrane is flat but its thickness is not constant and described by the order parameter $\hdim\left(\bfr\right)$. Due to hydrophobic mismatch of the anchor (difference between the equilibrium thickness of the membrane and size of the hydrophobic region on the anchor) the thickness of the membrane is alerted around the anchor, which modifies local composition of the lipids. }
\end{figure*}

When the membrane is almost flat, i.e., $\left|\nabla \hdim\left(\bfr\right)\right|\ll 1$, the mean curvature $\varkappa\left(\bfr\right)\approx \nabla^2 \hdim\left(\bfr\right)/2$. The external potential originates from many different interactions like steric repulsion or van der Waals attraction, and in cellular systems, from the properties of the glycocalyx. We assume that it has a minimum for $\hdim=\hdim_0$, and (since we are only interested in small excitations around the minimum) it can be approximated by a harmonic potential. If one, for simplicity, additionally assumes that its minimum is at $\hdim=0$, $V_\text{ext}\approx \gamma\hdim^2/2$, where the parameter $\gamma$ quantifies the strength of binding of the membrane.

The energy of a flat membrane with a variable thickness can also be approximated by Eq.~\eqref{secB:HelfrichModel} for $\left|\nabla \hdim\left(\bfr\right)\right|\ll 1$. In this picture, the order parameter $\hdim\left(\bfr\right)$ is defined as the difference between the local thickness and the thickness $\bar{d}_0$ of unperturbed membrane (i.e., the distance between two lipid layers). Of course, in this interpretation the meaning of $\kappa$ and $V_{\text{ext}}$ is different: the parameter $\kappa$, as shown in \cite{Bitbol2012}, is four times smaller than the true bending stiffness of the membrane, and the potential now describes the interaction between the two layers. Both possible interpretations of the order parameter $\hdim\left(\bfr\right)$ has been schematically illustrated in Fig.~\ref{secB:schem}.

A comprehensive description of a membrane definitely requires consideration of both the spatial configuration of the membrane and its thickness by taking into account two distinct configuration order parameters. In this paper, in order to keep the model as simple as possible, we use only a single field $\hdim\left(\bfr\right)$. Nevertheless, we keep two possible interpretations as they are both experimentally relevant. 

\subsection{Composition order parameter}

In order to describe the composition of the membrane we introduce a scalar order parameter $\phidim\left(\bfr\right)$. We define it as a difference between the local concentration of saturated lipids and their concentration at the critical demixing point. This way, above the critical temperature $\Tc$ the equilibrium value of the order parameter is zero and below $\Tc$ it has two possible equilibrium values: positive describing Lo phase and negative describing Ld phase.

In this paper we use only one composition order parameter to describe the membrane in order to keep the model simple; thorough description of the chemical composition would require the introduction of several order parameters (separately for each component and each layer), which typically complicates the phase diagram.

Around the critical point, the energy associated with the chemical composition can be approximated by the Landau--Ginzburg Hamiltonian 
\begin{multline}\label{secB:GLModel}
    \beta\Hamiltonian_{\text{LG}}=\\
    \int \dd\bfr \left[ \frac{\sigma}{2}\big(\nabla \phidim \left(\bfr\right)\big)^2+t\phidim^2\left(\bfr\right)+u\phidim^4\left(\bfr\right)-c \phidim\left(\bfr\right)\right],
\end{multline}
where $\sigma>0$, $t\propto T-\Tc$,  $u>0$, and $c$ is the ordering field proportional to the deviation of the chemical potential of the considered component from its critical value.

In our model, for the sake of simplicity, we assume that $c=0$ and $u=0$ in Eq.~\eqref{secB:GLModel}. The former means that the composition of the membrane is at its critical--point value, which seems to be in line with some of the experiments \cite{Honigmann2014}. The latter assumption make the Hamiltonian unbounded from below for $t<0$ and, therefore, restricts our analysis to the cases where the membrane is in a mixed state ($t>0$). For $u>0$ an analytical solution of our model is not known.

The two remaining terms in Hamiltonian \eqref{secB:GLModel} ($u=0$, $c=0$ and $t>0$) define the so--called Gaussian model \cite{Helfrich1973}.

\subsection{Coupling between order parameters}

In order to study the relation between the chemical composition and shape of the membrane, it is necessary to introduce coupling between the order parameters. 

The simplest, mathematical coupling arises from the observation that the integral in Landau--Ginzburg Hamiltonian \eqref{secB:GLModel} should be calculated on the curved manifold (given by $\hdim\left(\bfr\right)$) rather than on the flat reference plain. Close investigation shows that this effect gives corrections that are proportional to the powers of $\nabla \hdim$ and thus they are not relevant in the limit of almost flat surface $\left|\nabla \hdim\left(\bfr\right)\right|\ll 1$ assumed in our model.

The physical mechanism of coupling between the order parameter $\phidim$ and the mean curvature $\varkappa$ of the membrane was proposed and analyzed in Refs.~\cite{Leibler1986, Leibler1987}. In the leading order, it gives a coupling in Hamiltonian that is proportional to $\phidim \nabla^2 \hdim$, which can lead to curvature driven phase separation in the membrane.

In this paper we propose a different physical mechanism that leads to a direct coupling between the order parameters --- a term proportional to $\hdim \phidim$ in the Hamiltonian. 

When $\hdim$ describes spatial configuration of the membrane, its shape is fixed by an external potential $V_\text{ext}$, see Eq.~\eqref{secB:HelfrichModel}. The presence of $V_\text{ext}$ originates from an interaction of individual membrane lipids with surroundings and it is natural to assume that it depends on the composition $\phidim$. The simplest possible way to include this effect is to consider composition dependent equilibrium height of the membrane, which for small $\left|\phidim\right|$ gives
\begin{equation}\label{secB:coupling}
    V_{\text{ext}}=\frac{\gamma}{2}\left[\hdim\left(\bfr\right)-\alpha\phidim\left(\bfr\right)\right]^2,
\end{equation}
where we have introduced the proportionality coefficient $\alpha$. 

On the other hand, when $\hdim$ denotes the excess thickness of the membrane (over the reference value $\bar{d}_0$) we assume that different lipids have a different effective length of acyl chains. This makes the equilibrium thickness of the bilayer dependent on its chemical composition and justifies Eq.~\eqref{secB:coupling}.

The effect of the coupling between order parameters has been schematically shown in Fig.~\ref{secB:schem} for both possible interpretations of $\hdim$.

We note that from the point of view of the composition order parameter $\phidim$, the coupling term given by Eq.~\eqref{secB:coupling} has two effects: the term quadratic in $\phidim$ effectively shifts $t$ by $\alpha^2\gamma/2$ (and therefore pushes the system away from the critical point), and the term linear in $\phidim$ represents a position--dependent (via $\hdim\left(\bfr\right)$) chemical potential. 

\subsection{Inclusions}

Finally, we introduce inclusions that model the anchors (like proteins or lipids) immersed in the membrane. We assume that they are coupled to the configuration order parameter: If $\hdim$ is the height of the membrane above the reference plane, we assume that the inclusions are attached to some external structures (cytoskeleton) and deform locally the shape of the membrane. If $\hdim$ denotes the thickness of the membrane, we assume that the inclusions have a hydrophobic mismatch, i.e., the hydrophobic region on the inclusion has a different height than the height of the unperturbed membrane. As a result the membrane gets thicker or thinner close to the inclusion.

For simplicity we neglect the size of inclusions and assume that they are point--like~\cite{Netz,dommersnes1999n}. More realistic model would require to introduce some non--zero area covered by the inclusion; this area should be excluded from the integrals in Eqs.~\eqref{secB:HelfrichModel} and \eqref{secB:GLModel}, as the order parameters are undefined there. In the model, as a first approximation, we also neglect the possible different affinities of an inclusion to different components of the membrane, i.e., coupling of the inclusion directly to order parameter $\phidim$. This effect is worth to study separately, but we expect it to be subdominant in comparison with coupling to $\hdim$. Moreover, allowing for point--like inclusions coupled to $\phidim$ leads to some divergent integrals in our model and requires introducing a regularization scheme, we discuss this in more detail in Sec.~\ref{secE}.

We denote by $N$ the number of inclusions and label their positions by $\bfr_1, \bfr_2,\ldots, \bfr_N$. We approximate the interaction with harmonic potential:
\begin{equation}\label{secB:inclusions}
    \beta \Hamiltonian_\mathrm{I}=\frac{\lambda}{2}\sum_{i=1}^N\left[\hdim\left(\bfr_i\right)-\hdim_i\right]^2,
\end{equation}
where $\hdim_i$ denotes the value of $\hdim$ preferred by $i$--th inclusion, and the positive coefficient $\lambda$ defines the strength of the potential. In our calculations, for the sake of simplicity, we take the limit $\lambda\to\infty$, which enforces the relations $\hdim\left(\bfr_i\right)=\hdim_i$ for $i=1,2,\ldots,N$.

\subsection{Hamiltonian of the model}

We summarize this section by writing the full Hamiltonian of our model in the reduced, dimensionless variables.

We use the length scale
\begin{equation}\label{secC:lengthscale}
 \lsc=\left(\kappa/\gamma\right)^{1/4},
\end{equation}
as a unit of length, and define $h=\hdim/\lsc$ and $\bfrho=r/\lsc$. The length scale $\lsc$ is proportional \footnote{The parameter $\lsc$ is often called the correlation length, but in the model with no composition order parameter the length scale at which the correlation function decays is $\sqrt{2}\lsc$.} to the correlation length in a model of a tensionless membrane with a single order parameter $\hdim\left(\bfr\right)$ \cite{Bihr2015}. 
The unit of the composition is given by  $\sigma^{-1/2}$ and the unit of energy is $\kB T$. All the formulae for dimensionless variables are summarized in Table \ref{secC:dimensionless}.

\begin{table*}
\caption{\label{secC:dimensionless}Definition of dimensionless parameters used in the model. The unit of length is denoted by $L$, the unit of chemical composition (the unit of $\phidim$) is denoted by $C$, and the unit of energy is denoted by $E$. In the last column, the names of the reduced variables that we use in this article are given.}
\begin{ruledtabular}
\begin{tabular}{cccl}
variable & original unit & rescaled variable & name\\
\hline
$\hdim$ & $L$& $h=\hdim/\lsc$ & reduced height of the membrane\\
$\phidim$ & $C$ & $\phi=\phidim\, \sigma^{1/2} $ & reduced composition of the membrane\\
$\Adim$ & $L^2$ & $A=\Adim/\lsc^2$ & reduced area of the system\\
$\mathbf{r}$ & $L$ & $\bfrho=\mathbf{r}/\lsc$ & reduced distance/position\\
$\kappa$ & 1 & $\kappa$ & bending stiffness\\
t        & $L^{-2}C^{-2}$ & $\tau=t\,\lsc^2/\sigma$ & reduced temperature\\
$\alpha$   & $LC^{-1}$ & $\mu=\alpha\, \sigma^{-1/2}/\lsc $ & reduced coupling\\
$\lambda$  & $L^{-2}$ & $\nu=\lambda \lsc^2$ & reduced harmonic potential coefficient\\
$\hdim_i$   & $L$ & $h_i=\hdim_i/\lsc$ & reduced height of the membrane for proteins\\
\end{tabular}
\end{ruledtabular}
\end{table*}

The resulting model of a membrane depends on three dimensionless parameters: $\kappa$, $\tau$ and $\mu$. It is also convenient to introduce $\omega=\kappa \mu^2$ as this combination of parameters often appears in our formulae. The anchors immersed in the membrane introduce additional parameters: the positions of anchors $\bfrho_i$, the imposed reduced heights of the membrane $h_i=\bar{h}_i/\zeta$ for $i=1,\ldots, N$, and the harmonic potential constant $\nu$ (which we set infinite in our calculation). The Hamiltonian in the new variables takes the form
\begin{subequations}\label{secC:Hamiltonian}
\begin{align}
\label{secD:Hamiltoniansum}\Hamiltonian\left[h\left(\bfrho\right), \phi\left(\bfrho\right)\right]=&\Hamiltonian_\mathrm{MD}+\Hamiltonian_\mathrm{G}+\Hamiltonian_\mathrm{C}+\Hamiltonian_\mathrm{I},\\
\label{secD:HamiltonianH}\beta\Hamiltonian_\mathrm{MD}=&\int\dd\bfrho\, \frac{\kappa}{2}\left[\nabla^2 h\left(\bfrho\right)\right]^2,\\
\label{secD:HamiltonianG}\beta \Hamiltonian_\mathrm{G}=&\int\dd \bfrho\left(\frac{1}{2}\left[\nabla\phi\left(\bfrho\right)\right]^2+\tau \phi^2\left(\bfrho\right)\right),\\
\label{secD:HamiltonianC}\beta \Hamiltonian_\mathrm{C}=&\int\dd\bfrho\,\frac{\kappa}{2}\left[h\left(\bfrho\right)-\mu\phi\left(\bfrho\right)\right]^2, \\
\label{secD:HamiltonianP}\beta\Hamiltonian_\mathrm{I}=&\frac{\nu}{2}\sum_{i=1}^N\left[h\left(\bfrho_i\right)-h_i\right]^2,
\end{align}
\end{subequations}
where the symbol ``$\nabla$'' denotes now the gradient operator in dimensionless variable $\bfrho$.

In the above formulae: $\Hamiltonian_\mathrm{MD}$ describes the energy related to the curvature of the membrane, which is part of the membrane deformation Hamiltonian \eqref{secB:HelfrichModel} and, in  fact, it is the  Helfrich Hamiltonian expanded in small gradients of $h$ with vanishing surface tension; $\Hamiltonian_\mathrm{G}$ is the Hamiltonian of the Gaussian model, i.e., Eq.~\eqref{secB:GLModel} with $u=0$ and $c=0$; $\Hamiltonian_\mathrm{C}$, describing the coupling, originates from the second part of the membrane deformation Hamiltonian with the potential given by Eq.~\eqref{secB:coupling}; and $\Hamiltonian_\mathrm{I}$ is the rescaled version of  Eq.~\eqref{secB:inclusions}.

We note that in our model there is no external potential that keeps $h$ close to $0$. Instead, this condition is attained via the coupling term~\eqref{secD:HamiltonianC}, since the order parameter $\phi$ is kept close to zero (for $\tau >0$) by the term~\eqref{secD:HamiltonianG}. As we have checked, adding external potential proportional to $h^2\left(\bfr\right)$ is not changing basic properties of the system and, therefore, we do not include it in our model for the sake of simplicity.

We also note that in our model we allow the order parameters to take any real value. Physically, the height of the membrane above the reference plane is always restricted by some objects present in the system. Also, the excess thickness is bounded from below as the two layers of the membrane cannot intersect. Similarly, the concentration of one of the lipids $\phi$ is bounded by the finite values that describe membrane without or full of this lipid. Here we ignore this limits in order to solve the model analytically. As an \textit{a posteriori} justification of this assumption, we note that the boundary values of the order parameters are typically located far in the tail of the calculated Gaussian distributions and, therefore, the nonphysical values are highly improbable. 

\subsection{Method of calculation}\label{secC}




The partition function of the system is defined using the path integral over all configurations of the two fields
\begin{multline}
 \partf\left(\kappa, \tau, \mu, A; \left\{h_i, \bfrho_i\right\}_{i=1}^N\right)=\\
 \int \pathint h\left(\bfrho\right)\pathint \phi\left(\bfrho\right) \exp\left[-\beta\Hamiltonian\left(\kappa,\tau,\mu, A;\left\{h_i,\bfrho_i\right\}_{i=1}^N\right)\right].
\end{multline}
Since all terms in the Hamiltonian \eqref{secC:Hamiltonian} are quadratic in the order parameters and their derivatives, it is possible to calculate the partition function analytically --- we replace the fields $h\left(\bfrho\right)$ and $\phi\left(\bfrho\right)$ with their Fourier transforms and calculate the integrals separately for every wavevector. For the details of this procedure, as well as for the calculation of the correlation functions and order parameters profiles see  Appendix~\ref{appA}. 

\section{Correlation functions}\label{secD}


\subsection{Definition}

We start the investigation from studying the two--point correlation functions in the system without proteins. For such a system $N=0$, i.e., the Hamiltonian is
\begin{equation}\label{secD:Hamiltonian}
 \Hamiltonian=\Hamiltonian_\mathrm{MD}+\Hamiltonian_\mathrm{G}+\Hamiltonian_\mathrm{C},
\end{equation}
where the terms on the right hand side are given by Eq.~\eqref{secC:Hamiltonian}. Because of the symmetry of changing of sign of both order parameters, $\left<h\left(\bfrho\right)\right>=\left<\phi\left(\bfrho\right)\right>=0$ for any position $\bfrho$. Therefore, we define all the possible two--point correlation functions via
\begin{subequations}
\begin{align}
\corrfunc_{hh}\left(\rho;\kappa,\tau,\mu\right)&=\left<h\left(\bfrho_0\right)h\left(\bfrho_0+\bfrho\right)\right>,\\
\corrfunc_{h\phi}\left(\rho;\kappa,\tau,\mu\right)&=\left<h\left(\bfrho_0\right)\phi\left(\bfrho_0+\bfrho\right)\right>,\\
\corrfunc_{\phi\phi}\left(\rho;\kappa,\tau,\mu\right)&=\left<\phi\left(\bfrho_0\right)\phi\left(\bfrho_0+\bfrho\right)\right>.
\end{align}
\end{subequations}
Since the Hamiltonian \eqref{secD:Hamiltonian} is invariant under rotations and translations, these functions depend only on the length $\rho$ of the vector $\bfrho$ and they are independent of the reference point $\bfrho_0$.

\subsection{Integral formulae for the correlation functions}

Using path integral method, we obtain  the formulae for the correlation functions 
(see Appendix~\ref{appA:I}) 
\begin{subequations}\label{secD:integrals}
\begin{align}
 \nonumber &\corrfunc_{hh}\left(\rho;\kappa,\tau,\mu\right)=\\
 &\qquad\! \frac{1}{2\pi\kappa}\int_0^\infty \frac{x \left(x^2+\omega+2\tau\right) \BesselJ_0\left(\rho\, x\right)}{\left(x^4+1\right)\left(x^2+2\tau+\omega\right)-\omega}\, \dd x,\\
 \nonumber&\corrfunc_{h\phi}\left(\rho;\kappa,\tau,\mu\right)=\\
 &\qquad\! \frac{\mu}{2\pi}\int_0^\infty \frac{x\, \BesselJ_0\left(\rho\, x\right)}{\left(x^4+1\right)\left(x^2+2\tau+\omega\right)-\omega}\, \dd x,\\
 \nonumber&\corrfunc_{\phi\phi}\left(\rho;\kappa,\tau,\mu\right)=\\
  \label{secD:corrphiphi}&\qquad\! \frac{1}{2\pi}\int_0^\infty \frac{x \left(x^4+1\right) \BesselJ_0\left(\rho\, x\right)}{\left(x^4+1\right)\left(x^2+2\tau+\omega\right)-\omega}\, \dd x,
\end{align}
\end{subequations}
where $\BesselJ_i$ denotes the (unmodified) Bessel function of the first kind of order $i$, and $\omega=\kappa\mu^2$.

\subsection{Three zones}

\begin{figure}[t]
\begin{center}
 \includegraphics[width=0.45\textwidth]{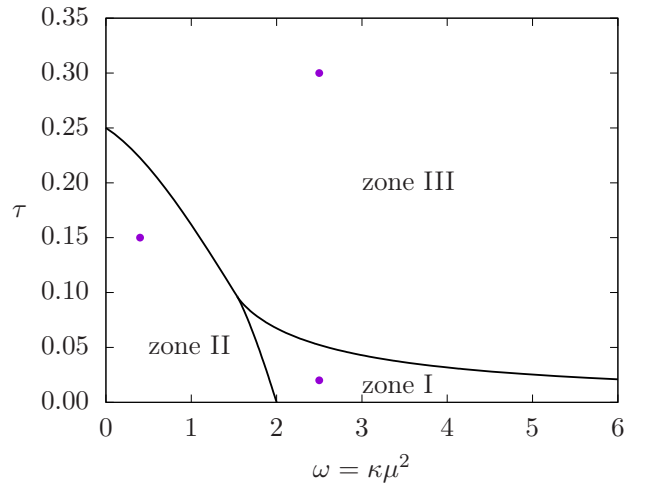}
\end{center}
\caption{\label{secD:figzones}Plot of the three zones in the space of parameters defined by the properties of roots of the polynomial $W\left(z;\tau,\kappa\mu^2\right)$, see Eq.~\eqref{secD:W}. The behaviour of the correlation functions for large $\rho$ is different in each zone. The blue dots denote the values of parameters used in Fig.~\ref{secD:figcorrfunc}.}
\end{figure}

The above integrals can be transformed to contour integrals on the complex plane and calculated using the residue theorem \cite{Lin2013}. The final formula depends on the form of complex roots of the polynomial in complex variable $z$
\begin{equation}\label{secD:W}
 W\left(z; \omega,\tau\right)=\left(z^4+1\right)\left(z^2+2\tau+\omega\right)-\omega,
\end{equation}
which is in the denominator of all the integrands in \eqref{secD:integrals}. Since $W\left(z;\omega,\tau\right)$ has real coefficients and only even powers of $z$, if $z_0$ is its root, then also $z_0^\ast$, $-z_0$ and $-z_0^\ast$ are the roots of $W$ (the symbol $z^\ast$ denotes the complex conjugate of a number $z$). This polynomial has no real roots.

The plane of the parameters $\left(\omega,\tau\right)$ is split into three different zones by the properties of the roots of the polynomial $W\left(z;\omega,\tau\right)$. In zone I the polynomial has only imaginary roots $\pm \iu t_1$, $\pm \iu t_2$ and $\pm \iu t_3$, where we assume $0<t_1<t_2<t_3$. In zone II there are two imaginary roots $\pm \iu t_1$ and four complex roots of a form $\pm a\pm \iu t_2$, with $a, t_1, t_2>0$ and $t_1<t_2$. Finally, in zone III the roots are of the same form as in zone II, $\pm \iu t_1$ and $\pm a\pm \iu t_2$ with $a, t_1, t_2>0$, but now $t_1>t_2$. The splitting of the parameter plane is presented in Fig.~\ref{secD:figzones}.

The behavior of the roots upon crossing the boundaries of the zones is discussed in Appendix~\ref{appB}.

\subsection{Explicit formulae in three zones}

Using the parameters describing roots of the polynomial $W$, we were able to calculate the integrals in Eq.~\eqref{secD:integrals} to obtain explicit formulae for the correlation functions.

In \textbf{zone I}, where all the roots of the polynomial \eqref{secD:W} are imaginary, the correlation functions are 
\begin{subequations}\label{secD:corrI}
\begin{align}
\nonumber&\corrfunc_{hh}\left(\rho;\kappa,\tau,\mu\right)=\frac{1}{2\pi\kappa}\left[
 \frac{\BesselKmod\left(\rho,t_1\right)\left(2\tau+\omega-t_1^2\right)}{\left(t_1^2-t_2^2\right)\left(t_1^2-t_3^2\right)}\right.\\
&\left.+\frac{\BesselKmod\left(\rho,t_2\right)\left(2\tau+\omega-t_2^2\right)}{\left(t_2^2-t_1^2\right)\left(t_2^2-t_3^2\right)}
+\frac{\BesselKmod\left(\rho,t_3\right)\left(2\tau+\omega-t_3^2\right)}{\left(t_3^2-t_1^2\right)\left(t_3^2-t_2^2\right)}
\right],\label{secD:corrIIIhh}\\
\nonumber&\corrfunc_{h\phi}\left(\rho;\kappa,\tau,\mu\right)=\frac{\mu}{2\pi}\left[
 \frac{\BesselKmod\left(\rho,t_1\right)}{\left(t_1^2-t_2^2\right)\left(t_1^2-t_3^2\right)}\right.\\ 
 &\quad\quad\left.
+\frac{\BesselKmod\left(\rho,t_2\right)}{\left(t_2^2-t_1^2\right)\left(t_2^2-t_3^2\right)}
+\frac{\BesselKmod\left(\rho,t_3\right)}{\left(t_3^2-t_1^2\right)\left(t_3^2-t_2^2\right)}
\right],\\
\nonumber&\corrfunc_{\phi\phi}\left(\rho;\kappa,\tau,\mu\right)=\frac{1}{2\pi}\left[
 \frac{\BesselK_0\left(\rho t_1\right)\left(t_1^4+1\right)}{\left(t_1^2-t_2^2\right)\left(t_1^2-t_3^2\right)}\right.\\
 &\quad\quad\left.
+\frac{\BesselK_0\left(\rho t_2\right)\left(t_2^4+1\right)}{\left(t_2^2-t_1^2\right)\left(t_2^2-t_3^2\right)}
+\frac{\BesselK_0\left(\rho t_3\right)\left(t_3^4+1\right)}{\left(t_3^2-t_1^2\right)\left(t_3^2-t_2^2\right)}
\right],
\end{align}
\end{subequations}
where we have introduced the function
\begin{equation}\label{secD:BesselKmod}
 \BesselKmod\left(\rho,t\right)=\begin{cases} \BesselK_0\left(\rho\, t\right) & \text{for } \rho>0,\\ -\ln t & \text{for } \rho=0,\end{cases}
\end{equation}
and $\BesselK_i$ is a modified Bessel function of the second kind of order $i$. Even though $\BesselKmod$ is not continuous in $\rho$, the correlation functions $\corrfunc_{hh}$ and $\corrfunc_{h\phi}$ are continuous at $\rho=0$ (the divergent terms in the expansion of Bessel functions cancel each other). The correlation function $\corrfunc_{\phi\phi}$ diverges logarithmically for small $\rho$
\begin{equation}\label{secD:phiphiasymptotics}
 \corrfunc_{\phi\phi}\left(\rho; \kappa, \tau, \mu\right)=-\frac{\ln \rho}{2\pi}+\mathrm{O}\left(1\right), \qquad \text{for }\rho\to 0,
\end{equation}
and the integral \eqref{secD:corrphiphi} is divergent for $\rho=0$. This properties of the correlation functions for $\rho\to 0$ are true in all zones; they are a reminiscence of the divergences present in the Gaussian model in two dimensions \cite{Goldenfeld1992}.

Each of the correlation functions in Eq.~\eqref{secD:corrI} consist of three terms. When $\rho$ is large these terms decay to zero exponentially with lengthscales, respectively, $1/t_1$, $1/t_2$ and $1/t_3$; and since in this zone $t_1<t_2<t_3$, the first term dominates over two other terms. Therefore, in this zone the correlation length $\xi=1/t_1$.

In \textbf{zone II} the correlation functions are
\begin{subequations}\label{secD:corrII}
 \begin{align}
 \nonumber &\corrfunc_{hh}\left(\rho;\kappa,\tau,\mu\right)=\frac{1}{2\pi \kappa}\left[\rule{0cm}{7mm}\frac{\BesselKmod\left(\rho, t_1\right)
  \left(2\tau+\omega-t_1^2\right)}{\left(t_1^2-t_2^2\right)^2+2a^2\left(t_1^2+t_2^2\right)+a^4}\right.\\
  &\quad\left.-\im\left(\frac{\BesselKmod\left(\rho,t_2+\iu a\right)\left[ \left(a-\iu t_2\right)^2+2\tau+\omega\right]}
  {2 a t_2 \left[t_1^2+\left(a-\iu t_2\right)^2\right]}\right)\right],\label{secD:corrIhh}\\
  \nonumber&\corrfunc_{h\phi}\left(\rho;\kappa,\tau,\mu\right)=\frac{\mu}{2\pi}\left[\rule{0cm}{7mm}\frac{\BesselKmod\left(\rho, t_1\right)}
  {\left(t_1^2-t_2^2\right)^2+2a^2\left(t_1^2+t_2^2\right)+a^4}\right.\\
  &\quad\quad \left.-\im\left(\frac{\BesselKmod\left(\rho,t_2+\iu a\right)}
  {2 a t_2 \left[t_1^2+\left(a-\iu t_2\right)^2\right]}\right)\right],\\
  \nonumber&\corrfunc_{\phi\phi}\left(\rho;\kappa,\tau,\mu\right)=\frac{1}{2\pi}\left[\rule{0cm}{7mm}\frac{\BesselK_0\left(\rho t_1\right)
  \left(t_1^4+1\right)}{\left(t_1^2-t_2^2\right)^2+2a^2\left(t_1^2+t_2^2\right)+a^4}\right.\\
  &\quad\quad \left.-\im\left(\frac{\BesselK_0\left[\rho\left(t_2+\iu a\right)\right]\left[\left(a-\iu t_2\right)^4+1\right]}{2 a t_2 \left[t_1^2+\left(a-\iu t_2\right)^2\right]}\right)\right],
 \end{align}
\end{subequations}
where the function $\BesselKmod\left(\rho,t\right)$ is given by \eqref{secD:BesselKmod}. Like in the previous case, the functions $\corrfunc_{hh}$ and $\corrfunc_{h\phi}$ are continuous for $\rho=0$ and $\corrfunc_{\phi\phi}$ diverges for $\rho\to 0$, see Eq.~\eqref{secD:phiphiasymptotics}.
\begin{figure*}[t]
\begin{center}
\begin{tabular}{ccc}
(a) & &\\
& \includegraphics[width=0.45\textwidth]{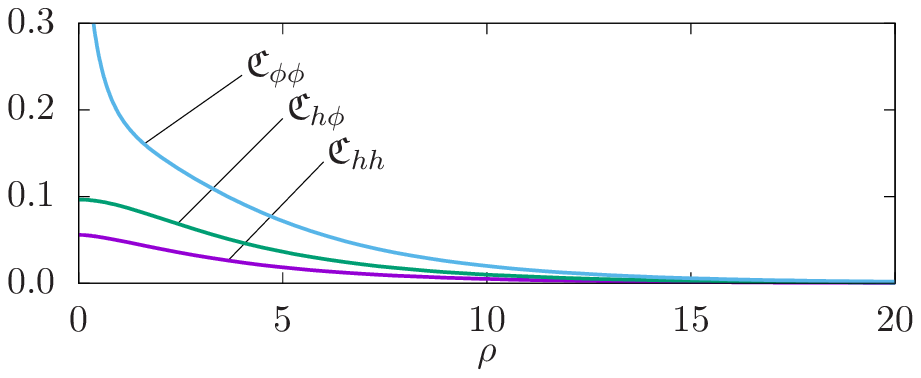} &\includegraphics[width=0.45\textwidth]{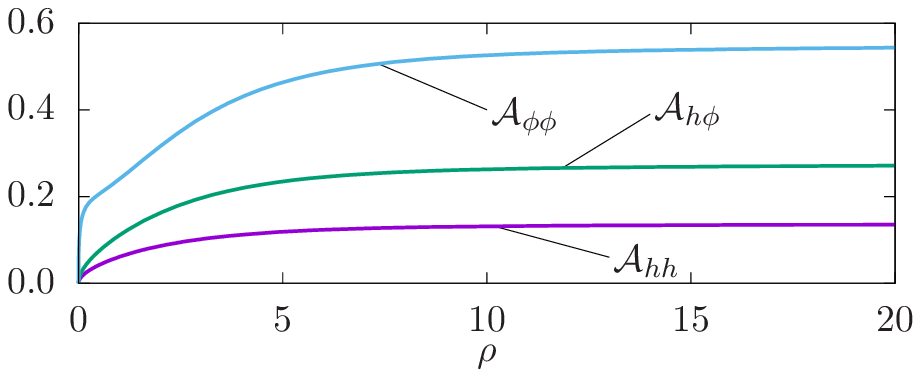}\\
(b) & & \\
& \includegraphics[width=0.45\textwidth]{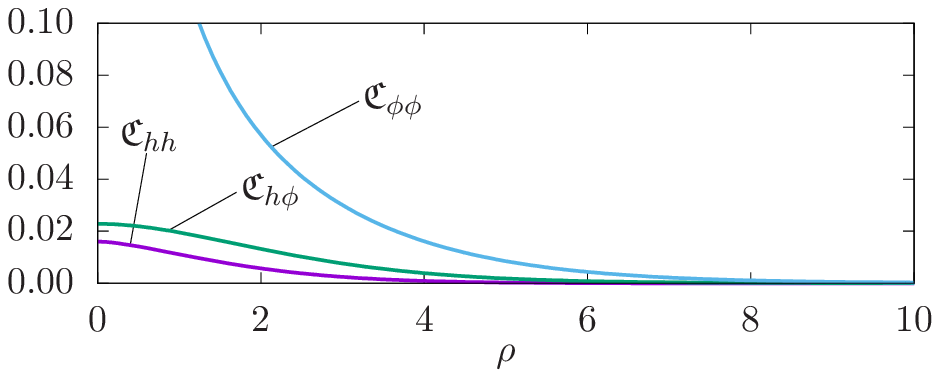}
& \includegraphics[width=0.45\textwidth]{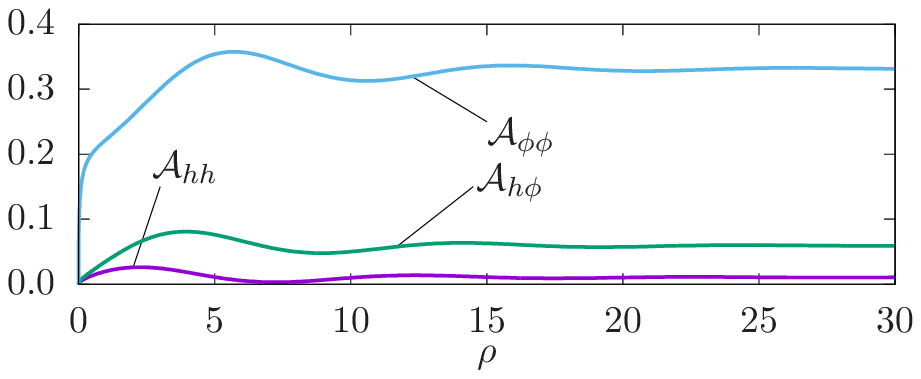}\\
(c) & &\\
& \includegraphics[width=0.45\textwidth]{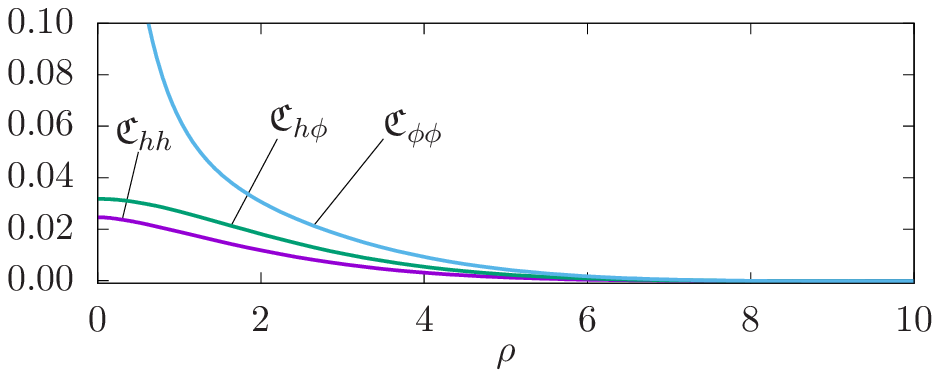}
& \includegraphics[width=0.45\textwidth]{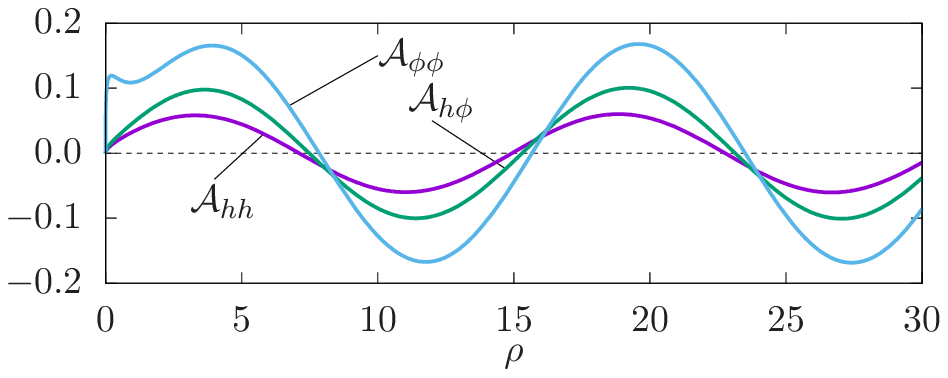}\\
\end{tabular}
\end{center}
\caption{\label{secD:figcorrfunc} Example plots of the correlation functions $\corrfunc_{hh}$, $\corrfunc_{h\phi}$ and $\corrfunc_{\phi\phi}$ and they amplitudes in different zones as functions of the distance $\rho$. (a) $\kappa=10$, $\tau=0.02$ and $\mu=0.5$ (zone I), (b) $\kappa=10$, $\tau=0.15$ and $\mu=0.2$ (zone II), (c) $\kappa=10$, $\tau=0.30$ and $\mu=0.5$ (zone III). The values of parameters used here have been marked with blue dots in Fig.~\ref{secD:figzones}. On each panel the plot on the left presents the correlation functions and the plot on the right their amplitudes (see Eq.~\eqref{secD:corrdecomposition}). The behavior of the amplitudes is different in each zone.}
\end{figure*}

For large $\rho$ the two terms present in all the formulae for the correlation functions \eqref{secD:corrII} decay to zero like $\exp\left(-t_1\rho\right)$ and $\exp\left(-t_2\rho\right)$, respectively. Because in this zone $t_1<t_2$, the first term dominates over the second one, and the correlation length $\xi=1/t_1$.

Finally, in \textbf{zone III} the correlation functions are given by Eq.~\eqref{secD:corrII}, the same as in the previously discussed zone II. The reason for this similarity is the same mathematical structure of the roots of the polynomial \eqref{secD:W} in these two zones. However, in this zone $t_1>t_2$ and, therefore, for large $\rho$ the second terms in the formulae~\eqref{secD:corrII} dominate; the correlation length $\xi$ is $1/t_2$. The difference between dominating terms justifies the distinction we have made between zones II and III.

It is convenient to decompose the correlation functions in all of the zones into a dominating decay and an amplitude
\begin{equation}\label{secD:corrdecomposition}
\corrfunc_{xx}\left(\rho; \kappa, \tau, \mu\right)= \mathcal{A}_{xx}\left(\rho; \kappa, \tau, \mu\right) \rho^{-1/2} \ee^{-\rho/\xi\left(\omega,\tau\right)},
\end{equation}
where ``$xx$'' denotes ``$hh$'', ``$h\phi$'' or ``$\phi\phi$'', $\mathcal{A}_{xx}$ are the amplitudes, and the correlation length
\begin{equation}\label{secD:xi}
\xi\left(\tau,\omega\right)=\begin{cases} 1/t_1\left(\tau,\omega\right) & \text{in zone I and II,}\\ 1/t_2\left(\tau,\omega\right) & \text{in zone III,} \end{cases}
\end{equation}
will be discussed in the next subsection.

The three amplitudes $\mathcal{A}_{xx}$ are zero for $\rho=0$ (in case of $\corrfunc_{\phi\phi}$, the divergence for $\rho\to 0$ is cured by the factor $\rho^{-1/2}$ present in Eq.~\eqref{secD:corrdecomposition}). The behavior of the amplitudes for $\rho>0$ is different in different zones. In zone I the amplitudes monotonically increase upon increasing $\rho$ and their value saturates at certain limiting values attained for $\rho\to\infty$. In zone II the amplitudes also have a well defined limit for $\rho\to\infty$ but they show some oscillations caused by the second term in each of the formulae~\eqref{secD:corrII}. These oscillations decay with a lengthscale $1/t_2$ and typically make the amplitude non--monotonic function of $\rho$. Only when $1/t_2$ is much smaller than $\xi$ the amplitude might stay monotonic but the oscillations are still visible. Finally, in zone III, where the oscillating term dominates  the amplitude has no limit for $\rho\to\infty$. Instead, it oscillates around zero. For large $\rho$ these oscillations have period $2\pi/a$, for smaller $\rho$ higher order terms perturb slightly the amplitudes. The full correlation functions (see \eqref{secD:corrdecomposition}) decay to zero exponentially in zones I and II, and like a damped oscillations in zone III.

The correlation functions, upon crossing the border of zones, show a smooth crossover between different asymptotic behaviors. Typically, two different length--scales become comparable or the period of oscillations diverges. This means that, close to the border, the size of the system necessary to observe characteristic behavior of the correlation functions becomes very large. We have not observed any phase transition associated with changing of the zones. Detailed analysis of the observed crossover is beyond the scope of this article.

The plots of the correlation functions and their amplitudes in different zones are presented in Fig.~\ref{secD:figcorrfunc}.

The phenomenon of different asymptotics of the correlation function is well--known in literature in the context of the theory of fluids; the line separating regions with exponential and damped oscillatory decay of the correlation functions (in our case borderline between the zone III and other zones) is called Fisher--Widom line \cite{Fisher1969, *Fisher2015, Evans1994, Janes2019}.

\subsection{Correlation length}\label{secD:E}

\begin{figure}[t]
\begin{center}
\begin{tabular}{ll}
(a) & \\
& \includegraphics[width=0.45\textwidth]{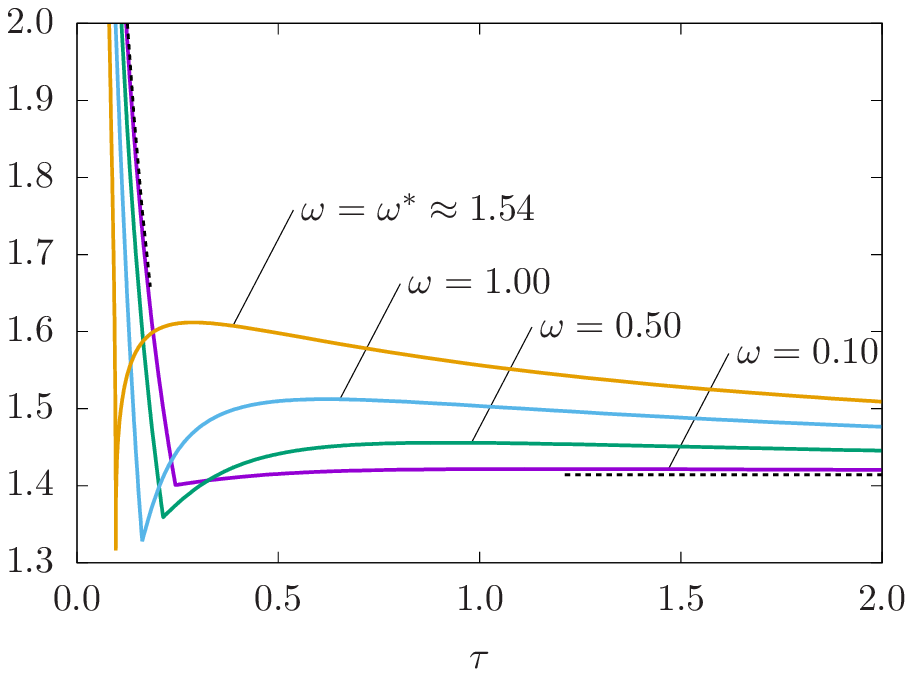} \\
(b) & \\
& \includegraphics[width=0.45\textwidth]{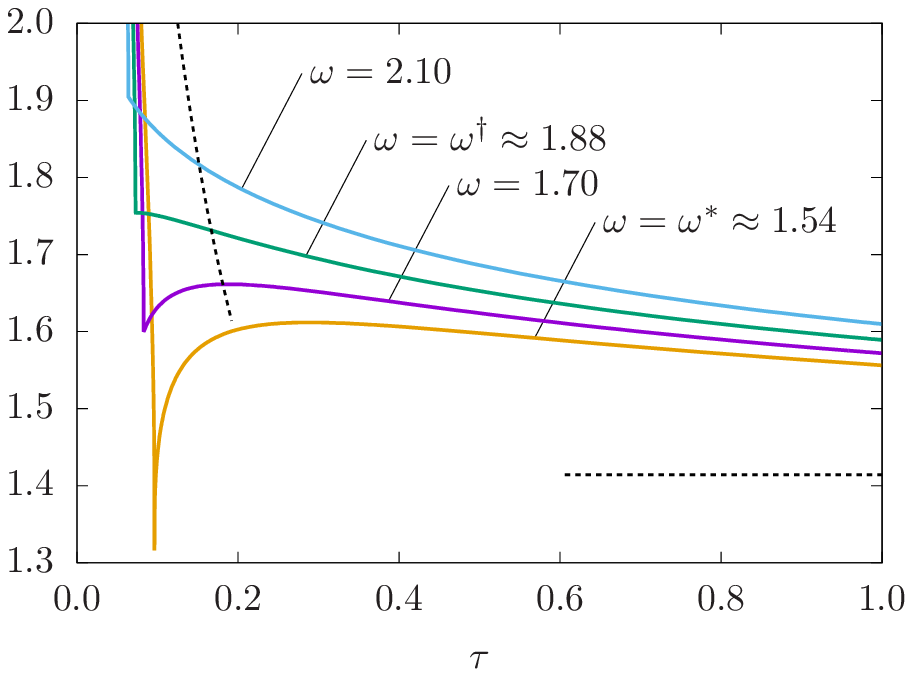}
\end{tabular}
\end{center}
\caption{\label{secD:figxi}Correlation length $\xi$ as a function of $\tau$ for different values of $\omega$. (a) For $\omega<\omega^\ast$ the minimum gets deeper upon increasing $\omega$; (b) for $\omega>\omega^\ast$ the value in the minimum increases upon increasing $\omega$ and, eventually, disappears for $\omega=\omega^\dagger$. The dashed lines denote the asymptotic behavior of the correlation length (which is independent of $\omega$): $\left(2\tau\right)^{-1/2}$ for small $\tau$, and $\sqrt{2}$ for large $\tau$.}
\end{figure}

In this subsection we discuss the properties of the correlation length $\xi\left(\tau,\omega\right)$ given by Eq.~\eqref{secD:xi} that describes the exponential decay of all the correlation functions.

Even though the correlation length $\xi$ has been defined separately in each of the zones, it is a continuous function of its parameters; upon crossing the border between zone I or II and zone III the derivative $\partial\xi/\partial \tau$ is discontinues and can even be divergent. On the contrary, there is no nonanalyticity associated with crossing the border between zone I and zone II, which is in line with the properties of the roots of the polynomial, see Fig.~\ref{secD:figroots}(b). We stress that the nonanalyticity appears only in the correlation length, it not present for the correlation functions.

For large values of the reduced temperature $\tau$
\begin{equation}\label{secD:xitauinf}
 \xi\left(\tau,\omega\right)=\sqrt{2}+\mathrm{O}\left(1/\tau\right), \quad \text{for }\tau\to\infty.
\end{equation}
This result agrees with the correlation length (measured in units $\lsc$, see \eqref{secC:lengthscale}) reported for a membrane without the composition order parameter $\phi\left(\bfrho\right)$ \cite{Bihr2015}. This limit is discussed in Sec.~\ref{secH:A}.

When the reduced temperature is close to zero, we have
\begin{equation}
 \xi\left(\tau,\omega\right)=\left(2\tau\right)^{-1/2}+\mathrm{O}\left(\tau^{1/2}\right),
\end{equation}
which means that for $\tau\to 0$ our model becomes critical and $\xi$ diverges with the critical exponent $\nu=1/2$. This value is characteristic for the Gaussian model. The properties of the correlation functions in this limit are presented in Sec.~\ref{secH:B}. 

In Fig.~\ref{secD:figxi} we present the plots of the correlation length $\xi$ as a function of $\tau$ for several fixed values of $\omega$. When $\omega$ is small, for small $\tau$ the correlation function is a decreasing function of $\tau$ and has a minimum exactly when the parameters are on the border between zone II and zone III (see Fig.~\ref{secD:figzones}). For this special value of $\tau$, $\xi$ is non--analytic and the derivative $\partial \xi/\partial \tau$ jumps from a finite negative value to a finite positive value. For larger values of $\tau$ the correlation length, upon increasing $\tau$, first increases, has a shallow maximum and decreases to the asymptotic value $\sqrt{2}$. Upon increasing $\omega$ (for small values of $\omega$, see Fig.~\ref{secD:figxi}(a)) the shape of the correlation function plotted as a function of $\tau$ changes only slightly. The value of $\tau$ for which there is a minimum is slowly decreasing (following the border between zones II and III) and gets deeper. At the same time the jump of the derivative $\partial \xi/\partial \tau$ in the minimum is increasing and the value in the minimum is decreasing.

The above picture changes when, upon increasing $\omega$, $\omega=\omega^\ast=8/\left(3\sqrt{3}\right)\approx 1.54$ is reached, i.e., the value in the point where all three zones meet. In the minimum (observed for $\tau=\tau^\ast$) the derivative $\partial \xi/\partial \tau$ is $-\infty$ from the left side and $+\infty$ from the right side. In this special point the correlation function has the smallest possible value
\begin{equation}
\xi_\text{min}=\xi\left(\tau^\ast,\omega^\ast\right)=3^{1/4}\approx 1.32 \quad \text{(in units of $\lsc$)}.
\end{equation}

Upon further increasing of $\omega$ (see Fig.~\ref{secD:figxi}(b)), the minimum moves towards smaller values of $\tau$, following the border between zones I and III. The derivative $\partial \xi/\partial \tau$ stays (minus) infinite from the left side, but the right side derivative is finite and decreasing. For $\omega=\omega^\dagger=\frac{4}{5}\left(10-2\sqrt{5}\right)^{1/2}\approx 1.88$ the right side derivative changes its sign; for $\omega\geqslant\omega^\dagger$ the correlation length $\xi$ does not have a minimum and, as a function of $\tau$, it monotonically decreases. At the border between zone I and III there is still a point of nonanalyticity with an infinite left side derivative $\partial \xi/\partial \tau$.

These properties of the correlation length $\xi$ can potentially be used to experimentally estimate the value of $\omega$ and, thus, the coupling between order parameters $\alpha$. The value of $\tau$ for which $\xi$ is non--analytic and (for $\omega<\omega^\dagger$) has a minimum is uniquely related to the value of $\omega$.

\subsection{Correlation functions in different limiting cases}

We conclude the analysis of the correlation functions in the model by studying their behavior in various limiting cases. In order to keep the text compact, the detailed analysis is reported in Appendix~\ref{appC}; here we provide only the most important results.

When the reduced temperature is big ($\tau\to\infty$ limit) the composition order parameter $\phi$ becomes negligible and thus the correlation functions $\corrfunc_{\phi\phi}$ and $\corrfunc_{h\phi}$ go to zero. In this limit only order parameter $h$ is relevant and the results known for the membrane deformation model are recovered.

When, on the other hand, the reduced temperature is small ($\tau\to 0$ limit) the system becomes critical and all the correlation functions are proportional to a single scaling function, the same as in the Gaussian model. Surprisingly, only the formula for $\corrfunc_{\phi\phi}$ is strictly universal, the other correlation functions still depend on the parameter $\mu$ describing the strength of the coupling.

In the case of weak coupling between order parameters ($\mu\to 0$ limit) the correlation function $\corrfunc_{h\phi}$ is vanishing. Surprisingly, the two order parameters stay coupled for and non--zero $\mu$ and decay with the same correlation length. Only when $\mu=0$ the decay for $\corrfunc_{hh}$ and $\corrfunc_{\phi\phi}$ happens on a different lengthscales.

When the coupling between the order parameters is strong ($\mu\to \infty$) the correlation length diverges and the system is again critical. We identify this limit with critical roughening present in the membrane deformation model without external potential. All the correlation functions can be described with a single scaling function.

We have also considered a limit of $\gamma\to\infty$ in the coupling term in the Hamiltonian \eqref{secB:coupling}. As we show in Appendix~\ref{appC:V}, it is equivalent to the limit of both $\tau\to 0$ and $\mu \to\infty$ with $\tau\mu^2$ fixed. Even though any of this limits alone implies criticality, when they are applied together the system in not critical anymore. In fact it can be mapped to a membrane deformation model with both elastic and surface energy included.

\section{Order parameter profiles and size of induced domains}\label{secE}

\begin{table*}
\caption{\label{secE:radiiasymp} Asymptotic behavior of six effective radii for small and large effective temperature $\tau$ and the coupling constant $\mu$ (the bending stiffness $\kappa$ is assumed to be constant). The table presents only leading order term in a given limit, the limits $\tau\to 0, \infty$ are calculated for fixed $\mu$, and the limits $\mu\to0,\infty$ are for fixed $\tau$. The special values $\rad^\ast$ and $\rad^\dagger$ are defined in Eq.~\eqref{secE:specrad}.}
\begin{ruledtabular}
\begin{tabular}{cccccc}
radius & definition of effective radius &  $\tau\to 0$ &  $\tau\to\infty$ &  $\mu\to 0$ & $\mu\to\infty$\\
\hline
$\rad_\Gamma$ & excess adsorption & $\sqrt{2}\left|\tau\ln\tau\right|^{-1/2}$ &  $\sqrt{8/\pi}\approx 1.596$   & $\tau$--dependent constant &  $\sqrt{8/\pi}\ \left[\kappa\mu^2/\left(2\tau\right)\right]^{1/4} $ \\
$\rad_V$ & excess volume & $\sqrt{2}\left|\tau\ln\tau\right|^{-1/2}$ & $\sqrt{8/\pi}\approx 1.596$ & $\sqrt{8/\pi}\approx 1.596$ &$\sqrt{8/\pi}\ \left[\kappa\mu^2/\left(2\tau\right)\right]^{1/4} $\\
$\rad^\phi_{\text{infl}}$ & inflection point of $\left<\phi\right>$ & $\omega$--dependent constant & $\rad^\ast\approx 0.8096$ & $\tau$--dependent constant & $\rad^\ast \left[\kappa\mu^2/\left(2\tau\right)\right]^{1/4}$ \\
$\rad^h_{\text{infl}}$ & inflection point of $\left<h\right>$ & $\omega$--dependent constant & $\rad^\ast\approx 0.8096$ & $\rad^\ast\approx 0.8096$ & $\rad^\ast \left[\kappa\mu^2/\left(2\tau\right)\right]^{1/4}$\\
$\rad_{1/2}^\phi$ & half of maximal value of $\left<\phi\right>$ & $f_1\left(\omega\right)\tau^{-1/4}$ & $\rad^\dagger\approx 1.302$ & $\tau$--dependent constant & $\rad^\dagger\left[\kappa\mu^2/\left(2\tau\right)\right]^{1/4}$\\
$\rad_{1/2}^h$ & half of maximal value of $\left<h\right>$ & $f_2\left(\omega\right)\tau^{-1/4}$ & $\rad^\dagger\approx 1.302$ & $\rad^\dagger\approx 1.302$ & $\rad^\dagger\left[\kappa\mu^2/\left(2\tau\right)\right]^{1/4}$
\end{tabular}
\end{ruledtabular}
\end{table*}

Now we introduce a single anchor into the membrane, i.e., we put $N=1$ in Eq.~\eqref{secC:Hamiltonian}, and  discuss the resulting order parameter profiles. For simplicity, we assume that the inclusion is located in the origin ($\bfrho_1=\bm{0}$), and thus the system has a rotational symmetry with respect to the origin. As a result, the equilibrium values of the order parameters depend on the distance from the origin $\rho$, and not on the exact position $\bfrho$. In Appendix \ref{appA:II}, using the method based on path integrals, we show that
\begin{subequations}\label{secD:profiles}
\begin{align}
\left<h\left(\rho\right)\right>&=h_1 \corrfunc_{hh}\left(\rho;\kappa, \tau, \mu\right)/\corrfunc_{hh}\left(0;\kappa, \tau, \mu\right),\\ \label{secD:profilesphi}
\left<\phi\left(\rho\right)\right>&=h_1 \corrfunc_{h\phi}\left(\rho;\kappa, \tau, \mu\right)/\corrfunc_{hh}\left(0;\kappa, \tau, \mu\right),
\end{align}
\end{subequations}
where $h_1$ denotes the value of the order parameter $h$ imposed on the membrane in the pinning point $\bfrho_1=\bfz$. Depending on the interpretation of $h$, this can be either fixed excess thickness that matches the hydrophobic mismatch of the anchoring protein or fixed position of the membrane imposed by the anchor attached to the cytoskeleton.
The formulae for the order parameters profiles $\left<h\left(\rho\right)\right>$ and $\left<\phi\left(\rho\right)\right>$ are equal to properly rescaled correlation functions $\corrfunc_{hh}$ and $\corrfunc_{h\phi}$, respectively. This implies that the order parameters have a different asymptotic behavior in the three zones identified in Sec.~\ref{secD}. 

Here, we skip the detailed analysis of $\left<h\left(\rho\right)\right>$ and $\left<\phi\left(\rho\right)\right>$ in different regimes as this would only repeat the discussion of the correlation functions presented in Sec.~\ref{secD}. Moreover, the plots of the profiles in different zones have already been presented in \cite{Stumpf2021}.

When the excess thickness  of the membrane is fixed to some nonzero value in the origin, the order parameter $h\left(\bfrho\right)$ is also nonzero in the region surrounding the pinning point, and, due to the coupling between order parameters, also $\phi\left(\bfrho\right)$ is nonzero there. Of course, the further away from the origin we go, the smaller the magnitudes of the order parameters are. We identify the perturbation in the order parameters caused by the presence of the anchor  with experimentally observed so--called domains of lipids that form around the pinning points \cite{Honigmann2014, Stumpf2021}. In order to facilitate  the comparison between our model and experimental systems, it is necessary to introduce an effective size $\rad$ of the induced domain. We note that in our model, the membrane is always above the critical temperature, where there is only a single bulk phase (with equilibrium average of both order parameters equal zero); therefore the observed perturbations cannot be the domains in the strict sense of the meaning. This explains why there is no straightforward, unique definition of the radius of such a domain. In this section we discuss six possible ways to define such a quantity.

\subsection{Effective radii based on integrated order parameters}

We define  the excess adsorption $\exads$ and the excess volume $\exvol$ via
\begin{subequations}
\begin{align}\label{secE:exadsdef}
    \exads\left(\kappa, \mu, \tau, \lambda_1\right)&=2\pi \int_0^\infty \left<\phi\left(\rho\right)\right>\rho\, \dd \rho,\\
    \exvol\left(\kappa, \mu, \tau, \lambda_1\right)&=2\pi \int_0^\infty \left<h\left(\rho\right)\right>\rho\, \dd \rho,
\end{align}
\end{subequations}
i.e., the area integral of the composition and thickness order parameter profiles. Excess adsorption is proportional to the additional amount of the component of the membrane preferred by the anchor that has gathered around it, and excess volume is equal to the additional volume (measured in $\lsc^3$ unit) of the membrane due to the hydrophobic mismatch of the anchor. We note that both quantities can be positive or negative: the sign of $\phi$, and hence $\exads$, determines the type of lipids that the anchor effectively prefers, whereas positive (negative) $\exvol$ corresponds to the hydrophobic core of the anchor larger (smaller) than the preferred distance between the leaflets of an unperturbed membrane. Using Eq.~\eqref{secD:profiles}, after some algebra, we get 
\begin{subequations}
\begin{align}
\label{secE:exads}    \exads\left(\kappa, \mu, \tau, \lambda_1\right)&=\frac{h_1 \mu}{2 \tau \corrfunc_{h\phi}\left(0; \kappa, \tau, \mu\right)},\\
\label{secE:exvol}    \exvol\left(\kappa, \mu, \tau, \lambda_1\right)&=\frac{h_1 \left(2\tau+\kappa\mu^2\right)}{2 \tau \kappa \corrfunc_{hh}\left(0; \kappa, \tau, \mu\right)}.
\end{align}
\end{subequations}
Both quantities are continuous functions of all their parameters and they are linear in the thickness of the membrane $h_1$ imposed by the inclusion. For small $\tau$ they have a divergence of a type $\left(\tau\ln\tau\right)^{-1}$. For $\tau\to\infty$, the excess adsorption $\exads$ decays to zero like $\tau^{-1}$, while the excess volume $\exvol$ has a finite limit $8h_1$.

 From definition \eqref{secE:exadsdef} and Eq.~\eqref{secD:profilesphi}
 it follows that, upon approaching the critical point, the excess adsorption is determined by the integral of the mixed correlation function $\corrfunc_{h\phi}$.  We recall that for systems described by a single order parameter, the integral of the (dimensionless) correlation function is equal to the susceptibility, and diverges as $\sim \left|t\right|^{-\gamma}$, where $t=\left(T-\Tc\right)/\Tc$ is the reduced temperature and $\gamma$
 is the critical exponent \cite{Hanke1999}. For the Gaussian model $\gamma=1$ and 
 logarithmic correction are expected, and  $\tau\sim t$, hence our results are in agreement with this general law.  Moreover, the same law applies to $\exvol$ which is based on the order parameter $h$, which shows that, due to the coupling, in the limit $\tau\to 0$ the critical behavior is relevant for both order parameters, as we have observed in Sec.~\ref{secH:B}.

The excess adsorption can be used to define the effective size of the induced domain. For a given set of parameters $\kappa$, $\mu$, $\tau$ and $h_1$, we consider a simplified circular deformation of the membrane of the radius $\rad_\Gamma$ with a constant chemical composition inside equal to $\left<\phi\left(0\right)\right>$, a zero composition order parameter outside, and the same excess adsorption as the real deformation. Using \eqref{secE:exads} we get
\begin{subequations}
\begin{equation}
    \rad_\Gamma\left(\kappa, \mu, \tau\right)=\left(\frac{\mu}{2\pi \tau \,\corrfunc_{hh}\left(0;\kappa,\tau,\mu\right)}\right)^{1/2}.
\end{equation}
Similarly, one can define the effective radius $\rad_V$ as the radius of the circular deformation of fixed thickness $h_1$ and the same volume as the real deformation:
\begin{equation}
    \rad_V\left(\kappa, \mu, \tau\right)=\left(\frac{2 \tau+\kappa\mu^2}{2\pi \tau \kappa \,\corrfunc_{hh}\left(0;\kappa,\tau,\mu\right)}\right)^{1/2}.
\end{equation}


\begin{figure}
\begin{center}
    \begin{tabular}{cc}
    (a) & \\
    & \includegraphics[width=0.4\textwidth]{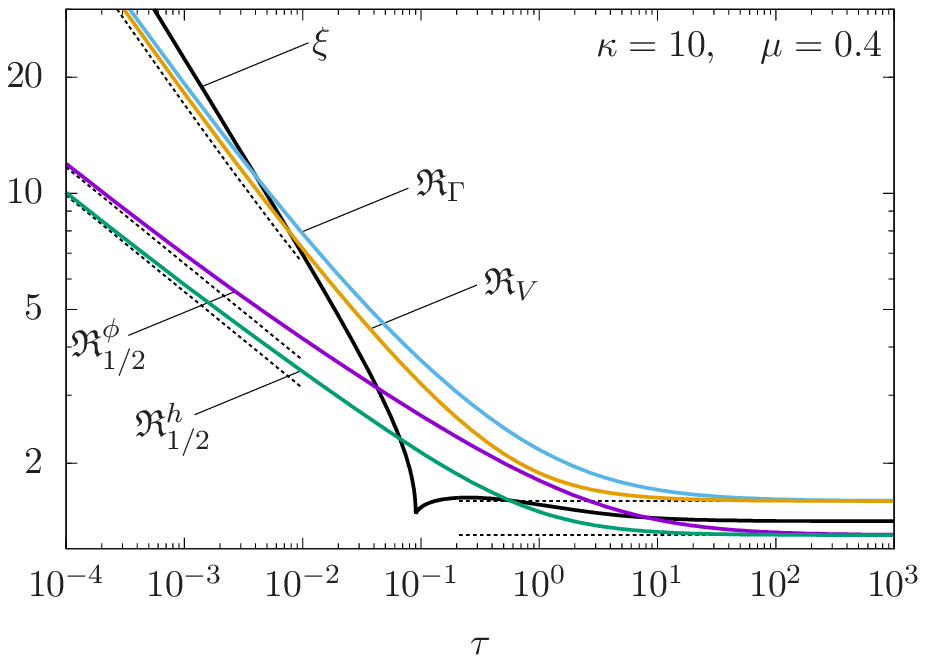}\\
    (b) & \\
    & \includegraphics[width=0.4\textwidth]{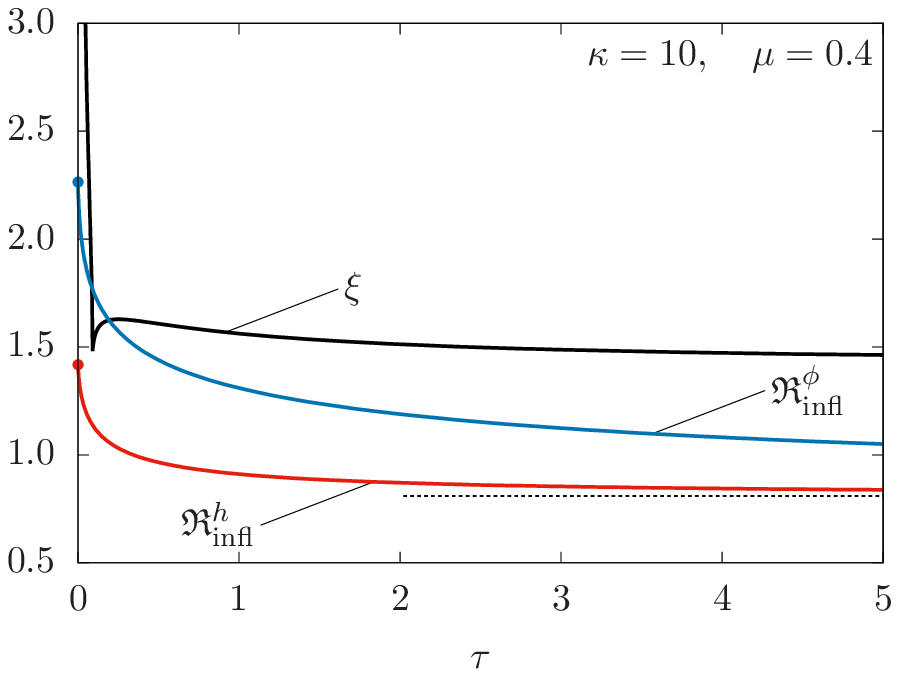}\\
    (c) &  \\
    & \includegraphics[width=0.4\textwidth]{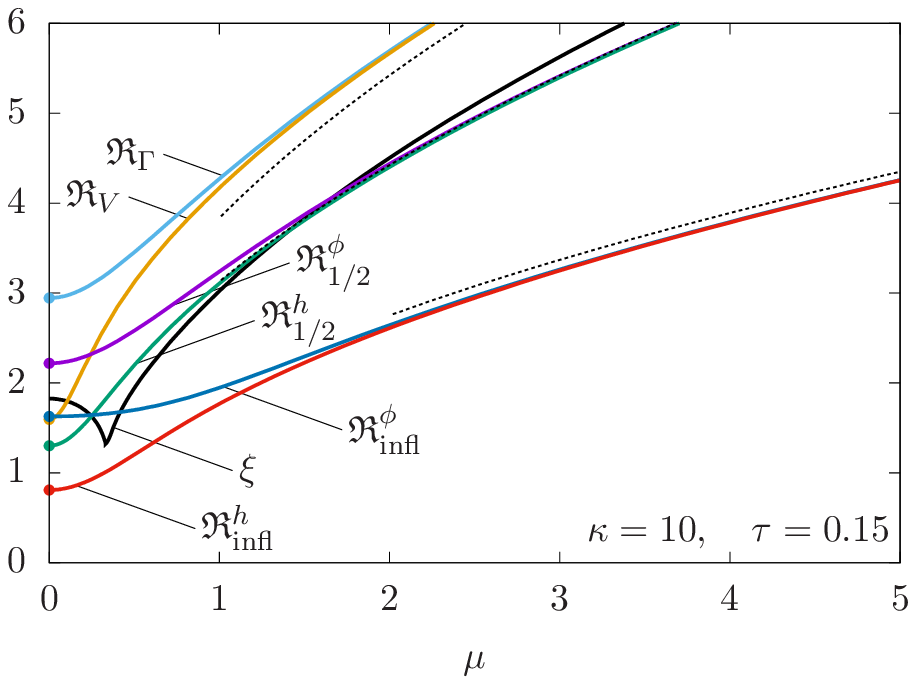}
    \end{tabular}
\end{center}
    \caption{Plots of the effective radii $\rad_\Gamma$, $\rad_V$, $\rad^\phi_\text{infl}$, $\rad^h_\text{infl}$, $\rad^\phi_{1/2}$, and $\rad^h_{1/2}$ describing the size of the domain induced by the defect. (a) and (b) Plot of six radii as functions of $\tau$ for $\kappa=10$ and $\mu=0.4$. (c) Plot of the radii as a function of $\mu$ for fixed $\kappa=10$ and $\tau=0.15$. In all panels, the dashed lines denote the asymptotic behaviour for $\tau$, $\mu\to 0$, $\infty$, and the dots for $\tau=0$ and $\mu=0$ denote the finite limits of the radii; cf.~Table \ref{secE:radiiasymp}.}
    \label{secE:figradii}

\end{figure}

\subsection{Effective radii based on local properties of the order parameters}

Another possible definitions of the effective radius are based on local properties of the order parameters. When observing the thickness of the membrane or its chemical composition it is natural to place the border of the deformation in the place where the observed quantity is changing in the fastest way, i.e., where the derivative is the biggest --- in the inflection point. Therefore, we introduce two effective radii $\rad^\phi_\text{infl}$ and $\rad^h_\text{infl}$ based on the inflection points:
\begin{align}
    \left.\frac{\partial^2 }{\partial \rho^2}\left<\phi\left(\rho\right)\right>\right|_{\rho=\rad^\phi_\text{infl}\left(\kappa, \mu, \tau\right)}&=0,\\
    \left.\frac{\partial^2 }{\partial \rho^2}\left<h\left(\rho\right)\right>\right|_{\rho=\rad^h_\text{infl}\left(\kappa, \mu, \tau\right)}&=0,
\end{align}
where, in the case of zone III, one has to take the smallest positive $\rho$ fulfilling the above condition. Unlike $\rad_\Gamma$ and $\rad_V$, these two radii are based on local properties of the order parameter which makes their analytical analysis more challenging.

Finally, the effective size of the domain can be defined as a distance at which the magnitude of the order parameter reaches half of its value in the center of the domain
\begin{align}
    \left<\phi\left(\rad^\phi_{1/2}\left(\kappa, \mu, \tau\right)\right)\right>&=\frac{1}{2}\left<\phi\left(0\right)\right>,\\
    \left<h\left(\rad^h_{1/2}\left(\kappa, \mu, \tau\right)\right)\right>&=\frac{1}{2}\left<h\left(0\right)\right>.
\end{align}
\end{subequations}
This definition is especially handful when working with microscope images of experimental system.

We note that all six radii introduced here, do not depend on the thickness $h_1$ of the membrane set by an inclusion. For $\rad_\Gamma$ and $\rad_V$ changing of $h_1$ is changing the reference value of the relevant order parameter at $\bfrho=\bf{0}$ together with the excess adsorption and the excess volume, and for $\rad^\phi_\text{infl}$, $\rad^h_\text{infl}$, $\rad^\phi_{1/2}$, and $\rad^h_{1/2}$ varying $h_1$ changes the magnitude of the order parameters, but does not  shift the inflection point or the midpoint.

\subsection{Comparison of effective radii in different limiting cases}

We have calculated rigorously the asymptotic behavior of all the radii for large and small values of their parameters --- the results are summarised in Table~\ref{secE:radiiasymp}. 

For $\tau\to 0$ the values of $\rad_\Gamma$ and $\rad_V$ diverge in exactly the same, independent of the value of $\omega=\kappa\mu^2$ way $\sim\left|\tau\ln\tau^{1/2}\right|$. The divergence is slower than the divergence of the correlation length $\xi\sim \tau^{-1/2}$ only by a logarithmic factor --- this seems to be a manifestation of the universality principle. The radii $\rad^\phi_{1/2}$ and $\rad^h_\text{1/2}$ also diverge for $\tau\to 0$ but like $\tau^{-1/4}$, i.e., much slower than $\xi$,  $\rad_\Gamma$ and $\rad_V$, and with $\omega$--dependent amplitude. In contrast, radii $\rad^\phi_\text{infl}$ and $\rad^h_\text{infl}$ in the limit $\tau\to 0$ approach two different, finite, $\omega$--dependent values. We note that, for $\tau=0$ our model is not well--defined, and therefore these limiting values cannot be reached. These results imply that for small values of $\tau$
\begin{multline}\label{secE:radsmalltau}
    \xi \gg \rad_\Gamma >\rad_V \gg \rad^\phi_{1/2}>\rad^h_\text{1/2}\gg\rad^\phi_\text{infl}>\rad^h_\text{infl},\\ \text{for }\tau\to 0.
\end{multline}
The example behavior of the radii for small $\tau$ is presented in Fig.~\ref{secE:figradii}(a) and (b).

In the limit $\tau\to\infty$ all radii approach finite, non--zero, $\omega$--independent values. In this limit
\begin{multline}
    \rad_\Gamma \gtrapprox \rad_V \gtrapprox \sqrt{8/\pi}>\xi\gtrapprox \sqrt{2} > \rad^\phi_{1/2}\gtrapprox\rad^h_\text{1/2}\\
    \gtrapprox \rad^\dagger>\rad^\phi_\text{infl}\gtrapprox\rad^h_\text{infl}\gtrapprox \rad^\ast,\qquad \text{for }\tau\to \infty,
\end{multline}
where
\begin{subequations}\label{secE:specrad}
\begin{align}
    \kei\left(\rad^\dagger\right)&=\frac{1}{2}\kei\left(0\right), & \rad^\dagger&\approx 1.302,\\
    \label{secE:radast}\left.\frac{\dd^2}{\dd \rho^2} \kei\left(\rho\right)\right|_{\rho=\rad^\ast}&=0, & \rad^\ast&\approx 0.8096,
\end{align}
\end{subequations}
(where $\rad^\ast$ is the smallest positive solution of \eqref{secE:radast}). The behavior of radii for large $\tau$ is illustrated in Fig.~\ref{secE:figradii}(a) and (b).

When $\mu\to 0$ (which for fixed $\kappa$ is equivalent to $\omega\to 0$) all radii approach finite values. For radii based on chemical composition $\phi$, this limit depends on $\tau$, while for radii based on $h$ it does not depend on any parameter, see Table~\ref{secE:radiiasymp}. As presented in Fig.~\ref{secE:figradii}(c), upon reducing $\mu$, depending on the values of other parameters, the relation between the radii can change. 

Finally, in the limit $\mu\to\infty$ all radii are proportional to $\left[\kappa\mu^2/\left(2\tau\right)\right]^{1/4}\sim \mu^{1/2}$. In this limit (like in the case of large values of $\tau$), we observe that the radii group into pairs that become asymptotically equal
\begin{multline}
    \rad_\Gamma,\rad_V\approx \sqrt{8/\pi} \left(\frac{\kappa\mu^2}{2\tau}\right)^{1/4} > \xi\approx \sqrt{2} \left(\frac{\kappa\mu^2}{2\tau}\right)^{1/4}\\ >\rad^\phi_{1/2}, \rad^h_\text{1/2}\approx \rad^\dagger \left(\frac{\kappa\mu^2}{2\tau}\right)^{1/4} \\>\rad^\phi_\text{infl},\rad^h_\text{infl}\approx \rad^\ast \left(\frac{\kappa\mu^2}{2\tau}\right)^{1/4},\qquad \text{for }\mu\to \infty,
\end{multline}
where we have used Eq.~\eqref{secH:xilargemu}. The above behavior of the radii has been illustrated in Fig.~\ref{secE:figradii}(c). Surprisingly, the amplitudes multiplying the dominant divergence $\left[\kappa\mu^2/\left(2\tau\right)\right]^{1/4}$ are identical to the limiting values of the radii for $\tau\to\infty$. We also note that, depending on the radius and parameters, the asymptotic formula can be approached both from below and above (which is in contrast with the limit $\tau\to \infty$, where the limiting value was always approached from above). 

The above rigorous analysis has been supported by numerical calculation of all the radii. We have not noted any non--analytical behavior of the radii upon crossing the borders of the three regimes discussed in Sec.~\ref{secD}, which is different from the behavior of $\xi$. In zone III, all the radii are of the same order as the correlation length $\xi$, while in zones I and II, in the limit $\tau\to 0$, the radii and the correlation length can be significantly different (cf.~Eq.~\eqref{secE:radsmalltau}). 
For all tested values of parameters, we have observed that all radii are decreasing upon increasing $\tau$ and they are increasing upon increasing $\mu$. We have also checked that
\begin{multline}
    \rad_\Gamma>\rad_V>\sqrt{8/\pi},\quad \rad^\phi_{1/2}>\rad^h_{1/2}>\rad^\dagger,\\ \text{and}\quad \rad^\phi_\text{infl}>\rad^h_\text{infl}>\rad^\ast,
\end{multline}
where $\rad^\dagger$ and $\rad^\ast$ are defined in Eq.~\eqref{secE:specrad}. For most of the tested values of parameters we have also observed $\rad_V>\rad^\phi_{1/2}$ and $\rad^h_{1/2}>\rad^\phi_\text{infl}$, however, when both $\tau$ and $\mu$ are small this relation does not hold.

Finally we note, that each of $\rad_V$, $\rad^h_{1/2}$, and $\rad^h_\text{infl}$ has the same limiting value for $\tau\to\infty$ and for $\mu\to 0$ (equal, respectively, to $\sqrt{8/\pi}$, $\rad^\dagger$, and $\rad^\ast$). Closer investigation shows that for small values of $\mu$ these functions are almost constant, except for a small region around $\tau=0$, for which their value is significantly higher. Upon reducing $\mu$ the size of this region is decreasing. Such a behavior suggests the existence of a scaling limit for $\mu\to 0$ and $\tau\to 0$. The detailed analysis of the model in this limit and the discussion of the physical relevance of the membrane with strongly fluctuating composition order parameter weakly coupled to spacial degrees of freedom goes beyond the scope of the manuscript.

\section{Discussion}\label{secG}

The main goal of this paper was to investigate the structure properties of a simple  model that couples  thickness deformations  of the two--component lipid membrane  to its composition. This coupling  is relevant in the context of domain formation in cell membranes and model membranes, where it has been observed that  membrane  lipids segregate near anchors linking  the membrane to the cytoskeletal actin filaments. Typically, these anchors  have a hydrophobic part with a thickness slightly different from that of the hydrophobic part of the membrane. Due to this hydrophobic mismatch, the hydrophobic core of the membrane locally deforms, effectively attracting to this region lipids with appropriate length of the hydrophobic part.

The advantage of our model is that it can be solved exactly. 
Using the path integral approach, we have calculated analytically  correlation functions of all three pairs of order parameters. Our model has three independent (dimensionless) parameters. One of them, i.e.,  the reduced deviation from the critical temperature of membrane demixing $\tau$, is a natural control parameter in experiments. The parameter describing membrane elasticity $\kappa$ can in principle be measured. This is not the case for the third parameter, which is  the strength of coupling between the local deformation of the membrane thickness (or the membrane height) and the local change in the lipid concentration $\mu$. However, as we argued in Sec.~\ref{secD:E}, its value can be estimated from the behavior of the correlation length, which is measurable. 

In the phase space spanned by these parameters we have distinguished three zones of distinct functional forms of the   two--point correlation functions. In all zones the leading asymptotic decay is exponential, however, it is multiplied by different prefactors: in zone I and II by a constant number and by oscillating function in zone III. Close to the critical point, in zones I and II this behavior is very similar to the one observed in Landau--Ginzburg model, whereas away from criticality, in zone III, it resembles the membrane deformation model behavior. 

These correlations are responsible for enhanced concentration order parameter near an inclusion embedded in the membrane, which locally change the thickness (height) of a membrane. This phenomenon is an analog of critical adsorption occurring in binary liquid mixtures upon approaching critical point of demixing from a homogeneous phase. We have found that excess adsorption of membrane lipids of one kind diverges in the same way as predicted for two--dimensional Ising--like systems near 
symmetry--breaking point--like inclusions~\cite{Hanke1999}. In order to facilitate comparisons with experiments, we have proposed several definitions of the size of domains rich in the lipid effectively attracted to the inclusion, and we have discussed their universal aspects, advantages and disadvantages. 
For example, for a study of coalescence of two domains in comparison with images from microscope, it is convenient to use  $\rad^{\phi}_{1/2}$. However, like for all other proposed radii, this definition does not depend on the hydrophobic mismatch of the inclusion, so the size of the domain is not changing upon increasing the mismatch, which is counterintuitive.

The current work has several very natural extensions.
First of all, one can include the  $\phi^4$ term in the Hamiltonian. This allows for studying the model at and below the critical temperature but requires numerical calculations. Second, one can introduce several  
inclusions in arrangement that mimics anchors linking the membrane to the actin network, and  compare the lipid concentration field  with experimental images. Due to the presence of quenched disorder along the lines following the filaments of the actin, such a model can be treated only numerically.
Third, one can also add to our model a coupling between the curvature of a membrane and its composition, and study the combine effect of the two mechanisms of domain formation.
Finally, this model can also be used to study membrane mediated Casimir--like interactions between floating inclusions. These effective forces could be very strong and might be a dominant factor in the process of formation of clusters of proteins on the membrane. 

The authors thank P.~Jakubczyk, M.~Napi\'orkowski, and A.~Parry for inspiring discussions and suggestions.

\appendix

\section{Method of calculations}\label{appA}

In the Appendix we present the details of the calculation of the formulae used in the paper. The calculation is based on \cite{Bihr2015}, where the membrane with a single order parameter $h\left(\bfr\right)$ is studied. We have decided to include the derivations here, because the additional order parameter $\phi\left(\bfr\right)$, described by the Hamiltonian of the Gaussian model, makes it necessary to regularize the integrals by introducing the cutoff, which has not been necessary in \cite{Bihr2015}.

\subsection{Correlation functions}\label{appA:I}

To calculate the correlation functions for the membrane without pinning points we calculate the probability density of having $h\left(\bfrho_\mathrm{a}\right)=h_\mathrm{a}$, $\phi\left(\bfrho_\mathrm{a}\right)=\phi_\mathrm{a}$, $h\left(\bfrho_\mathrm{b}\right)=h_\mathrm{b}$, and $\phi\left(\bfrho_\mathrm{b}\right)=\phi_\mathrm{b}$. In canonical ensemble it is given by 
\begin{multline}\label{appA:pA}
\prob\left(\bfrho_\mathrm{a}, h_\mathrm{a}, \phi_\mathrm{a}, \bfrho_\mathrm{b}, h_\mathrm{b}, \phi_\mathrm{b};\kappa,\mu,\tau\right)=\\
\const \int \mathcal{D} h\left(\bfrho\right) \int \mathcal{D} \phi\left(\bfrho\right) \delta\left[h\left(\bfrho_\mathrm{a}\right)-h_\mathrm{a}\right]\delta\left[\phi\left(\bfrho_\mathrm{a}\right)-\phi_\mathrm{a}\right]\\\times \delta\left[h\left(\bfrho_\mathrm{b}\right)-h_\mathrm{b}\right]\delta\left[\phi\left(\bfrho_\mathrm{b}\right)-\phi_\mathrm{b}\right]\exp\left(-\beta \Hamiltonian \left[h\left(\bfrho\right),\phi\left(\bfrho\right)\right]\right),
\end{multline}
where the term $\const$ denotes a constant prefactor (its exact value is not relevant, the final value is determined using the normalization condition), $\int \mathcal{D} h\left(\bfrho\right) \int \mathcal{D} \phi\left(\bfrho\right)$ is the path integral over all possible configurations of the two order parameters, $\delta$ denotes the Dirac delta function (it is used to fix the values of the order parameters in $\bfrho_\mathrm{a}$ and $\bfrho_\mathrm{b}$), and the Hamiltonian $\beta \mathcal{H}$ is given by Eq.~\eqref{secC:Hamiltonian} without the pinning part ($N=0$).

Using the the relation
\begin{equation}\label{appA:delta}
\delta\left(x\right)=\frac{1}{2\pi}\int_{-\infty}^{\infty}\dd\psi\, \ee^{\iu \psi x},
\end{equation}
we transform Eq.~\eqref{appA:pA} into
\begin{multline}\label{appA:pB}
\prob\left(\bfrho_\mathrm{a}, h_\mathrm{a}, \phi_\mathrm{a}, \bfrho_\mathrm{b}, h_\mathrm{b}, \phi_\mathrm{b};\kappa,\mu,\tau\right)=\const \int \mathcal{D} h\left(\bfrho\right)\\
 \int \mathcal{D} \phi\left(\bfrho\right) \int_{-\infty}^\infty \dd \psi_1\int_{-\infty}^\infty \dd \psi_2\int_{-\infty}^\infty \dd \psi_3\int_{-\infty}^\infty \dd \psi_4\\
\exp\Big[-\beta \Hamiltonian\left[h\left(\bfrho\right),\phi\left(\bfrho\right)\right]+\iu \psi_1 h\left(\bfrho_\mathrm{a}\right)+\iu \psi_2 \phi\left(\bfrho_\mathrm{a}\right)\\
+\iu \psi_3 h\left(\bfrho_\mathrm{b}\right)+\iu \psi_4 \phi\left(\bfrho_\mathrm{b}\right)-\iu \psi_1 h_\mathrm{a}-\iu\psi_2 \phi_\mathrm{a}-\iu \psi_3 h_\mathrm{b}-\iu\psi_4 \phi_\mathrm{b}\Big].
\end{multline}
Next step is to introduce the Fourier transform of the order parameters
\begin{equation}\label{appA:Fourier}
h\left(\bfrho\right)= \sum_\qcc h_\bfq \ee^{\iu \bfq \bfrho},\quad \phi\left(\bfrho\right)= \sum_\qcc \phi_\bfq \ee^{\iu \bfq \bfrho},
\end{equation}
where, in order to regularize some prefactors and integrals, we have introduced a cutoff $\Lambda$; to obtain the final results we apply the limit $\Lambda\to\infty$.
The allowed values of the wavevector $\bfq$, assuming square shape of the membrane of the area $A$ and periodic boundary conditions, are
\begin{equation}\label{appA:spectrum}
\bfq=\frac{2\pi}{\sqrt{A}}\left(n,m\right), \quad n,m\in \mathbb{Z}, \quad \qcc.
\end{equation}
Using \eqref{appA:Fourier} we transform all the terms of Hamiltonian \eqref{secC:Hamiltonian}. After some algebra we get
\begin{subequations}
\begin{align}
\beta\Hamiltonian_\mathrm{H}&=A \sum_\qcc \frac{\kappa}{2} q^4 \left|h_\bfq\right|^2,\\
\beta\Hamiltonian_\mathrm{G}&=A \sum_\qcc\left(\frac{1}{2} q^2+\tau\right)\left|\phi_\bfq\right|^2,\\
\beta\Hamiltonian_\mathrm{C}&=A \sum_\qcc \frac{\kappa}{2} \left(\left| h_\bfq\right|^2+\mu^2\left|\phi_\bfq\right|^2-2\mu h_\bfq \phi^\ast_\bfq\right),
\end{align}
\end{subequations}
where we have used identities $h_{-\bfq}=h_\bfq^\ast$ and $\phi_{-\bfq}=\phi_\bfq^\ast$ which follow from the fact that the order parameters $h\left(\bfrho\right)$ and $\phi\left(\bfrho\right)$ are real--valued functions. Using the above result in Eq.~\eqref{appA:pB}, after some reordering of terms in the exponent we get
\begin{multline}\label{appA:pC}
\prob\left(\bfrho_\mathrm{a}, h_\mathrm{a}, \phi_\mathrm{a}, \bfrho_\mathrm{b}, h_\mathrm{b}, \phi_\mathrm{b}; \kappa,\mu,\tau\right)=\const \int \mathcal{D} h\left(\bfrho\right)\\
 \int \mathcal{D} \phi\left(\bfrho\right) \int_{-\infty}^\infty \dd \psi_1\int_{-\infty}^\infty \dd \psi_2\int_{-\infty}^\infty \dd \psi_3\int_{-\infty}^\infty \dd \psi_4\\
\exp\Big[-\sum_\qcc \Big(A \frac{\kappa}{2}\left(q^4+1 \right)\left|h_\bfq\right|^2+\frac{A}{2}\left(q^2+2\tau+\omega\right)\left|\phi_\bfq\right|^2\\
-A \kappa\mu h_\bfq \phi_\bfq^\ast-\iu \psi_1 h_\bfq \ee^{\iu \bfq \bfrho_\mathrm{a}}-\iu \psi_2 \phi_\bfq \ee^{\iu \bfq \bfrho_\mathrm{a}}-\iu \psi_3 h_\bfq \ee^{\iu \bfq \bfrho_\mathrm{b}}\\
-\iu \psi_4 \phi_\bfq \ee^{\iu \bfq \bfrho_\mathrm{b}} \Big)-\iu \psi_1 h_\mathrm{a}-\iu\psi_2 \phi_\mathrm{a}-\iu \psi_3 h_\mathrm{b}-\iu\psi_4 \phi_\mathrm{b}\Big].
\end{multline}
In order to simplify the above formula it is necessary to define what exactly is meant by the path integral $\int\mathcal{D}x\left(\bfrho\right)$, where $x$ denotes one of the order parameters $h$ or $\phi$. Clearly, one has to integrate over all degrees of freedom $x_\bfq$ describing the function $x\left(\bfrho\right)$, but they are not independent variables since $x_{-\bfq}=x_\bfq^\ast$. Therefore, we group together all terms containing $x_\bfq$ and $x_{-\bfq}$, and integrate them separately over real and imaginary part of $x_\bfq$. Since
\begin{subequations}\label{appA:GaussianInt}
\begin{multline}
\int_{-\infty}^\infty \dd\re x_\bfq\int_{-\infty}^\infty \dd\im x_\bfq\\ \ee^{-a\left(\bfq\right)\left|x_\bfq\right|^2-2b\left(\bfq\right)x_\bfq}\,\ee^{-a\left(-\bfq\right)\left|x_{-\bfq}\right|^2-2b\left(-\bfq\right)x_{-\bfq}}\\
=\left[\sqrt{\frac{\pi}{2a\left(\bfq\right)}} \exp\left(\frac{b\left(\bfq\right)b\left(-\bfq\right)}{a\left(\bfq\right)}\right)\right]^2,
\end{multline}
and
\begin{equation}
\int_{-\infty}^\infty \dd x_\bfz\,\ee^{-a\left(\bfz\right)\left|x_\bfz\right|^2-2b\left(\bfz\right)x_\bfz}=\sqrt{\frac{\pi}{a\left(\bfz\right)}} \exp\left(\frac{b\left(\bfz\right)^2}{a\left(\bfz\right)}\right),
\end{equation}
\end{subequations}
where we have used the fact that $x_\bfz$ is real and assumed that $a\left(-\bfq\right)=a\left(\bfq\right)$; we can use the following rule for calculating the path integrals
\begin{multline}\label{appA:path}
\int\mathcal{D}x\left(\bfrho\right)\exp\left[-\sum_\qcc \left(a\left(\bfq\right)\left|x_\bfq\right|^2+2b\left(\bfq\right)x_\bfq\right)\right]\\
= \const \exp\left[\sum_\qcc \frac{b\left(\bfq\right)b\left(-\bfq\right)}{a\left(\bfq\right)}\right].
\end{multline}
We note that the omitted constant depends on the parameters present in $a\left(\bfq\right)$, which in our case are $\kappa$, $\mu$ and $\tau$. The strict calculation of the constant standing in front of the integral for the probability $\prob$ requires not only including all the prefactors from Eq.~\eqref{appA:GaussianInt} but also Jacobian coming from the change of variables made with the Fourier transform. Moreover, such a constant is clearly cutoff--dependent and may diverge in the limit $\Lambda\to\infty$ or $A\to\infty$. Here, we avoid all these problems by calculating the constant via the normalization of the probability distribution.

Applying the transformation~\eqref{appA:path} for $h\left(\bfrho\right)$ in Eq.~\eqref{appA:pC} with
\begin{subequations}
\begin{align}
\label{appA:aI}a\left(\bfq\right)&=\frac{A\kappa}{2}\left(q^4+1\right),\\
2b\left(\bfq\right)&=-A \kappa \mu \phi^\ast_\bfq-\iu \psi_1 \ee^{\iu \bfq\bfrho_\mathrm{a}}-\iu \psi_3 \ee^{\iu\bfq\bfrho_\mathrm{b}},
\end{align}
\end{subequations}
after short derivation gives
\begin{multline}\label{appA:pD}
\prob\left(\bfrho_\mathrm{a}, h_\mathrm{a}, \phi_\mathrm{a}, \bfrho_\mathrm{b}, h_\mathrm{b}, \phi_\mathrm{b}; \kappa,\mu,\tau\right)=\const  \int \mathcal{D} \phi\left(\bfrho\right)\\
 \int_{-\infty}^\infty \dd \psi_1\int_{-\infty}^\infty \dd \psi_2\int_{-\infty}^\infty \dd \psi_3\int_{-\infty}^\infty \dd \psi_4\\
\exp\bigg[-\sum_\qcc \bigg(\frac{A}{2}\,\frac{\left(q^4+1\right)\left(q^2+2\tau+\omega\right)-\omega}{q^4+1}\left|\phi_\bfq\right|^2\\
+\frac{\psi_1^2+\psi_3^2}{2 A \kappa\left(q^4+1\right)}-\frac{\iu\, \mu}{q^4+1}\ee^{\iu \bfq\bfrho_\mathrm{a}}\psi_1 \phi_\bfq\\
-\frac{\iu\, \mu}{q^4+1}\ee^{\iu \bfq\bfrho_\mathrm{b}}\psi_3 \phi_\bfq+\frac{\cos\bfq\left(\bfrho_\mathrm{a}-\bfrho_\mathrm{b}\right)}{A\kappa\left(q^4+1\right)}\psi_1\psi_3\\
-\iu \psi_2 \phi_\bfq \ee^{\iu \bfq \bfrho_\mathrm{a}}-\iu \psi_4 \phi_\bfq \ee^{\iu \bfq \bfrho_\mathrm{b}} \bigg)-\iu \psi_1 h_\mathrm{a}-\iu\psi_2 \phi_\mathrm{a}\\
-\iu \psi_3 h_\mathrm{b}-\iu\psi_4 \phi_\mathrm{b}\bigg].
\end{multline}
In the resulting formula one of the factors has the exact form of the polynomial from Eq.~\eqref{secD:W}: $W\left(q\right)=\left(q^4+1\right)\left(q^2+2\tau+\omega\right)-\omega$. The transformation \eqref{appA:path} can now be applied to Eq.~\eqref{appA:pD} in order to calculate the path integral over $\phi\left(\bfrho\right)$. From Eq.~\eqref{appA:pD} we read
\begin{subequations}
\begin{align}
\label{appA:aII}a\left(\bfq\right)=&\,\frac{A}{2}\,\frac{W\left(q\right)}{q^4+1},\\
\nonumber2b\left(\bfq\right)=&-\frac{\iu\, \mu}{q^4+1}\ee^{\iu \bfq\bfrho_\mathrm{a}}\psi_1-\frac{\iu\, \mu}{q^4+1}\ee^{\iu \bfq\bfrho_\mathrm{b}}\psi_3\\
&-\iu \psi_2  \ee^{\iu \bfq \bfrho_\mathrm{a}}-\iu \psi_4 \ee^{\iu \bfq \bfrho_\mathrm{b}}.
\end{align}
\end{subequations}
After some algebra we derive
\begin{multline}\label{appA:pE}
\prob\left(\bfrho_\mathrm{a}, h_\mathrm{a}, \phi_\mathrm{a}, \bfrho_\mathrm{b}, h_\mathrm{b}, \phi_\mathrm{b}; \kappa,\mu,\tau\right)=\\
\const \int_{-\infty}^\infty \dd \psi_1\int_{-\infty}^\infty \dd \psi_2\int_{-\infty}^\infty \dd \psi_3\int_{-\infty}^\infty \dd \psi_4\\
\exp\bigg[-\frac{1}{A}\sum_\qcc \bigg(\frac{q^2+2\tau+\omega}{2\kappa W\left(q\right)}\left(\psi_1^2+\psi_3^2\right)\\
+\frac{q^4+1}{2 W\left(q\right)}\left(\psi_2^2+\psi_4^2\right)+\frac{\mu}{ W\left(q\right)}\left(\psi_1\psi_2+\psi_3\psi_4\right)\\
+\frac{\mu\cos\bfq\left(\bfrho_\mathrm{a}-\bfrho_\mathrm{b}\right)}{ W\left(q\right)}\left(\psi_1\psi_4+\psi_2\psi_3\right)\\
+\frac{\left(q^2+2\tau+\omega\right)\cos\bfq\left(\bfrho_\mathrm{a}-\bfrho_\mathrm{b}\right)}{\kappa W\left(q\right)}\psi_1\psi_3\\
+\frac{\left(q^4+1\right)\cos\bfq\left(\bfrho_\mathrm{a}-\bfrho_\mathrm{b}\right)}{ W\left(q\right)}\psi_2\psi_4\bigg)\\
-\iu \psi_1 h_\mathrm{a}-\iu\psi_2 \phi_\mathrm{a}-\iu \psi_3 h_\mathrm{b}-\iu\psi_4 \phi_\mathrm{b}\bigg].
\end{multline}
To simplify the above formula it is convenient to take the limit $A\to\infty$. From Eq.~\eqref{appA:spectrum} it follows that, upon increasing $A$, the allowed values of $\bfq$ are getting closer to each other, and in the limit of infinite area, the sum is replaced with an integral following a formula
\begin{equation}
\lim_{A\to\infty} \frac{1}{A} \sum_\qcc f\left(\bfq\right)=\frac{1}{4\pi^2}\int_\qcc \dd\bfq f\left(\bfq\right),
\end{equation}
valid for any function $f\left(\bfq\right)$ that decays sufficiently fast for large $q$.

We introduce  the three functions
\begin{widetext}
\begin{subequations}\label{appA:corrfunc}
\begin{align}
\bar{\corrfunc}_{hh}\left(\rho_\mathrm{ab}\right)&=\frac{1}{4\pi^2}\int_\qcc \dd \bfq \frac{\left(q^2+2\tau+\omega\right)\cos\bfq\left(\bfrho_\mathrm{a}-\bfrho_\mathrm{b}\right)}{\kappa W\left(q\right)}=\frac{1}{2\pi\kappa}\int_0^\Lambda\dd q \frac{q\left(q^2+2\tau+\omega\right)\BesselJ_0\left(q \rho_\mathrm{ab}\right)}{W\left(q\right)},\\
\bar{\corrfunc}_{h\phi}\left(\rho_\mathrm{ab}\right)&=\frac{1}{4\pi^2}\int_\qcc \dd \bfq \frac{\mu\cos\bfq\left(\bfrho_\mathrm{a}-\bfrho_\mathrm{b}\right)}{ W\left(q\right)}=\frac{\mu}{2\pi}\int_0^\Lambda\dd q \frac{q\BesselJ_0\left(q \rho_\mathrm{ab}\right)}{ W\left(q\right)},\\
\bar{\corrfunc}_{\phi\phi}\left(\rho_\mathrm{ab}\right)&=\frac{1}{4\pi^2}\int_\qcc \dd \bfq \frac{\left(q^4+1\right)\cos\bfq\left(\bfrho_\mathrm{a}-\bfrho_\mathrm{b}\right)}{ W\left(q\right)}=\frac{1}{2\pi}\int_0^\Lambda\dd q \frac{q\left(q^4+1\right)\BesselJ_0\left(q \rho_\mathrm{ab}\right)}{ W\left(q\right)},
\end{align}
\end{subequations}
\end{widetext}
where we have defined $\rho_\mathrm{ab}=\left|\bfrho_\mathrm{a}-\bfrho_\mathrm{b}\right|$, and simplified the integrals by performing the integral over the direction of vector $\bfq$. We note that the functions have been decorated with bar to distinguish them from the correlation functions (see Eq.~\eqref{secD:integrals}). Our goal in this subsection is to prove that, in the limit $\Lambda\to\infty$, they are actually equal to the correlation functions.

Using Eq.~\eqref{appA:corrfunc}, formula for the probability in Eq.~\eqref{appA:pE} can be written in the form
\begin{multline}\label{appA:pF}
\prob\left(\bfrho_\mathrm{a}, h_\mathrm{a}, \phi_\mathrm{a}, \bfrho_\mathrm{b}, h_\mathrm{b}, \phi_\mathrm{b}; \kappa,\mu,\tau\right)=\\
\const \int_{-\infty}^\infty \dd \psi_1\int_{-\infty}^\infty \dd \psi_2\int_{-\infty}^\infty \dd \psi_3\int_{-\infty}^\infty \dd \psi_4\\
\exp\bigg[-\frac{1}{2} \bar{\corrfunc}_{hh}\left(0\right)\left(\psi_1^2+\psi_3^2\right)-\frac{1}{2}
\bar{\corrfunc}_{\phi\phi}\left(0\right)\left(\psi_2^2+\psi_4^2\right)\\
-\bar{\corrfunc}_{h\phi}\left(0\right)\left(\psi_1\psi_2+\psi_3\psi_4\right)
-\bar{\corrfunc}_{h\phi}\left(\rho_\mathrm{ab}\right)\left(\psi_1\psi_4+\psi_2\psi_3\right)\\
-\bar{\corrfunc}_{hh}\left(\rho_\mathrm{ab}\right)\psi_1\psi_3-\bar{\corrfunc}_{\phi\phi}\left(\rho_\mathrm{ab}\right)\psi_2\psi_4\\
-\iu \psi_1 h_\mathrm{a}-\iu\psi_2 \phi_\mathrm{a}-\iu \psi_3 h_\mathrm{b}-\iu\psi_4 \phi_\mathrm{b}\bigg].
\end{multline}
We note that the coefficient $\bar{\corrfunc}_{\phi\phi}\left(0\right)$ diverges for $\Lambda\to \infty$. This is the main motivation for introducing the regularization.

Using the matrix notation, the formula in Eq.~\eqref{appA:pF} for the probability can be rewritten in a form
\begin{multline}\label{appA:pG}
\prob\left(\bfrho_\mathrm{a}, h_\mathrm{a}, \phi_\mathrm{a}, \bfrho_\mathrm{b}, h_\mathrm{b}, \phi_\mathrm{b}; \kappa,\mu,\tau\right)=\\
\const\int \dd \bm{\Psi}\exp\left(-\frac{1}{2}\bm{\Psi}^\transpose\mathbb{M}\bm{\Psi}-\iu \bm{\Phi}^\transpose \bm{\Psi}\right),
\end{multline}
where the four--dimensional vector $\bm{\Psi}=\left[\psi_1,\psi_2,\psi_3,\psi_4\right]^\transpose$, $\bm{\Phi}=\left[h_\mathrm{a},\phi_\mathrm{a}, h_\mathrm{b},\phi_\mathrm{b}\right]^\transpose$, the symbol ``$\transpose$'' denotes transposition of the vector or matrix, and the symmetric matrix $\mathbb{M}$ is given by
\begin{equation}
\mathbb{M}=\begin{bmatrix} 
\bar{\corrfunc}_{hh}\left(0\right) & \bar{\corrfunc}_{h\phi}\left(0\right) & \bar{\corrfunc}_{hh}\left(\rho_\mathrm{ab}\right) & \bar{\corrfunc}_{h\phi}\left(\rho_\mathrm{ab}\right) \\
\bar{\corrfunc}_{h\phi}\left(0\right) & \bar{\corrfunc}_{\phi\phi}\left(0\right) & \bar{\corrfunc}_{h\phi}\left(\rho_\mathrm{ab}\right) & \bar{\corrfunc}_{\phi\phi}\left(\rho_\mathrm{ab}\right) \\
\bar{\corrfunc}_{hh}\left(\rho_\mathrm{ab}\right) & \bar{\corrfunc}_{h\phi}\left(\rho_\mathrm{ab}\right) & \bar{\corrfunc}_{hh}\left(0\right) & \bar{\corrfunc}_{h\phi}\left(0\right) \\
\bar{\corrfunc}_{h\phi}\left(\rho_\mathrm{ab}\right) & \bar{\corrfunc}_{\phi\phi}\left(\rho_\mathrm{ab}\right) & \bar{\corrfunc}_{h\phi}\left(0\right) & \bar{\corrfunc}_{\phi\phi}\left(0\right)
\end{bmatrix}.
\end{equation}
This reveals the simple structure of the formula for the probability in Eq.~\eqref{appA:pG}. We have imposed four conditions on the membrane: 1) in point $\bfrho_\mathrm{a}$ the order parameter $h$ is equal to $h_\mathrm{a}$, 2) in point $\bfrho_\mathrm{a}$ the order parameter $\phi$ is equal to $\phi_\mathrm{a}$, 3) in point $\bfrho_\mathrm{b}$ the order parameter $h$ is equal to $h_\mathrm{b}$, and finally, 4) in point $\bfrho_\mathrm{b}$ the order parameter $\phi$ is equal to $h_\mathrm{b}$. The element of the matrix $\mathbb{M}_{ij}$ is solely related to the conditions $i$) and $j$). The type of the correlation function is selected based on the fields that are fixed and the argument of the correlation function is the distance between the points where the conditions are imposed. The second term in the exponent in Eq.~\eqref{appA:pG} sets the relation between variables $\psi_i$ and the values of fixed order parameters for the four conditions imposed on the membrane. We note that, this rule can easily be extended to an arbitrary number of points where the order parameters are fixed and, therefore, allows one to skip the part of the calculations with path integrals. We leave the mathematical proof of correctness of this general procedure as a simple exercise.

The integral in Eq.~\eqref{appA:pG} can be calculated using the matrix analog of Eq.~\eqref{appA:GaussianInt}
\begin{multline}\label{appA:GaussianMatrix}
\int \dd \bm{v} \exp\left(-\frac{1}{2} \bm{v}^\transpose \mathbb{A}\bm{v}+\bm{w}^\transpose\bm{v}\right)=\\
\frac{\left(2\pi\right)^{n/2}}{\sqrt{\det \mathbb{A}}} \exp\left(\frac{1}{2}\bm{w}^\transpose \mathbb{A}^{-1}\bm{w}\right),
\end{multline}
where $\mathbb{A}$ is an arbitrary, $n\times n$, symmetric, positive--definite matrix. After simple derivation, from Eq.~\eqref{appA:pF} we get
\begin{multline}\label{appA:pH}
\prob\left(\bfrho_\mathrm{a}, h_\mathrm{a}, \phi_\mathrm{a}, \bfrho_\mathrm{b}, h_\mathrm{b}, \phi_\mathrm{b};\kappa,\mu,\tau\right)=\\
\left(4\pi^2 \sqrt{\det \mathbb{M}}\right)^{-1}\exp\left(-\frac{1}{2} \bm{\Phi}^\transpose\mathbb{M}^{-1}\bm\Phi\right),
\end{multline}
where the prefactor is calculated from the normalization condition 
\begin{multline}\label{appA:norm}
\int_{-\infty}^\infty \dd h_\mathrm{a}\int_{-\infty}^\infty \dd \phi_\mathrm{a}\int_{-\infty}^\infty \dd h_\mathrm{b}\int_{-\infty}^\infty \dd \phi_\mathrm{b}\\
\prob\left(\kappa,\mu,\tau; \bfrho_\mathrm{a}, h_\mathrm{a}, \phi_\mathrm{a}, \bfrho_\mathrm{b}, h_\mathrm{b}, \phi_\mathrm{b}\right)=1.
\end{multline}
using Eq.~\eqref{appA:GaussianMatrix}. Finally, we calculate the correlation functions using the relations
\begin{subequations}
\begin{align}
\int \dd \bm{v}\, v_i \exp\left(-\frac{1}{2} \bm{v}^\transpose \mathbb{A}\bm{v}\right)&=0,\\
\int \dd \bm{v}\, v_i v_j \exp\left(-\frac{1}{2} \bm{v}^\transpose \mathbb{A}\bm{v}\right)&=\frac{\left(2\pi\right)^{n/2}}{\sqrt{\det \mathbb{A}}} \left(\mathbb{A}^{-1}\right)_{ij},
\end{align}
\end{subequations}
where $v_i$ and $v_j$ denote components of $n$--dimensional vector $\bm{v}$; $\mathbb{A}$ is a symmetric, $n\times n$, positive--definite matrix; and $\left(\mathbb{A}^{-1}\right)_{ij}$ denotes a component of the matrix $\mathbb{A}^{-1}$, i.e., inverse of $\mathbb{A}$. After straightforward calculation we get
\begin{subequations}
\begin{align}
\corrfunc_{hh}\left(\rho_\mathrm{ab}\right)&=\lim_{\Lambda\to\infty}\left(\left<h_\mathrm{a}h_\mathrm{b}\right>-\left<h_\mathrm{a}\right>\left<h_\mathrm{b}\right>\right)=\lim_{\Lambda\to\infty}\bar{\corrfunc}_{hh}\left(\rho_\mathrm{ab}\right),\\
\corrfunc_{h\phi}\left(\rho_\mathrm{ab}\right)&=\lim_{\Lambda\to\infty}\left(\left<h_\mathrm{a}\phi_\mathrm{b}\right>-\left<h_\mathrm{a}\right>\left<\phi_\mathrm{b}\right>\right)=\lim_{\Lambda\to\infty}\bar{\corrfunc}_{h\phi}\left(\rho_\mathrm{ab}\right),\\
\corrfunc_{\phi\phi}\left(\rho_\mathrm{ab}\right)&=\lim_{\Lambda\to\infty}\left(\left<\phi_\mathrm{a}\phi_\mathrm{b}\right>-\left<\phi_\mathrm{a}\right>\left<\phi_\mathrm{b}\right>\right)=\lim_{\Lambda\to\infty}\bar{\corrfunc}_{\phi\phi}\left(\rho_\mathrm{ab}\right).
\end{align}
\end{subequations}
The result, together with Eq.~\eqref{appA:corrfunc}, proves the formulae in Eq.~\eqref{secD:integrals}.

\subsection{Order parameters profiles} \label{appA:II}

We now move to the problem of finding the order parameter profiles in the system with $N$ pinning points. The part of the Hamiltonian responsible for the pinning is $\Hamiltonian_\mathrm{P}$, see Eq.~\eqref{secD:HamiltonianP}. In order to include this term in our calculation of the path integrals we use the Hubbard--Stratonovich transformation
\begin{multline}
\exp\left(-\beta\Hamiltonian_\mathrm{P}\right)=\prod_{i=1}^N\exp\left[-\frac{\nu}{2}\left[h\left(\rho_i\right)-h_i\right]^2\right]=\\
\prod_{i=1}^N\frac{1}{\sqrt{2\pi \nu}}\int_{-\infty}^\infty\dd \psi_i \exp\left[-\frac{\psi_i^2}{2 \nu}+\iu \psi_i\left[h\left(\rho_i\right)-h_i\right]\right],
\end{multline}
which in the limit $\nu\to\infty$ produces the same factor in the exponent as the Dirac delta function, see Eq.~\eqref{appA:delta}. Therefore, neglecting the prefactor, the calculation of the path integrals goes along the same line as in the previous section and gives formula similar to \eqref{appA:pG}.

In order to make the calculation as general as possible, we denote by $x$ the order parameter for which we want to calculate the average ($x=h$ for height order parameter and $x=\phi$ for the composition order parameter), and denote its value in the point of interest $\bfrho$ by $x_0$, i.e., $x\left(\bfrho\right)=x_0$. If the membrane, following the pinning Hamiltonian $\Hamiltonian_\mathrm{P}$, is pinned in $N$ additional points $\bfrho_1, \bfrho_2,\ldots \bfrho_N$, where its height is fixed to $h_1, h_2, \ldots, h_N$, respectively, then, following the discussion presented in the section \ref{appA:I}, the probability is given by
\begin{multline}
\prob_x\left(\bfrho,x_0,\left\{\bfrho_i,h_i\right\}_{i=1}^{N};\kappa,\mu,\tau\right)=\\
\const \prod_{i=0}^N \left(\int_{-\infty}^\infty\dd \psi_i\right)\exp\Bigg[-\frac{1}{2}\bar{\corrfunc}_{xx}\left(0\right)\psi_0^2\\
-\sum_{i=1}^N \bar{\corrfunc}_{hx}\left(\left|\bfrho_i-\bfrho\right|\right)\psi_0\psi_i-\frac{1}{2}\sum_{i,j=1}^N \mathbb{B}_{ij}\psi_i \psi_j\\
-\iu \psi_0 x_0-\iu\sum_{i=1}^N \psi_i h_i\Bigg],
\end{multline}
where the matrix $\mathbb{B}_{ij}=\bar{\corrfunc}_{hh}\left(\left|\bfrho_i-\bfrho_j\right|\right)$. We now use Eq.~\eqref{appA:GaussianMatrix} to calculate the integrals over $\psi_1, \psi_2,\ldots,\psi_N$ in the above formula, and then Eq.~\eqref{appA:GaussianInt} to integrate over $\psi_0$. After straightforward calculation we get
\begin{widetext}
\begin{multline}
\prob_x\left(\bfrho,x_0,\left\{\bfrho_i,h_i\right\}_{i=1}^{N};\kappa,\mu,\tau\right)=
\const \exp\Bigg[-\frac{1}{2}\Bigg(x_0^2-2x_0\sum_{i,j=1}^N h_i \bar{\corrfunc}_{hx}\left(\left|\bfrho_j-\bfrho\right|\right)\left(\mathbb{B}^{-1}\right)_{ij}\Bigg)\\
\times\Bigg(\bar{\corrfunc}_{xx}\left(0\right)-\sum_{i,j=1}^N \bar{\corrfunc}_{hx}\left(\left|\bfrho_i-\bfrho\right|\right)\bar{\corrfunc}_{hx}\left(\left|\bfrho_j-\bfrho\right|\right)\left(\mathbb{B}^{-1}\right)_{ij}\Bigg)^{-1}\Bigg].
\end{multline} 
\end{widetext}
After finding the constant from the normalization condition, we derive
\begin{multline}\label{appA:opprofile}
\left<x\left(\bfrho\right)\right>=\lim_{\Lambda\to\infty} \left<x_0\right>=\\
\lim_{\Lambda\to\infty}\int_{-\infty}^\infty \dd x_0 \, x_0\, \prob_x\left(\bfrho,x_0,\left\{\bfrho_i,h_i\right\}_{i=1}^{N};\kappa,\mu,\tau\right)=\\
\lim_{\Lambda\to \infty}\sum_{i,j=1}^N h_i \bar{\corrfunc}_{hx}\left(\bfrho_j-\bfrho\right)\left(\mathbb{B}^{-1}\right)_{ij}.
\end{multline}
We note that in the final formula for the order parameter profile, the limit $\Lambda\to\infty$ exists, as the correlation functions present in Eq.~\eqref{appA:opprofile} are all well defined in this limit.

In the case of $N=1$ and $\rho_1=\bfz$, we have $\mathbb{B}=\left[\bar{\corrfunc}_{hh}\left(0\right)\right]$, and the formulae in Eq.~\eqref{secD:profiles} are recovered. In a different case of $h_1=h_2=\ldots=h_N\equiv h_0$ the formulae for the profile from \cite{Stumpf2021} are verified.

\section{Behavior of the roots of the polynomial $W$}\label{appB}

In this appendix we discuss the complex roots of the polynomial $W\left(z; \tau, \omega\right)$ given by Eq.~\eqref{secD:W}. As we have pointed in the main text, the properties of the roots define three zones present in the model, see Fig.~\ref{secD:figzones}. In zone I the roots have the form $\pm\iu t_1$, $\pm\iu t_2$, and $\pm\iu t_3$, with $t_1<t_2<t_3$. In zones II and III the six roots have the form $\pm\iu t_1$ and $\pm a\pm\iu t_2$, where $a,t_1,t_2>0$ and $t_1<t_2$ in zone II and $t_1>t_2$ in zone III. 

We note, that all the above coefficients can in principle be calculated analytically with the help of Cardano's formula, as $W$ is a third order polynomial of $z^2$. Nevertheless, we keep the roots as parameters for the sake of simplicity.

\begin{figure*}[t]
\begin{center}
    \begin{tabular}{llll}
         (a) & & (b) &  \\
         & \includegraphics[width=0.45\textwidth]{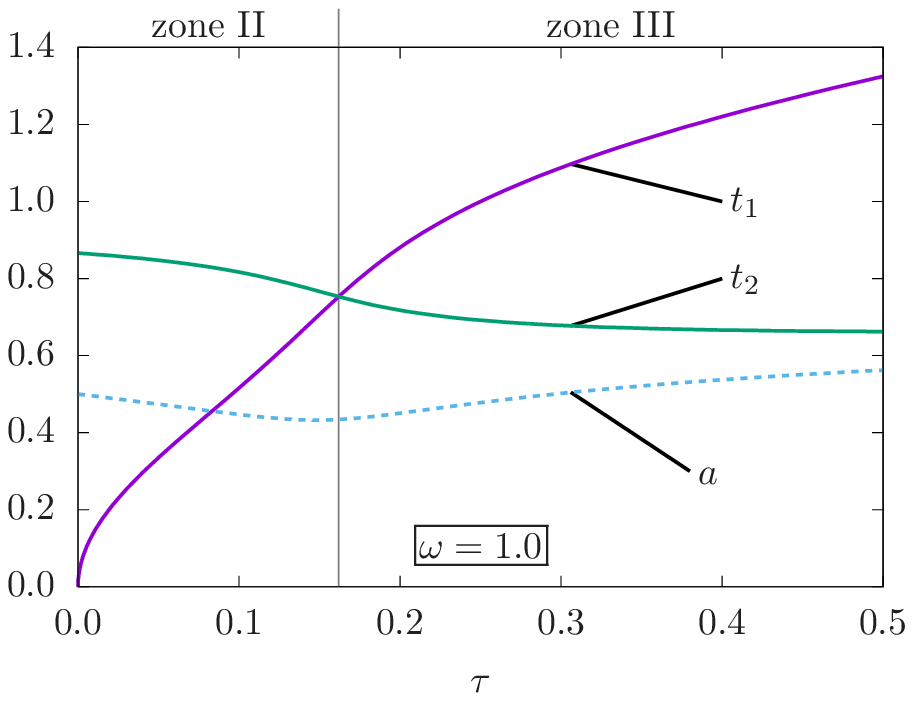} & & \includegraphics[width=0.45\textwidth]{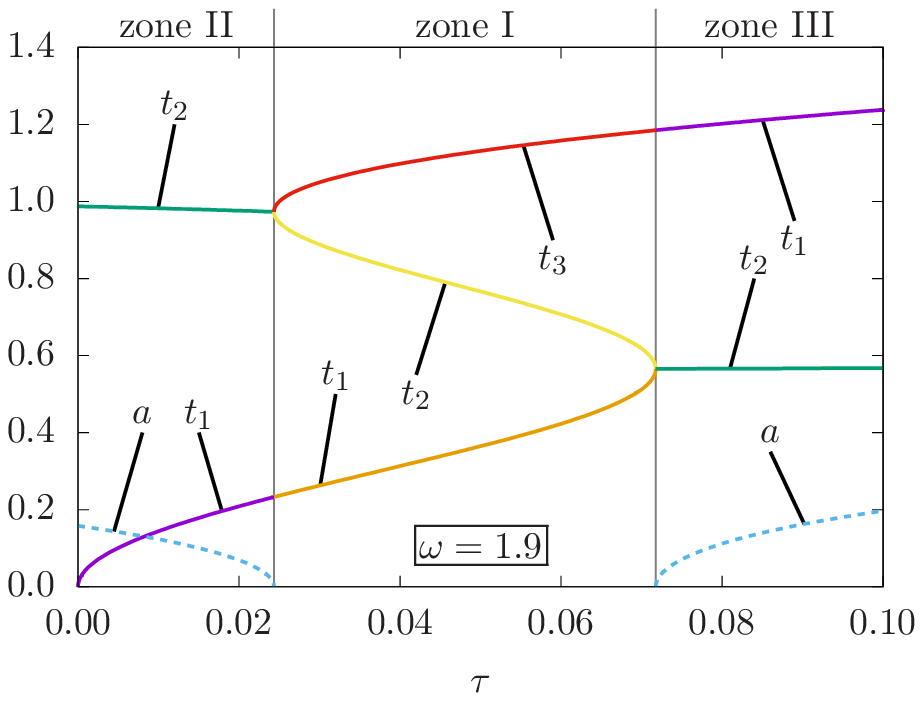}
    \end{tabular}
\end{center}
\caption{\label{secD:figroots} Parameters $t_1$, $t_2$, $t_3$ and $a$ that describe the complex roots of the polynomial $W\left(z;\omega,\tau\right)$ as a function of $\tau$ for fixed (a) $\omega=1.0$, and (b) $\omega=1.9$. The zones on the plane of parameters have been marked on top of the graphs and gray vertical lines denote the borders between zones (see Fig.~\ref{secD:figzones}).}
\end{figure*}

The behavior of the parameters upon crossing the borders of the zones has been illustrated in Fig~\ref{secD:figroots}. Upon going between zones II and III, the parameters $t_1\left(\omega,\tau\right)$, $t_2\left(\omega,\tau\right)$ and $a\left(\omega,\tau\right)$ are analytic functions. This is because the only difference between this zones is the relation between $t_1$ and $t_2$. On the contrary, going from zone I to zone II or III is accompanied with a rapid change of $t_1$ or $t_2$ as this functions have been differently defined in zone I. Upon approaching  the border of zone I from zone II (Fig.~\ref{secD:figroots}(b)) $t_1$ and $t_2$ stays finite and $a$ approaches to zero. Exactly at the border, $a$ is zero and $t_2$ (from zone II) splits into the parameters $t_2$ and $t_3$ (from zone I). The parameter $t_1$ stays analytic upon crossing this border. Upon approaching the border between zone I and III from the side of zone I, the parameters $t_1$ and $t_2$ approach each other. Exactly at the border they become equal and become $t_2$ from zone III. The parameter $a$ from zone III is zero at the border and growths upon entering inside the zone III. The parameter $t_3$ in zone I is analytic and is renamed to $t_1$ in zone III. The behavior of the roots upon crossing the borders of zone I is similar to what happens to the roots of the polynomial $z^2+c$ when $c$ is continuously changed from positive to negative values.

In the special point $\omega^\ast=8/\left(3\sqrt{3}\right)\approx 1.54 $ and $\tau^\ast=1/\left(6\sqrt{3}\right)\approx 0.0962 $, where all three zones meet, the polynomial $W$ has a pair of triple--degenerate roots equal to $\pm 3^{-1/4}\,\iu\approx \pm 1.32\,\iu$. This means that upon approaching to this point from zone I, the parameters $t_1$, $t_2$ and $t_3$ become equal; and upon approaching from zone II or III, parameters $t_1$ and $t_2$ become equal and $a$ decays to zero.

\section{Limiting cases for the correlation functions}\label{appC}

This appendix is devoted to study the properties of the correlation functions, defined in Sec.~\ref{secD}, in various limiting cases. Where possible, we relate our case to other, already known models. 

\subsection{Limit $\tau\to \infty$ with $\kappa$ and $\mu$ fixed}\label{secH:A}

The parameter $\tau$ appears in the Hamiltonian of the model \eqref{secC:Hamiltonian} in the term $\Hamiltonian_\mathrm{G}$, where it is multiplied by $\phi^2\left(\bfrho\right)$. Therefore, the limit $\tau\to\infty$, with other parameters fixed, implies that $\phi\left(\bfrho\right)\to 0$, which reduces the interaction to the membrane deformation Hamiltonian for the field $h\left(\bfrho\right)$. This observation is in line with the limiting value of the correlation length given by Eq.~\eqref{secD:xitauinf}: in the membrane deformation model the correlation length (in units of $\lsc$) is equal to $\sqrt{2}$.

In order to calculate the correlation functions in this limit we first note that for large values of $\tau$ the system is in zone III, and therefore, the formulae given in Eq.~\eqref{secD:corrII} must be used. In the second step we calculate the roots of the polynomial \eqref{secD:W} in this limit and parameters associated with them. After some algebra we get for $\tau\to\infty$
\begin{equation}
t_1=\sqrt{2\tau}+\mathrm{O}\left(\tau^{-1/2}\right), \quad t_2=a=\frac{1}{\sqrt{2}}+\mathrm{O}\left(\tau^{-1}\right).
\end{equation}
Finally, we use the above result in Eq.~\eqref{secD:corrII}. The first term in each of the formulae decays to zero for large $\tau$ like $\exp\left(-\sqrt{2\tau}\rho\right)$ and, for $\rho\neq 0$, it can be neglected in comparison with the second term. After simple calculation we derive
\begin{subequations}
\begin{align}
\corrfunc_{hh}\left(\rho;\kappa,\tau\to\infty,\mu\right)&=-\frac{1}{2\pi\kappa}\kei \left(\rho\right)+\mathrm{O}\left(\tau^{-1}\right),\label{secH:ChhHelfrich}\\
\corrfunc_{h\phi}\left(\rho;\kappa,\tau\to\infty,\mu\right)&=-\frac{\mu}{4\pi\tau} \kei\left(\rho\right)+\mathrm{O}\left(\tau^{-2}\right),\\
\corrfunc_{\phi\phi}\left(\rho;\kappa,\tau\to\infty,\mu\right)&=-\frac{\kappa\mu^2}{8\pi \tau^2}\kei\left(\rho\right)+\mathrm{O}\left(\tau^{-3}\right),\label{secH:CffHelfrich}
\end{align}
\end{subequations}
where $\kei\left(\rho\right)=\im\BesselK_0\left[\rho\left(1+\iu\right)/\sqrt{2}\right]$ denotes the Kelvin function $\kei$.

The result for the correlation function $\corrfunc_{hh}$ (Eq.~\eqref{secH:ChhHelfrich}) is in a full agreement with the result know for the membrane deformation model \cite{Bihr2015,Nelson2004}.  The disappearing of the correlation functions $\corrfunc_{h\phi}$ and $\corrfunc_{\phi\phi}$ in the limit $\tau\to\infty$ is caused by the vanishing of the order parameter $\phi$ in this limit.

We note that the limiting correlation function $\corrfunc_{\phi\phi}$ (given by Eq.~\eqref{secH:CffHelfrich}) is finite for $\rho=0$, even though for any finite $\tau$ the function diverges logarithmically (see Eq.~\eqref{secD:phiphiasymptotics}). The disagreement shows that in this case the limit $\tau\to \infty$ is not uniform --- for any finite $\tau$ there is a region around $\rho=0$, where the value of $\corrfunc_{\phi\phi}\left(\rho;\kappa,\tau,\mu\right)$ is essentially different from $\corrfunc_{\phi\phi}\left(\rho;\kappa,\tau\to\infty,\mu\right)$, but the size of this region shrinks $\sim\tau^{-1/2}$ upon increasing $\tau$.

\subsection{Limit $\tau\to 0$ with $\kappa$ and $\mu$ fixed}\label{secH:B}

We now move to the opposite limit $\tau\to 0$. In this regime, the Gaussian model is known to be critical \cite{Goldenfeld1992}. Since, as discussed in Sec.~\ref{secD:E}, in our model the correlation length diverges for small values of $\tau$, here we also expect criticality.

We start the analysis by noting that for $\tau\to 0$ the system can be either in zone I or in zone II. For $\omega\geqslant 2$, the system is in zone I, the parameters describing the roots of polynomial \eqref{secD:W} are
\begin{subequations}
\begin{align}
t_1&=\sqrt{2\tau}+\mathrm{O}\left(\tau^{3/2}\right),\\
t_2&=\left(\omega-\sqrt{\omega^2-4}\right)^{1/2}/\sqrt{2}+\mathrm{O}\left(\tau\right),\\
t_3&=\left(\omega+\sqrt{\omega^2-4}\right)^{1/2}/\sqrt{2}+\mathrm{O}\left(\tau\right),
\end{align}
\end{subequations}
and the correlation functions are given by Eq.~\eqref{secD:corrI}. For $\omega<2$, the system is in zone II, the parameters are
\begin{subequations}
\begin{align}
t_1&=\sqrt{2\tau}+\mathrm{O}\left(\tau^{3/2}\right),\\
t_2&=\sqrt{2+\omega}/2+\mathrm{O}\left(\tau\right),\\
a&=\sqrt{2-\omega}/2+\mathrm{O}\left(\tau\right),
\end{align}
\end{subequations}
and the correlation functions are given by Eq.~\eqref{secD:corrII}. The special case of $\omega=2$ has been incorporated into the first case above, because for small nonzero $\tau$ and $\omega=2$, the system is in zone I, just like for $\omega>2$. 

For fixed $\rho$ and $\tau\to 0$ we derive the following formulae for the correlation functions (the calculations were done separately for the system in zone I and in zone II giving the same results)
\begin{subequations}
\begin{align}
\corrfunc_{hh}\left(\rho; \kappa, \tau\to 0, \mu\right)&=-\frac{\mu^2 \ln\tau}{4\pi}+\mathrm{O}\left(1\right),\\
\corrfunc_{h\phi}\left(\rho; \kappa, \tau\to 0, \mu\right)&=-\frac{\mu \ln\tau}{4\pi}+\mathrm{O}\left(1\right),\\
\corrfunc_{\phi\phi}\left(\rho; \kappa, \tau\to 0, \mu\right)&=-\frac{\ln\tau}{4\pi}+\mathrm{O}\left(1\right).
\end{align}
\end{subequations}
All the functions do not depend on $\rho$, which means that the fluctuations keep both order parameters constant. This is not surprising since for $\tau\to 0$ the correlation length $\xi$ diverges, and thus keeping $\rho$ fixed implies the regime $\rho\ll\xi$ in which the correlation function is expected to stay almost constant. Moreover, the relation $\corrfunc_{hh}=\mu \corrfunc_{h\phi}=\mu^2 \corrfunc_{\phi\phi}$ implies that the coupling between order parameters given by Eq.~\eqref{secD:HamiltonianC} is strictly fulfilled (at least in the leading order). We note that, all the correlation functions diverge logarithmically for $\tau\to 0$.

To gain more insight into the behavior of the correlation functions for small $\tau$, it is useful to introduce the scaling limit $\tau\to 0$, $\rho\to\infty$ with the scaling variable $u=\sqrt{2\tau}\rho\approx \rho/\xi$ fixed. In this limit, after some algebra, we get
\begin{subequations}\label{secH:corrfuncSL}
\begin{align}
\corrfunc_{hh}\left(u; \kappa, \tau\to 0, \mu\right)&=\frac{\mu^2\BesselK_0\left(u\right)}{2\pi}+\mathrm{O}\left(\tau\right),\\
\corrfunc_{h\phi}\left(u; \kappa, \tau\to 0, \mu\right)&=\frac{\mu\BesselK_0\left(u\right)}{2\pi}+\mathrm{O}\left(\tau\right),\\
\corrfunc_{\phi\phi}\left(u; \kappa, \tau\to 0, \mu\right)&=\frac{\BesselK_0\left(u\right)}{2\pi}+\mathrm{O}\left(\tau\right).
\end{align}
\end{subequations}
This result should be compared with the prediction for systems in the vicinity of a critical point based on the scaling hypothesis \cite{Kadanoff1993} 
\begin{equation}\label{secH:scaling}
\corrfunc=\rho^{-\left(d-2+\eta\right)}\mathcal{C}\left(u\right),
\end{equation}
where $d=2$ is the dimensionality of the system, $\eta$ is a critical exponent, and $\mathcal{C}\left(u\right)$ is a universal scaling function. We note that only $\corrfunc_{\phi\phi}$ is strictly following Eq.~\eqref{secH:scaling}, with $\eta=0$ (the same value as in Gaussian model) and $\mathcal{C}\left(u\right)=\BesselK_0\left(u\right)/\left(2\pi\right)$ (see, e.g., \cite{Goldenfeld1992}); the scaling formulae for $\corrfunc_{hh}$ and $\corrfunc_{h\phi}$ contain additional non--universal (depending on the coupling $\mu$) factor. Like in the case of fixed $\rho$, the correlation functions in the leading order differ only by the power of $\mu$, which implies that the order parameters are strongly coupled. This explains why both order parameters become critical in the limit $\tau\to 0$, and the divergence of all the correlation functions \eqref{secH:corrfuncSL} for $u\to 0$.

\subsection{Limit $\mu\to 0$ with $\kappa$ and $\tau$ fixed}

We now consider the case of $\mu\to 0$, i.e., when the two order parameters are weakly coupled. In this limit, depending on the value of $\tau$, the system is in zone II or zone III, and therefore, the correlation functions are given by Eq.~\eqref{secD:corrII}.

We start from expanding the roots of the polynomial~\eqref{secD:W} in the limit of small $\omega=\kappa\mu^2$. After some algebra we get
\begin{subequations}\label{secH:smallmuroots}
\begin{align}
t_1&=\sqrt{2\tau}+\frac{\sqrt{2} \tau^{3/2}\omega}{1+4\tau^2}+\mathrm{O}\left(\omega^2\right),\\
t_2&=\frac{1}{\sqrt{2}}+\frac{\left(1-2\tau\right)\omega}{4\sqrt{2}\left(1+4\tau^2\right)}+\mathrm{O}\left(\omega^2\right),\\
a&=\frac{1}{\sqrt{2}}-\frac{\left(1+2\tau\right)\omega}{4\sqrt{2}\left(1+4\tau^2\right)}+\mathrm{O}\left(\omega^2\right),
\end{align}
\end{subequations}
and the same expansion is valid for both zone II and zone III. This allows us to calculate the correlation length in this limit
\begin{equation}\label{secH:smallmuxi}
\xi\left(\tau,\omega\to 0\right)=\max\left(\left(2\tau\right)^{-1/2},\sqrt{2}\right)+\mathrm{O}\left(\omega\right),
\end{equation}
which is in agreement with the plots of the correlation length in Fig.~\ref{secD:figxi}(a).

Using Eq.~\eqref{secH:smallmuroots} in Eq.~\eqref{secD:corrII} we derive the asymptotic form of the correlation functions:
\begin{subequations}\label{secH:smallmu}
\begin{align}
\label{secH:smallmuhh}\corrfunc_{hh}\left(\rho;\kappa,\tau,\mu\to 0\right)=&\,\frac{\mu^2 \BesselK_0\left(\rho\sqrt{2\tau}\right)}{2\pi\left(1+4\tau^2\right)^2}-\frac{\kei\left(\rho\right)}{2\pi\kappa},\\
\nonumber\corrfunc_{h\phi}\left(\rho;\kappa,\tau,\mu\to 0\right)=&\,\frac{\mu\,\BesselK_0\left(\rho\sqrt{2\tau}\right)}{2\pi \left(1+4\tau^2\right)}\\
\label{secH:smallmuhf}&-\frac{\mu \left[2\tau\kei\left(\rho\right)+\ker\left(\rho\right)\right]}{2\pi\left(1+4\tau^2\right)},\\
\nonumber\corrfunc_{\phi\phi}\left(\rho;\kappa,\tau,\mu\to 0\right)=&\,\frac{\BesselK_0\left(\rho\sqrt{2\tau}\right)}{2\pi}\\&
\label{secH:smallmuff}\hspace{-1.7cm}+\frac{\kappa\mu^2\left[\left(1-4\tau^2\right)\kei\left(\rho\right)-4\tau\ker\left(\rho\right)\right]}{2\pi\left(1+4\tau^2\right)^2},
\end{align}
\end{subequations}
where we have introduced another Kelvin function $\ker\left(\rho\right)=\re\BesselK_0\left[\rho\left(1+\iu\right)/\sqrt{2}\right]$. In Eqs.~\eqref{secH:smallmu} we have calculated the leading term separately for the two terms present in the formula for each of the correlation functions in Eq.~\eqref{secD:corrII}; the neglected, higher order terms were always smaller at least by a factor of $\mu^2$.

Each of the formulae for the correlation functions \eqref{secH:smallmu} consists of two terms: first term, proportional to $\BesselK_0\left(\rho\sqrt{2\tau}\right)$, for large $\rho$ decays exponentially to zero with a lengthscale $\left(2\tau\right)^{-1/2}$; the second term, proportional to the combination of $\kei\left(\rho\right)$ and $\ker\left(\rho\right)$, for large $\rho$ decays exponentially to zero with a lengthscale $\sqrt{2}$. For $\tau<1/4$ (i.e., in zone II) the former lengthscale is bigger and, therefore, for $\mu\neq 0$ all correlation functions decay with the same lengthscale $\left(2\tau\right)^{-1/2}$, in line with Eq.~\eqref{secH:smallmuxi}. Nevertheless, upon decreasing $\mu$ to $0$, the amplitudes multiplying the first term in Eq.~\eqref{secH:smallmuhh}, both terms in Eq.~\eqref{secH:smallmuhf}, and second term in Eq.~\eqref{secH:smallmuff} are decaying to $0$. As a result for $\mu=0$ the function $\corrfunc_{hh}$ decays to zero with a lengthscale $\sqrt{2}$, $\corrfunc_{h\phi}$ is zero and $\corrfunc_{\phi\phi}$ decays to zero with the original lengthscale $\left(2\tau\right)^{-1/2}$. For $\tau\geqslant 1/4$ (i.e., in zone III) for $\mu> 0$ second terms in formulae \eqref{secH:smallmu} dominate and all correlation functions decay with a lengthscale $\sqrt{2}$. For $\mu=0$, due to zeroing of some amplitudes, the same result as for $\tau<1/4$ is recovered.

The above analysis shows that, even though, the correlation functions change for $\mu\to 0$ in a continuous manner, the correlation length is discontinues at $\mu=0$: For $\mu>0$ the correlation length is the same for all correlation functions and it is given by Eq.~\eqref{secH:smallmuxi}. For $\mu=0$, we have
\begin{subequations}
\begin{align}
\label{secH:corrh}\corrfunc_{hh}\left(\rho;\kappa, \tau,\mu=0\right)&=-\frac{\kei\left(\rho\right)}{2\pi\kappa},\\
\corrfunc_{h\phi}\left(\rho;\kappa, \tau,\mu=0\right)&=0,\\
\label{secH:corrphi}\corrfunc_{\phi\phi}\left(\rho;\kappa, \tau,\mu=0\right)&=\frac{\BesselK_0\left(\rho\sqrt{2\tau}\right)}{2\pi},
\end{align}
\end{subequations}
i.e., the correlation function $\corrfunc_{hh}$ decays to zero with a lengthscale $\sqrt{2}$, the correlation function $\corrfunc_{\phi\phi}$ with a lengthscale $\left(2\tau\right)^{-1/2}$, and there is no correlation between the order parameters. We note that, Eq.~\eqref{secH:corrh} agrees with the correlation function in the membrane deformation model and Eq.~\eqref{secH:corrphi} with the correlation function in Gaussian model (see Eq.~\eqref{secH:ChhHelfrich} and Refs.~\cite{Bihr2015, Goldenfeld1992}).

\subsection{Limit $\mu\to\infty$ with $\kappa$ and $\tau$ fixed}

The effect of increasing the coupling $\mu$ between the order parameters in the Hamiltonian~\eqref{secD:Hamiltonian} is not evident. In order to study and explain the behavior of the system, we first note that in the limit $\mu\to\infty$ the system is for $\tau>0$ in zone III, see Fig.~\ref{secD:figzones}. In this zone the roots of the polynomial \eqref{secD:W} are described by three parameters $t_1$, $t_2$ and $a$. For large $\mu$ we have calculated
\begin{subequations}\label{secH:bm}
\begin{align}
t_1&=\mu\sqrt{\kappa}+\frac{\tau}{\mu\sqrt{\kappa}}+\mathrm{O}\left(\mu^{-3}\right),\\
t_2&=\left(\frac{\tau}{2\kappa}\right)^{1/4}\mu^{-1/2}+\mathrm{O}\left(\mu^{-3/2}\right),\\
a&=\left(2\tau/\omega\right)^{1/4}/\sqrt{2}+\mathrm{O}\left(\omega^{-3/4}\right),
\end{align}
\end{subequations}
which implies
\begin{equation}\label{secH:xilargemu}
\xi=1/t_2=\left(\frac{2\kappa}{\tau}\right)^{1/4}\sqrt{\mu}+\mathrm{O}\left(\mu^{-1/2}\right)\to \infty,
\end{equation}
and, therefore, in the limit $\mu\to\infty$ the system becomes critical. By substituting Eq.~\eqref{secH:bm} into Eq.~\eqref{secD:corrII}, for fixed $\rho$ (i.e. for $\rho\ll\xi$ when $\mu\to
\infty$) we get
\begin{subequations}
\begin{align}
\corrfunc_{hh}\left(\rho;\kappa,\tau,\mu\to\infty\right)&=\frac{\mu}{8\sqrt{2\tau\kappa}}+\mathrm{O}\left(\ln \mu\right),\\
\corrfunc_{h\phi}\left(\rho;\kappa,\tau,\mu\to\infty\right)&=\frac{1}{8\sqrt{2\tau\kappa}}+\mathrm{O}\left(\frac{\ln \mu}{\mu}\right),\\
\corrfunc_{\phi\phi}\left(\rho;\kappa,\tau,\mu\to\infty\right)&=\frac{1}{8\mu\sqrt{2\tau\kappa}}+\mathrm{O}\left(\frac{\ln \mu}{\mu^2}\right),
\end{align}
\end{subequations}
which shows that in this limit both order parameters stay constant on the lengthscale $\xi$ (the same behavior we have seen in the limit $\tau\to 0$, see Sec.~\ref{secH:B}). The value of field $h\left(\bfrho\right)$ fluctuates with a standard deviation that scales like $\sqrt{\mu}$ and the field $\phi\left(\bfrho\right)$ has a standard deviation $\sim 1/\sqrt{\mu}$. Therefore, for large $\mu$ the Gaussian term \eqref{secD:HamiltonianG} in the Hamiltonian \eqref{secD:Hamiltoniansum} becomes negligible. This, in turn, allows the field $\phi$ to adapt to changes of the field $h$ at no additional energy cost and makes the coupling term \eqref{secD:HamiltonianC} also negligible. Without pinning, the only relevant term left in the Hamiltonian is \eqref{secD:HamiltonianH} which allows for large scale fluctuations of the membrane \cite{Lipowsky1990}. The behavior of our system in the limit $\mu\to\infty$ can thus be identified with the critical roughening. 

Finally, to describe the correlation functions in the limit $\mu\to\infty$ it is useful to introduce the scaling variable $v=\rho/\xi=\rho\, t_2\sim \rho/\sqrt{\mu}$. As we have checked, in the scaling limit $\mu\to\infty$ with $v$ fixed we get
\begin{subequations}
\begin{align}
\corrfunc_{hh}\left(v;\kappa,\tau,\mu\to\infty\right)&=\frac{\kei\left(\sqrt{2}v\right)\mu}{\pi\sqrt{8\tau\kappa}}+\mathrm{O}\left(\mu^{-3}\right),\\
\corrfunc_{h\phi}\left(v;\kappa,\tau,\mu\to\infty\right)&=\frac{\kei\left(\sqrt{2}v\right)}{\pi\sqrt{8\tau\kappa}}+\mathrm{O}\left(\mu^{-2}\right),\\
\corrfunc_{\phi\phi}\left(v;\kappa,\tau,\mu\to\infty\right)&=\frac{\kei\left(\sqrt{2}v\right)}{\pi\sqrt{8\tau\kappa}\mu}+\mathrm{O}\left(\mu^{-3}\right).
\end{align}
\end{subequations}

\subsection{Limit of strong binding between order parameters}\label{appC:V}

Another interesting limiting case is when the binding $\gamma$ between two order parameters is going to $\infty$ (see Eq.~\eqref{secB:coupling}). As a result, the two order parameters must strictly fulfill the relation $\hdim\left(\mathbf{r}\right)=\alpha \phidim\left(\mathbf{r}\right)$ and the system is effectively described by a single order parameter. In this case, the Hamiltonian reduces to
\begin{multline}
    \beta\tilde{\mathcal{H}}=\int \dd \mathbf{r}\Bigg[\frac{\kappa}{2}\left(\nabla^2 \hdim\left(\mathbf{r}\right)\right)^2+\frac{\tilde{\sigma}}{2}\left(\nabla \hdim\left(\mathbf{r}\right)\right)^2\\
    +\frac{\tilde{\gamma}}{2}\hdim^2\left(\mathbf{r}\right)+\frac{\lambda}{2}\sum_{i=1}^N\left(\hdim\left(\mathbf{r}_i\right)-\ell_i\right)^2\Bigg],
\end{multline}
with $\tilde{\sigma}=\sigma/\alpha^2$ and $\tilde{\gamma}=2t/\alpha^2$. The above Hamiltonian has already been discussed in literature \cite{Janes2019}, as it describes a membrane with binding stiffness $\kappa$, surface tension $\tilde{\sigma}$ in an external harmonic potential of the strength $\tilde{\gamma}$. The results of our model, after taking the proper limit (as described below), are in a full agreement with \cite{Janes2019}.

The limit $\gamma\to\infty$ implies that $\zeta=\left(\kappa/\gamma\right)^{1/4}\to 0$, and thus, $\tau\to 0$ and $\mu\to\infty$ (see Table~\ref{secC:dimensionless}) with fixed 
\begin{equation}
   \kappa \tau\mu^2=\tau \omega=\frac{t \kappa \alpha^2}{\sigma^2}\equiv \frac{1}{8}\chi^2,
\end{equation}
where, in order to simplify the notation, we have added an extra factor $\kappa$ and we have defined $\chi\geqslant 0$. The vanishing of the unit of length $\zeta$ makes the analysis of this special limit challenging.

In Fig.~\ref{secD:figzones} the limit considered here is located in the region where the border between zones I and III asymptotically touches the line $\tau=0$. Closer analysis shows that in this limit the system is in zone I for $\chi<1$ (and the correlation function decays exponentially) and in zone III for $\chi>1$ (where $\corrfunc_{hh}$ shows damped oscillations). The same change of asymptotics has been reported in \cite{Janes2019} for $\sigma/\lambda_\mathrm{m}^0=1/4$, which is in an agreement with our results since $\sigma/\lambda_\mathrm{m}^0$ in \cite{Janes2019} is equivalent to $\left(4\chi\right)^{-1}$ in our paper.

Detailed analysis of the roots of the polynomial \eqref{secD:W} shows that in this limit for $\chi<1$ (zone I)
\begin{subequations}
\begin{align}
    t_1&=\frac{2}{\chi}\sqrt{1-\sqrt{1-\chi^2}}\, \tau^{1/2}+\mathrm{O}\left(\tau^{3/2}\right),\\
    t_2&=\frac{2}{\chi}\sqrt{1+\sqrt{1-\chi^2}}\, \tau^{1/2}+\mathrm{O}\left(\tau^{3/2}\right),\\
    t_3&=\frac{\chi}{2\sqrt{2\tau}}+\mathrm{O}\left(\tau^{1/2}\right),
\end{align}
\end{subequations}
and for $\chi>1$ (zone III)
\begin{subequations}
\begin{align}
a&=\frac{\sqrt{2\tau}}{\chi}\sqrt{\chi-1}+\mathrm{O}\left(\tau^{3/2}\right),\\
t_1&=\frac{\chi}{2\sqrt{2\tau}}+\mathrm{O}\left(\tau^{1/2}\right),\\
t_2&=\frac{\sqrt{2\tau}}{\chi}\sqrt{\chi+1}+\mathrm{O}\left(\tau^{3/2}\right).
\end{align}
\end{subequations}
Therefore, the (dimensional) correlation length $\bar{\xi}$ is for $\chi<1$ (zone I)
\begin{equation}
    \bar{\xi}=\zeta \xi=\frac{\zeta}{t_1}=\left(\frac{\chi}{1-\sqrt{1-\chi^2}}\right)^{1/2}\tilde{\zeta},
\end{equation}
and for $\chi>1$ (zone III)
\begin{equation}
    \bar{\xi}=\zeta \xi=\frac{\zeta}{t_2}=\left(\frac{2\chi}{\left(\chi+1\right)}\right)^{1/2}\tilde{\zeta},
\end{equation}
with $\tilde{\zeta}=\left(\kappa/\tilde{\gamma}\right)^{1/4}$. The above formula for the correlation length can be shown to be identical with Eq.~(36) in \cite{Janes2019}. We note that, even though in this limit two different critical regimes $\tau\to 0$ and $\mu\to\infty$ overlap, only the dimensionless correlation length $\xi$ is infinite. Together with increasing $\xi$, the unit of length $\zeta\to 0$ which keeps the dimensional correlation length $\bar{\xi}=\zeta\xi$ finite, and the system is actually not critical.

Because our notation makes the analysis of this limit unnecessary complicated, we refrain from a detailed study of our model for $\gamma\to 0$. Properties of the model in this special case have been discussed in \cite{Janes2019}. 

\providecommand*\hyphen{-}


\begin{thebibliography}{47}%
\makeatletter
\providecommand \@ifxundefined [1]{%
 \@ifx{#1\undefined}
}%
\providecommand \@ifnum [1]{%
 \ifnum #1\expandafter \@firstoftwo
 \else \expandafter \@secondoftwo
 \fi
}%
\providecommand \@ifx [1]{%
 \ifx #1\expandafter \@firstoftwo
 \else \expandafter \@secondoftwo
 \fi
}%
\providecommand \natexlab [1]{#1}%
\providecommand \enquote  [1]{``#1''}%
\providecommand \bibnamefont  [1]{#1}%
\providecommand \bibfnamefont [1]{#1}%
\providecommand \citenamefont [1]{#1}%
\providecommand \href@noop [0]{\@secondoftwo}%
\providecommand \href [0]{\begingroup \@sanitize@url \@href}%
\providecommand \@href[1]{\@@startlink{#1}\@@href}%
\providecommand \@@href[1]{\endgroup#1\@@endlink}%
\providecommand \@sanitize@url [0]{\catcode `\\12\catcode `\$12\catcode
  `\&12\catcode `\#12\catcode `\^12\catcode `\_12\catcode `\%12\relax}%
\providecommand \@@startlink[1]{}%
\providecommand \@@endlink[0]{}%
\providecommand \url  [0]{\begingroup\@sanitize@url \@url }%
\providecommand \@url [1]{\endgroup\@href {#1}{\urlprefix }}%
\providecommand \urlprefix  [0]{URL }%
\providecommand \Eprint [0]{\href }%
\providecommand \doibase [0]{https://doi.org/}%
\providecommand \selectlanguage [0]{\@gobble}%
\providecommand \bibinfo  [0]{\@secondoftwo}%
\providecommand \bibfield  [0]{\@secondoftwo}%
\providecommand \translation [1]{[#1]}%
\providecommand \BibitemOpen [0]{}%
\providecommand \bibitemStop [0]{}%
\providecommand \bibitemNoStop [0]{.\EOS\space}%
\providecommand \EOS [0]{\spacefactor3000\relax}%
\providecommand \BibitemShut  [1]{\csname bibitem#1\endcsname}%
\let\auto@bib@innerbib\@empty
\bibitem [{\citenamefont {Dietrich}\ \emph {et~al.}(2001)\citenamefont
  {Dietrich}, \citenamefont {Bagatolli}, \citenamefont {Volovyk}, \citenamefont
  {Thompson}, \citenamefont {Levi}, \citenamefont {Jacobson},\ and\
  \citenamefont {Gratton}}]{Dietrich2001}%
  \BibitemOpen
  \bibfield  {author} {\bibinfo {author} {\bibfnamefont {C.}~\bibnamefont
  {Dietrich}}, \bibinfo {author} {\bibfnamefont {L.~A.}\ \bibnamefont
  {Bagatolli}}, \bibinfo {author} {\bibfnamefont {Z.~N.}\ \bibnamefont
  {Volovyk}}, \bibinfo {author} {\bibfnamefont {N.~L.}\ \bibnamefont
  {Thompson}}, \bibinfo {author} {\bibfnamefont {M.}~\bibnamefont {Levi}},
  \bibinfo {author} {\bibfnamefont {K.}~\bibnamefont {Jacobson}},\ and\
  \bibinfo {author} {\bibfnamefont {E.}~\bibnamefont {Gratton}},\ }\bibfield
  {title} {\bibinfo {title} {Lipid rafts reconstituted in model membranes},\
  }\href {https://doi.org/doi:10.1016/S0006-3495(01)76114-0} {\bibfield
  {journal} {\bibinfo  {journal} {Biophys. J.}\ }\textbf {\bibinfo {volume}
  {80}},\ \bibinfo {pages} {1417} (\bibinfo {year} {2001})}\BibitemShut
  {NoStop}%
\bibitem [{\citenamefont {Veatch}\ \emph {et~al.}(2007)\citenamefont {Veatch},
  \citenamefont {Soubias}, \citenamefont {Keller},\ and\ \citenamefont
  {Gawrisch}}]{Veatch2007}%
  \BibitemOpen
  \bibfield  {author} {\bibinfo {author} {\bibfnamefont {S.~L.}\ \bibnamefont
  {Veatch}}, \bibinfo {author} {\bibfnamefont {O.}~\bibnamefont {Soubias}},
  \bibinfo {author} {\bibfnamefont {S.~L.}\ \bibnamefont {Keller}},\ and\
  \bibinfo {author} {\bibfnamefont {K.}~\bibnamefont {Gawrisch}},\ }\bibfield
  {title} {\bibinfo {title} {Critical fluctuations in domain-forming lipid
  mixtures},\ }\href {https://doi.org/10.1073/pnas.0703513104} {\bibfield
  {journal} {\bibinfo  {journal} {Proc. Natl. Acad. Sci. U.S.A.}\ }\textbf
  {\bibinfo {volume} {104}},\ \bibinfo {pages} {17650} (\bibinfo {year}
  {2007})}\BibitemShut {NoStop}%
\bibitem [{\citenamefont {Honerkamp-Smith}\ \emph {et~al.}(2009)\citenamefont
  {Honerkamp-Smith}, \citenamefont {Veatch},\ and\ \citenamefont
  {Keller}}]{HonerkampSmith2009}%
  \BibitemOpen
  \bibfield  {author} {\bibinfo {author} {\bibfnamefont {A.~R.}\ \bibnamefont
  {Honerkamp-Smith}}, \bibinfo {author} {\bibfnamefont {S.~L.}\ \bibnamefont
  {Veatch}},\ and\ \bibinfo {author} {\bibfnamefont {S.~L.}\ \bibnamefont
  {Keller}},\ }\bibfield  {title} {\bibinfo {title} {An introduction to
  critical points for biophysicists; observations of compositional
  heterogeneity in lipid membranes},\ }\href@noop {} {\bibfield  {journal}
  {\bibinfo  {journal} {Biochim. Biophys. Acta, Biomembr.}\ }\textbf {\bibinfo
  {volume} {1788}},\ \bibinfo {pages} {53} (\bibinfo {year}
  {2009})}\BibitemShut {NoStop}%
\bibitem [{\citenamefont {Stone}\ \emph
  {et~al.}(2017{\natexlab{a}})\citenamefont {Stone}, \citenamefont {Shelby},
  \citenamefont {N{\'{u}}{\~{n}}ez}, \citenamefont {Wisser},\ and\
  \citenamefont {Veatch}}]{Stone2017a}%
  \BibitemOpen
  \bibfield  {author} {\bibinfo {author} {\bibfnamefont {M.~B.}\ \bibnamefont
  {Stone}}, \bibinfo {author} {\bibfnamefont {S.~A.}\ \bibnamefont {Shelby}},
  \bibinfo {author} {\bibfnamefont {M.~F.}\ \bibnamefont {N{\'{u}}{\~{n}}ez}},
  \bibinfo {author} {\bibfnamefont {K.}~\bibnamefont {Wisser}},\ and\ \bibinfo
  {author} {\bibfnamefont {S.~L.}\ \bibnamefont {Veatch}},\ }\bibfield  {title}
  {\bibinfo {title} {{Protein sorting by lipid phase-like domains supports
  emergent signaling function in B lymphocyte plasma membranes}},\ }\href
  {https://doi.org/10.7554/eLife.19891} {\bibfield  {journal} {\bibinfo
  {journal} {eLife}\ }\textbf {\bibinfo {volume} {6}},\ \bibinfo {pages}
  {e19891} (\bibinfo {year} {2017}{\natexlab{a}})}\BibitemShut {NoStop}%
\bibitem [{\citenamefont {Stone}\ \emph
  {et~al.}(2017{\natexlab{b}})\citenamefont {Stone}, \citenamefont {Shelby},\
  and\ \citenamefont {Veatch}}]{Stone2017b}%
  \BibitemOpen
  \bibfield  {author} {\bibinfo {author} {\bibfnamefont {M.~B.}\ \bibnamefont
  {Stone}}, \bibinfo {author} {\bibfnamefont {S.~A.}\ \bibnamefont {Shelby}},\
  and\ \bibinfo {author} {\bibfnamefont {S.~L.}\ \bibnamefont {Veatch}},\
  }\bibfield  {title} {\bibinfo {title} {Super-resolution microscopy: Shedding
  light on the cellular plasma membrane},\ }\href
  {https://doi.org/10.1021/acs.chemrev.6b00716} {\bibfield  {journal} {\bibinfo
   {journal} {Chem. Rev.}\ }\textbf {\bibinfo {volume} {117}},\ \bibinfo
  {pages} {7457} (\bibinfo {year} {2017}{\natexlab{b}})}\BibitemShut {NoStop}%
\bibitem [{\citenamefont {Voci}\ \emph {et~al.}(2018)\citenamefont {Voci},
  \citenamefont {Goudeau}, \citenamefont {Valenti}, \citenamefont {Lesch},
  \citenamefont {Jović}, \citenamefont {Rapino}, \citenamefont {Paolucci},
  \citenamefont {Arbault},\ and\ \citenamefont {Sojic}}]{Voci2018}%
  \BibitemOpen
  \bibfield  {author} {\bibinfo {author} {\bibfnamefont {S.}~\bibnamefont
  {Voci}}, \bibinfo {author} {\bibfnamefont {B.}~\bibnamefont {Goudeau}},
  \bibinfo {author} {\bibfnamefont {G.}~\bibnamefont {Valenti}}, \bibinfo
  {author} {\bibfnamefont {A.}~\bibnamefont {Lesch}}, \bibinfo {author}
  {\bibfnamefont {M.}~\bibnamefont {Jović}}, \bibinfo {author} {\bibfnamefont
  {S.}~\bibnamefont {Rapino}}, \bibinfo {author} {\bibfnamefont
  {F.}~\bibnamefont {Paolucci}}, \bibinfo {author} {\bibfnamefont
  {S.}~\bibnamefont {Arbault}},\ and\ \bibinfo {author} {\bibfnamefont
  {N.}~\bibnamefont {Sojic}},\ }\bibfield  {title} {\bibinfo {title}
  {Surface-confined electrochemiluminescence microscopy of cell membranes},\
  }\href {https://doi.org/10.1021/jacs.8b08080} {\bibfield  {journal} {\bibinfo
   {journal} {J. Am. Chem. Soc.}\ }\textbf {\bibinfo {volume} {140}},\ \bibinfo
  {pages} {14753} (\bibinfo {year} {2018})}\BibitemShut {NoStop}%
\bibitem [{\citenamefont {Roobala}\ \emph {et~al.}(2018)\citenamefont
  {Roobala}, \citenamefont {Ilanila},\ and\ \citenamefont
  {Basu}}]{Roobala2018}%
  \BibitemOpen
  \bibfield  {author} {\bibinfo {author} {\bibfnamefont {C.}~\bibnamefont
  {Roobala}}, \bibinfo {author} {\bibfnamefont {I.}~\bibnamefont {Ilanila}},\
  and\ \bibinfo {author} {\bibfnamefont {J.}~\bibnamefont {Basu}},\ }\bibfield
  {title} {\bibinfo {title} {Applications of sted fluorescence nanoscopy in
  unravelling nanoscale structure and dynamics of biological systems},\
  }\href@noop {} {\bibfield  {journal} {\bibinfo  {journal} {J. Biosci.}\
  }\textbf {\bibinfo {volume} {43}},\ \bibinfo {pages} {471} (\bibinfo {year}
  {2018})}\BibitemShut {NoStop}%
\bibitem [{\citenamefont {Meder}\ \emph {et~al.}(2006)\citenamefont {Meder},
  \citenamefont {Moreno}, \citenamefont {Verkade}, \citenamefont {Vaz},\ and\
  \citenamefont {Simons}}]{Meder2006}%
  \BibitemOpen
  \bibfield  {author} {\bibinfo {author} {\bibfnamefont {D.}~\bibnamefont
  {Meder}}, \bibinfo {author} {\bibfnamefont {M.~J.}\ \bibnamefont {Moreno}},
  \bibinfo {author} {\bibfnamefont {P.}~\bibnamefont {Verkade}}, \bibinfo
  {author} {\bibfnamefont {W.~L.~C.}\ \bibnamefont {Vaz}},\ and\ \bibinfo
  {author} {\bibfnamefont {K.}~\bibnamefont {Simons}},\ }\href@noop {}
  {\bibfield  {journal} {\bibinfo  {journal} {Proc. Natl. Acad. Sci. USA}\
  }\textbf {\bibinfo {volume} {103}},\ \bibinfo {pages} {329} (\bibinfo {year}
  {2006})}\BibitemShut {NoStop}%
\bibitem [{\citenamefont {Pralle}\ \emph {et~al.}(2000)\citenamefont {Pralle},
  \citenamefont {Keller}, \citenamefont {Florin}, \citenamefont {Simons},\ and\
  \citenamefont {H{\"{o}}rber}}]{Pralle2000}%
  \BibitemOpen
  \bibfield  {author} {\bibinfo {author} {\bibfnamefont {A.}~\bibnamefont
  {Pralle}}, \bibinfo {author} {\bibfnamefont {P.}~\bibnamefont {Keller}},
  \bibinfo {author} {\bibfnamefont {E.-L.}\ \bibnamefont {Florin}}, \bibinfo
  {author} {\bibfnamefont {K.}~\bibnamefont {Simons}},\ and\ \bibinfo {author}
  {\bibfnamefont {J.~K.~H.}\ \bibnamefont {H{\"{o}}rber}},\ }\href@noop {}
  {\bibfield  {journal} {\bibinfo  {journal} {J. Cell Biol.}\ }\textbf
  {\bibinfo {volume} {148}},\ \bibinfo {pages} {997} (\bibinfo {year}
  {2000})}\BibitemShut {NoStop}%
\bibitem [{\citenamefont {Levental}\ \emph {et~al.}(2020)\citenamefont
  {Levental}, \citenamefont {Levental},\ and\ \citenamefont
  {Heberle}}]{Levental2020}%
  \BibitemOpen
  \bibfield  {author} {\bibinfo {author} {\bibfnamefont {I.}~\bibnamefont
  {Levental}}, \bibinfo {author} {\bibfnamefont {K.~R.}\ \bibnamefont
  {Levental}},\ and\ \bibinfo {author} {\bibfnamefont {F.~A.}\ \bibnamefont
  {Heberle}},\ }\href@noop {} {\bibfield  {journal} {\bibinfo  {journal}
  {Trends Cell Biol.}\ }\textbf {\bibinfo {volume} {30}},\ \bibinfo {pages}
  {341} (\bibinfo {year} {2020})}\BibitemShut {NoStop}%
\bibitem [{\citenamefont {Lenne}\ and\ \citenamefont
  {Nicolas}(2009)}]{B822956B}%
  \BibitemOpen
  \bibfield  {author} {\bibinfo {author} {\bibfnamefont {P.-F.}\ \bibnamefont
  {Lenne}}\ and\ \bibinfo {author} {\bibfnamefont {A.}~\bibnamefont
  {Nicolas}},\ }\bibfield  {title} {\bibinfo {title} {Physics puzzles on
  membrane domains posed by cell biology},\ }\href
  {https://doi.org/10.1039/B822956B} {\bibfield  {journal} {\bibinfo  {journal}
  {Soft Matter}\ }\textbf {\bibinfo {volume} {5}},\ \bibinfo {pages} {2841}
  (\bibinfo {year} {2009})}\BibitemShut {NoStop}%
\bibitem [{\citenamefont {Destainville}\ \emph {et~al.}(2018)\citenamefont
  {Destainville}, \citenamefont {Manghi},\ and\ \citenamefont
  {Cornet}}]{destainville2018rationale}%
  \BibitemOpen
  \bibfield  {author} {\bibinfo {author} {\bibfnamefont {N.}~\bibnamefont
  {Destainville}}, \bibinfo {author} {\bibfnamefont {M.}~\bibnamefont
  {Manghi}},\ and\ \bibinfo {author} {\bibfnamefont {J.}~\bibnamefont
  {Cornet}},\ }\bibfield  {title} {\bibinfo {title} {A rationale for mesoscopic
  domain formation in biomembranes},\ }\href@noop {} {\bibfield  {journal}
  {\bibinfo  {journal} {Biomolecules}\ }\textbf {\bibinfo {volume} {8}},\
  \bibinfo {pages} {104} (\bibinfo {year} {2018})}\BibitemShut {NoStop}%
\bibitem [{\citenamefont {Hanke}\ and\ \citenamefont
  {Dietrich}(1999)}]{Hanke1999}%
  \BibitemOpen
  \bibfield  {author} {\bibinfo {author} {\bibfnamefont {A.}~\bibnamefont
  {Hanke}}\ and\ \bibinfo {author} {\bibfnamefont {S.}~\bibnamefont
  {Dietrich}},\ }\bibfield  {title} {\bibinfo {title} {Critical adsorption on
  curved objects},\ }\href {https://doi.org/10.1103/PhysRevE.59.5081}
  {\bibfield  {journal} {\bibinfo  {journal} {Phys. Rev. E}\ }\textbf {\bibinfo
  {volume} {59}},\ \bibinfo {pages} {5081} (\bibinfo {year}
  {1999})}\BibitemShut {NoStop}%
\bibitem [{\citenamefont {Honerkamp-Smith}\ \emph {et~al.}(2008)\citenamefont
  {Honerkamp-Smith}, \citenamefont {Cicuta}, \citenamefont {Collins},
  \citenamefont {Veatch}, \citenamefont {Den~Nijs}, \citenamefont {Schick},\
  and\ \citenamefont {Keller}}]{honerkamp2008line}%
  \BibitemOpen
  \bibfield  {author} {\bibinfo {author} {\bibfnamefont {A.~R.}\ \bibnamefont
  {Honerkamp-Smith}}, \bibinfo {author} {\bibfnamefont {P.}~\bibnamefont
  {Cicuta}}, \bibinfo {author} {\bibfnamefont {M.~D.}\ \bibnamefont {Collins}},
  \bibinfo {author} {\bibfnamefont {S.~L.}\ \bibnamefont {Veatch}}, \bibinfo
  {author} {\bibfnamefont {M.}~\bibnamefont {Den~Nijs}}, \bibinfo {author}
  {\bibfnamefont {M.}~\bibnamefont {Schick}},\ and\ \bibinfo {author}
  {\bibfnamefont {S.~L.}\ \bibnamefont {Keller}},\ }\bibfield  {title}
  {\bibinfo {title} {Line tensions, correlation lengths, and critical exponents
  in lipid membranes near critical points},\ }\href@noop {} {\bibfield
  {journal} {\bibinfo  {journal} {Biophysical journal}\ }\textbf {\bibinfo
  {volume} {95}},\ \bibinfo {pages} {236} (\bibinfo {year} {2008})}\BibitemShut
  {NoStop}%
\bibitem [{\citenamefont {Venturoli}\ \emph {et~al.}(2005)\citenamefont
  {Venturoli}, \citenamefont {Smit},\ and\ \citenamefont
  {Sperotto}}]{Venturoli2005}%
  \BibitemOpen
  \bibfield  {author} {\bibinfo {author} {\bibfnamefont {M.}~\bibnamefont
  {Venturoli}}, \bibinfo {author} {\bibfnamefont {B.}~\bibnamefont {Smit}},\
  and\ \bibinfo {author} {\bibfnamefont {M.~M.}\ \bibnamefont {Sperotto}},\
  }\bibfield  {title} {\bibinfo {title} {Simulation studies of protein-induced
  bilayer deformations, and lipid-induced protein tilting, on a mesoscopic
  model for lipid bilayers with embedded proteins},\ }\href
  {https://doi.org/10.1529/biophysj.104.050849} {\bibfield  {journal} {\bibinfo
   {journal} {Biophys. J.}\ }\textbf {\bibinfo {volume} {88}},\ \bibinfo
  {pages} {1778} (\bibinfo {year} {2005})}\BibitemShut {NoStop}%
\bibitem [{\citenamefont {Bitbol}\ \emph {et~al.}(2012)\citenamefont {Bitbol},
  \citenamefont {Constantin},\ and\ \citenamefont {Fournier}}]{Bitbol2012}%
  \BibitemOpen
  \bibfield  {author} {\bibinfo {author} {\bibfnamefont {A.-F.}\ \bibnamefont
  {Bitbol}}, \bibinfo {author} {\bibfnamefont {D.}~\bibnamefont {Constantin}},\
  and\ \bibinfo {author} {\bibfnamefont {J.-B.}\ \bibnamefont {Fournier}},\
  }\bibfield  {title} {\bibinfo {title} {Bilayer elasticity at the nanoscale:
  the need for new terms},\ }\href@noop {} {\bibfield  {journal} {\bibinfo
  {journal} {PLoS One}\ }\textbf {\bibinfo {volume} {7}},\ \bibinfo {pages}
  {e48306} (\bibinfo {year} {2012})}\BibitemShut {NoStop}%
\bibitem [{\citenamefont {{Leibler, S.}}\ and\ \citenamefont {{Andelman,
  D.}}(1987)}]{Leibler1987}%
  \BibitemOpen
  \bibfield  {author} {\bibinfo {author} {\bibnamefont {{Leibler, S.}}}\ and\
  \bibinfo {author} {\bibnamefont {{Andelman, D.}}},\ }\bibfield  {title}
  {\bibinfo {title} {Ordered and curved meso-structures in membranes and
  amphiphilic films},\ }\href
  {https://doi.org/10.1051/jphys:0198700480110201300} {\bibfield  {journal}
  {\bibinfo  {journal} {J. Phys. France}\ }\textbf {\bibinfo {volume} {48}},\
  \bibinfo {pages} {2013} (\bibinfo {year} {1987})}\BibitemShut {NoStop}%
\bibitem [{\citenamefont {Sens}\ and\ \citenamefont {Safran}(2000)}]{Sens2000}%
  \BibitemOpen
  \bibfield  {author} {\bibinfo {author} {\bibfnamefont {P.}~\bibnamefont
  {Sens}}\ and\ \bibinfo {author} {\bibfnamefont {S.~A.}\ \bibnamefont
  {Safran}},\ }\bibfield  {title} {\bibinfo {title} {Inclusions induced phase
  separation in mixed lipid film},\ }\href
  {https://doi.org/10.1007/s101890050026} {\bibfield  {journal} {\bibinfo
  {journal} {Eur. Phys. J. E}\ }\textbf {\bibinfo {volume} {1}},\ \bibinfo
  {pages} {237} (\bibinfo {year} {2000})}\BibitemShut {NoStop}%
\bibitem [{\citenamefont {Brown}\ and\ \citenamefont
  {London}(1998)}]{Brown1998}%
  \BibitemOpen
  \bibfield  {author} {\bibinfo {author} {\bibfnamefont {D.}~\bibnamefont
  {Brown}}\ and\ \bibinfo {author} {\bibfnamefont {E.}~\bibnamefont {London}},\
  }\bibfield  {title} {\bibinfo {title} {Structure and origin of ordered lipid
  domains in biological membranes},\ }\href@noop {} {\bibfield  {journal}
  {\bibinfo  {journal} {The Journal of membrane biology}\ }\textbf {\bibinfo
  {volume} {164}},\ \bibinfo {pages} {103} (\bibinfo {year}
  {1998})}\BibitemShut {NoStop}%
\bibitem [{\citenamefont {Shrestha}\ \emph {et~al.}(2020)\citenamefont
  {Shrestha}, \citenamefont {Kahraman},\ and\ \citenamefont
  {Haselwandter}}]{PhysRevE.102.060401}%
  \BibitemOpen
  \bibfield  {author} {\bibinfo {author} {\bibfnamefont {A.}~\bibnamefont
  {Shrestha}}, \bibinfo {author} {\bibfnamefont {O.}~\bibnamefont {Kahraman}},\
  and\ \bibinfo {author} {\bibfnamefont {C.~A.}\ \bibnamefont {Haselwandter}},\
  }\bibfield  {title} {\bibinfo {title} {Regulation of membrane proteins
  through local heterogeneity in lipid bilayer thickness},\ }\href
  {https://doi.org/10.1103/PhysRevE.102.060401} {\bibfield  {journal} {\bibinfo
   {journal} {Phys. Rev. E}\ }\textbf {\bibinfo {volume} {102}},\ \bibinfo
  {pages} {060401} (\bibinfo {year} {2020})}\BibitemShut {NoStop}%
\bibitem [{\citenamefont {Stumpf}\ \emph {et~al.}(2021)\citenamefont {Stumpf},
  \citenamefont {Nowakowski}, \citenamefont {Eggeling}, \citenamefont
  {Macio\l{}ek},\ and\ \citenamefont {Smith}}]{Stumpf2021}%
  \BibitemOpen
  \bibfield  {author} {\bibinfo {author} {\bibfnamefont {B.~H.}\ \bibnamefont
  {Stumpf}}, \bibinfo {author} {\bibfnamefont {P.}~\bibnamefont {Nowakowski}},
  \bibinfo {author} {\bibfnamefont {C.}~\bibnamefont {Eggeling}}, \bibinfo
  {author} {\bibfnamefont {A.}~\bibnamefont {Macio\l{}ek}},\ and\ \bibinfo
  {author} {\bibfnamefont {A.-S.}\ \bibnamefont {Smith}},\ }\bibfield  {title}
  {\bibinfo {title} {Protein induced lipid demixing in homogeneous membranes},\
  }\href {https://doi.org/10.1103/PhysRevResearch.3.L042013} {\bibfield
  {journal} {\bibinfo  {journal} {Phys. Rev. Research}\ }\textbf {\bibinfo
  {volume} {3}},\ \bibinfo {pages} {L042013} (\bibinfo {year}
  {2021})}\BibitemShut {NoStop}%
\bibitem [{\citenamefont {Honigmann}\ \emph {et~al.}(2014)\citenamefont
  {Honigmann}, \citenamefont {Sadeghi}, \citenamefont {Keller}, \citenamefont
  {Hell}, \citenamefont {Eggeling},\ and\ \citenamefont
  {Vink}}]{Honigmann2014}%
  \BibitemOpen
  \bibfield  {author} {\bibinfo {author} {\bibfnamefont {A.}~\bibnamefont
  {Honigmann}}, \bibinfo {author} {\bibfnamefont {S.}~\bibnamefont {Sadeghi}},
  \bibinfo {author} {\bibfnamefont {J.}~\bibnamefont {Keller}}, \bibinfo
  {author} {\bibfnamefont {S.~W.}\ \bibnamefont {Hell}}, \bibinfo {author}
  {\bibfnamefont {C.}~\bibnamefont {Eggeling}},\ and\ \bibinfo {author}
  {\bibfnamefont {R.}~\bibnamefont {Vink}},\ }\bibfield  {title} {\bibinfo
  {title} {A lipid bound actin meshwork organizes liquid phase separation in
  model membranes},\ }\href {https://doi.org/10.7554/eLife.01671} {\bibfield
  {journal} {\bibinfo  {journal} {eLife}\ }\textbf {\bibinfo {volume} {3}},\
  \bibinfo {pages} {e01671} (\bibinfo {year} {2014})}\BibitemShut {NoStop}%
\bibitem [{\citenamefont {Rautu}\ \emph {et~al.}(2015)\citenamefont {Rautu},
  \citenamefont {Rowlands},\ and\ \citenamefont {Turner}}]{Rautu2015}%
  \BibitemOpen
  \bibfield  {author} {\bibinfo {author} {\bibfnamefont {S.~A.}\ \bibnamefont
  {Rautu}}, \bibinfo {author} {\bibfnamefont {G.}~\bibnamefont {Rowlands}},\
  and\ \bibinfo {author} {\bibfnamefont {M.~S.}\ \bibnamefont {Turner}},\
  }\bibfield  {title} {\bibinfo {title} {Membrane composition variation and
  underdamped mechanics near transmembrane proteins and coats},\ }\href
  {https://doi.org/10.1103/physrevlett.114.098101} {\bibfield  {journal}
  {\bibinfo  {journal} {Phys. Rev. Lett.}\ }\textbf {\bibinfo {volume} {114}},\
  \bibinfo {pages} {098101} (\bibinfo {year} {2015})}\BibitemShut {NoStop}%
\bibitem [{\citenamefont {Ayton}\ \emph {et~al.}(2005)\citenamefont {Ayton},
  \citenamefont {McWhirter}, \citenamefont {McMurtry},\ and\ \citenamefont
  {Voth}}]{Ayton2005}%
  \BibitemOpen
  \bibfield  {author} {\bibinfo {author} {\bibfnamefont {G.~S.}\ \bibnamefont
  {Ayton}}, \bibinfo {author} {\bibfnamefont {J.~L.}\ \bibnamefont
  {McWhirter}}, \bibinfo {author} {\bibfnamefont {P.}~\bibnamefont
  {McMurtry}},\ and\ \bibinfo {author} {\bibfnamefont {G.~A.}\ \bibnamefont
  {Voth}},\ }\bibfield  {title} {\bibinfo {title} {Coupling field theory with
  continuum mechanics: A simulation of domain formation in giant unilamellar
  vesicles},\ }\href {https://doi.org/10.1529/biophysj.105.059436} {\bibfield
  {journal} {\bibinfo  {journal} {Biophys. J.}\ }\textbf {\bibinfo {volume}
  {88}},\ \bibinfo {pages} {3855} (\bibinfo {year} {2005})}\BibitemShut
  {NoStop}%
\bibitem [{\citenamefont {Veksler}\ and\ \citenamefont
  {Gov}(2007)}]{Veksler2007}%
  \BibitemOpen
  \bibfield  {author} {\bibinfo {author} {\bibfnamefont {A.}~\bibnamefont
  {Veksler}}\ and\ \bibinfo {author} {\bibfnamefont {N.~S.}\ \bibnamefont
  {Gov}},\ }\bibfield  {title} {\bibinfo {title} {Phase transitions of the
  coupled membrane-cytoskeleton modify cellular shape},\ }\href
  {https://doi.org/10.1529/biophysj.107.113282} {\bibfield  {journal} {\bibinfo
   {journal} {Biophys. J.}\ }\textbf {\bibinfo {volume} {93}},\ \bibinfo
  {pages} {3798} (\bibinfo {year} {2007})}\BibitemShut {NoStop}%
\bibitem [{\citenamefont {Sadeghi}\ \emph {et~al.}(2014)\citenamefont
  {Sadeghi}, \citenamefont {M{\"{u}}ller},\ and\ \citenamefont
  {Vink}}]{Sadeghi2014}%
  \BibitemOpen
  \bibfield  {author} {\bibinfo {author} {\bibfnamefont {S.}~\bibnamefont
  {Sadeghi}}, \bibinfo {author} {\bibfnamefont {M.}~\bibnamefont
  {M{\"{u}}ller}},\ and\ \bibinfo {author} {\bibfnamefont {R.~L.~C.}\
  \bibnamefont {Vink}},\ }\bibfield  {title} {\bibinfo {title} {Raft formation
  in lipid bilayers coupled to curvature},\ }\href
  {https://doi.org/10.1016/j.bpj.2014.07.072} {\bibfield  {journal} {\bibinfo
  {journal} {Biophys. J.}\ }\textbf {\bibinfo {volume} {107}},\ \bibinfo
  {pages} {1591} (\bibinfo {year} {2014})}\BibitemShut {NoStop}%
\bibitem [{\citenamefont {Simunovic}\ \emph {et~al.}(2016)\citenamefont
  {Simunovic}, \citenamefont {Evergren}, \citenamefont {Golushko},
  \citenamefont {Pr{\'e}vost}, \citenamefont {Renard}, \citenamefont
  {Johannes}, \citenamefont {McMahon}, \citenamefont {Lorman}, \citenamefont
  {Voth},\ and\ \citenamefont {Bassereau}}]{Simunovic11226}%
  \BibitemOpen
  \bibfield  {author} {\bibinfo {author} {\bibfnamefont {M.}~\bibnamefont
  {Simunovic}}, \bibinfo {author} {\bibfnamefont {E.}~\bibnamefont {Evergren}},
  \bibinfo {author} {\bibfnamefont {I.}~\bibnamefont {Golushko}}, \bibinfo
  {author} {\bibfnamefont {C.}~\bibnamefont {Pr{\'e}vost}}, \bibinfo {author}
  {\bibfnamefont {H.-F.}\ \bibnamefont {Renard}}, \bibinfo {author}
  {\bibfnamefont {L.}~\bibnamefont {Johannes}}, \bibinfo {author}
  {\bibfnamefont {H.~T.}\ \bibnamefont {McMahon}}, \bibinfo {author}
  {\bibfnamefont {V.}~\bibnamefont {Lorman}}, \bibinfo {author} {\bibfnamefont
  {G.~A.}\ \bibnamefont {Voth}},\ and\ \bibinfo {author} {\bibfnamefont
  {P.}~\bibnamefont {Bassereau}},\ }\bibfield  {title} {\bibinfo {title} {How
  curvature-generating proteins build scaffolds on membrane nanotubes},\ }\href
  {https://doi.org/10.1073/pnas.1606943113} {\bibfield  {journal} {\bibinfo
  {journal} {Proc. Natl. Acad. Sci. USA}\ }\textbf {\bibinfo {volume} {113}},\
  \bibinfo {pages} {11226} (\bibinfo {year} {2016})}\BibitemShut {NoStop}%
\bibitem [{\citenamefont {Pr\'evost}\ \emph {et~al.}(2015)\citenamefont
  {Pr\'evost}, \citenamefont {Zhao}, \citenamefont {Manzi}, \citenamefont
  {Lemichez}, \citenamefont {Lappalainen}, \citenamefont {Callan-Jones},\ and\
  \citenamefont {Bassereau}}]{Prevost2015}%
  \BibitemOpen
  \bibfield  {author} {\bibinfo {author} {\bibfnamefont {C.}~\bibnamefont
  {Pr\'evost}}, \bibinfo {author} {\bibfnamefont {H.}~\bibnamefont {Zhao}},
  \bibinfo {author} {\bibfnamefont {J.}~\bibnamefont {Manzi}}, \bibinfo
  {author} {\bibfnamefont {E.}~\bibnamefont {Lemichez}}, \bibinfo {author}
  {\bibfnamefont {P.}~\bibnamefont {Lappalainen}}, \bibinfo {author}
  {\bibfnamefont {A.}~\bibnamefont {Callan-Jones}},\ and\ \bibinfo {author}
  {\bibfnamefont {P.}~\bibnamefont {Bassereau}},\ }\bibfield  {title} {\bibinfo
  {title} {{IRSp53} senses negative membrane curvature and phase separates
  along membrane tubules},\ }\href {https://doi.org/10.1038/ncomms9529}
  {\bibfield  {journal} {\bibinfo  {journal} {Nat. Commun.}\ }\textbf {\bibinfo
  {volume} {6}},\ \bibinfo {pages} {8529} (\bibinfo {year} {2015})}\BibitemShut
  {NoStop}%
\bibitem [{\citenamefont {Dan}\ \emph {et~al.}(1993)\citenamefont {Dan},
  \citenamefont {Pincus},\ and\ \citenamefont {Safran}}]{dan1993membrane}%
  \BibitemOpen
  \bibfield  {author} {\bibinfo {author} {\bibfnamefont {N.}~\bibnamefont
  {Dan}}, \bibinfo {author} {\bibfnamefont {P.}~\bibnamefont {Pincus}},\ and\
  \bibinfo {author} {\bibfnamefont {S.}~\bibnamefont {Safran}},\ }\bibfield
  {title} {\bibinfo {title} {Membrane-induced interactions between
  inclusions},\ }\href@noop {} {\bibfield  {journal} {\bibinfo  {journal}
  {langmuir}\ }\textbf {\bibinfo {volume} {9}},\ \bibinfo {pages} {2768}
  (\bibinfo {year} {1993})}\BibitemShut {NoStop}%
\bibitem [{\citenamefont {{N. Dan}}\ \emph {et~al.}(1994)\citenamefont {{N.
  Dan}}, \citenamefont {{A. Berman}}, \citenamefont {{P. Pincus}},\ and\
  \citenamefont {{S. A. Safran}}}]{refId0}%
  \BibitemOpen
  \bibfield  {author} {\bibinfo {author} {\bibnamefont {{N. Dan}}}, \bibinfo
  {author} {\bibnamefont {{A. Berman}}}, \bibinfo {author} {\bibnamefont {{P.
  Pincus}}},\ and\ \bibinfo {author} {\bibnamefont {{S. A. Safran}}},\
  }\bibfield  {title} {\bibinfo {title} {Membrane-induced interactions between
  inclusions},\ }\href {https://doi.org/10.1051/jp2:1994227} {\bibfield
  {journal} {\bibinfo  {journal} {J. Phys. II France}\ }\textbf {\bibinfo
  {volume} {4}},\ \bibinfo {pages} {1713} (\bibinfo {year} {1994})}\BibitemShut
  {NoStop}%
\bibitem [{\citenamefont {Nelson}\ \emph {et~al.}(2004)\citenamefont {Nelson},
  \citenamefont {Piran},\ and\ \citenamefont {Weinberg}}]{Nelson2004}%
  \BibitemOpen
  \bibfield  {author} {\bibinfo {author} {\bibfnamefont {D.}~\bibnamefont
  {Nelson}}, \bibinfo {author} {\bibfnamefont {T.}~\bibnamefont {Piran}},\ and\
  \bibinfo {author} {\bibfnamefont {S.}~\bibnamefont {Weinberg}},\ }\href@noop
  {} {\emph {\bibinfo {title} {Statistical mechanics of membranes and
  surfaces}}}\ (\bibinfo  {publisher} {World Scientific},\ \bibinfo {year}
  {2004})\BibitemShut {NoStop}%
\bibitem [{\citenamefont {Lipowsky}\ and\ \citenamefont
  {Sackmann}(1995)}]{lipowsky1995structure}%
  \BibitemOpen
  \bibfield  {author} {\bibinfo {author} {\bibfnamefont {R.}~\bibnamefont
  {Lipowsky}}\ and\ \bibinfo {author} {\bibfnamefont {E.}~\bibnamefont
  {Sackmann}},\ }\href@noop {} {\bibinfo {title} {Structure and dynamics of
  membranes—from cells to vesicles (handbook of biological physics vol 1)}}
  (\bibinfo {year} {1995})\BibitemShut {NoStop}%
\bibitem [{\citenamefont {Helfrich}(1973)}]{Helfrich1973}%
  \BibitemOpen
  \bibfield  {author} {\bibinfo {author} {\bibfnamefont {W.}~\bibnamefont
  {Helfrich}},\ }\bibfield  {title} {\bibinfo {title} {Elastic properties of
  lipid bilayers: theory and possible experiments},\ }\href@noop {} {\bibfield
  {journal} {\bibinfo  {journal} {Z. Naturforsch., C: Biosci.}\ }\textbf
  {\bibinfo {volume} {28}},\ \bibinfo {pages} {693} (\bibinfo {year}
  {1973})}\BibitemShut {NoStop}%
\bibitem [{\citenamefont {Bruinsma}\ \emph {et~al.}(1994)\citenamefont
  {Bruinsma}, \citenamefont {Goulian},\ and\ \citenamefont
  {Pincus}}]{Bruinsma1994}%
  \BibitemOpen
  \bibfield  {author} {\bibinfo {author} {\bibfnamefont {R.}~\bibnamefont
  {Bruinsma}}, \bibinfo {author} {\bibfnamefont {M.}~\bibnamefont {Goulian}},\
  and\ \bibinfo {author} {\bibfnamefont {P.}~\bibnamefont {Pincus}},\
  }\bibfield  {title} {\bibinfo {title} {Self--assembly of membrane
  junctions.},\ }\href@noop {} {\bibfield  {journal} {\bibinfo  {journal}
  {Biophys. J.}\ }\textbf {\bibinfo {volume} {67}},\ \bibinfo {pages} {746}
  (\bibinfo {year} {1994})}\BibitemShut {NoStop}%
\bibitem [{\citenamefont {{Leibler, S.}}(1986)}]{Leibler1986}%
  \BibitemOpen
  \bibfield  {author} {\bibinfo {author} {\bibnamefont {{Leibler, S.}}},\
  }\bibfield  {title} {\bibinfo {title} {Curvature instability in membranes},\
  }\href {https://doi.org/10.1051/jphys:01986004703050700} {\bibfield
  {journal} {\bibinfo  {journal} {J. Phys. France}\ }\textbf {\bibinfo {volume}
  {47}},\ \bibinfo {pages} {507} (\bibinfo {year} {1986})}\BibitemShut
  {NoStop}%
\bibitem [{\citenamefont {{Roland R. Netz}}(1997)}]{Netz}%
  \BibitemOpen
  \bibfield  {author} {\bibinfo {author} {\bibnamefont {{Roland R. Netz}}},\
  }\bibfield  {title} {\bibinfo {title} {Inclusions in fluctuating membranes:
  Exact results},\ }\href {https://doi.org/10.1051/jp1:1997205} {\bibfield
  {journal} {\bibinfo  {journal} {J. Phys. I France}\ }\textbf {\bibinfo
  {volume} {7}},\ \bibinfo {pages} {833} (\bibinfo {year} {1997})}\BibitemShut
  {NoStop}%
\bibitem [{\citenamefont {Dommersnes}\ and\ \citenamefont
  {Fournier}(1999)}]{dommersnes1999n}%
  \BibitemOpen
  \bibfield  {author} {\bibinfo {author} {\bibfnamefont {P.}~\bibnamefont
  {Dommersnes}}\ and\ \bibinfo {author} {\bibfnamefont {J.-B.}\ \bibnamefont
  {Fournier}},\ }\bibfield  {title} {\bibinfo {title} {N-body study of
  anisotropic membrane inclusions: Membrane mediated interactions and ordered
  aggregation},\ }\href@noop {} {\bibfield  {journal} {\bibinfo  {journal} {The
  European Physical Journal B-Condensed Matter and Complex Systems}\ }\textbf
  {\bibinfo {volume} {12}},\ \bibinfo {pages} {9} (\bibinfo {year}
  {1999})}\BibitemShut {NoStop}%
\bibitem [{Note1()}]{Note1}%
  \BibitemOpen
  \bibinfo {note} {The parameter $\zeta $ is often called the correlation
  length, but in the model with no composition order parameter the length scale
  at which the correlation function decays is $\protect \sqrt {2}\zeta
  $.}\BibitemShut {Stop}%
\bibitem [{\citenamefont {Bihr}\ \emph {et~al.}(2015)\citenamefont {Bihr},
  \citenamefont {Seifert},\ and\ \citenamefont {Smith}}]{Bihr2015}%
  \BibitemOpen
  \bibfield  {author} {\bibinfo {author} {\bibfnamefont {T.}~\bibnamefont
  {Bihr}}, \bibinfo {author} {\bibfnamefont {U.}~\bibnamefont {Seifert}},\ and\
  \bibinfo {author} {\bibfnamefont {A.-S.}\ \bibnamefont {Smith}},\ }\bibfield
  {title} {\bibinfo {title} {Multiscale approaches to protein-mediated
  interactions between membranes—relating microscopic and macroscopic
  dynamics in radially growing adhesions},\ }\href@noop {} {\bibfield
  {journal} {\bibinfo  {journal} {New J. Phys.}\ }\textbf {\bibinfo {volume}
  {17}},\ \bibinfo {pages} {083016} (\bibinfo {year} {2015})}\BibitemShut
  {NoStop}%
\bibitem [{\citenamefont {Lin}(2013)}]{Lin2013}%
  \BibitemOpen
  \bibfield  {author} {\bibinfo {author} {\bibfnamefont {Q.-G.}\ \bibnamefont
  {Lin}},\ }\bibfield  {title} {\bibinfo {title} {Infinite integrals involving
  bessel functions by contour integration},\ }\href
  {https://doi.org/10.1080/10652469.2012.758119} {\bibfield  {journal}
  {\bibinfo  {journal} {Integr. Transforms Special Funct.}\ }\textbf {\bibinfo
  {volume} {24}},\ \bibinfo {pages} {783} (\bibinfo {year} {2013})}\BibitemShut
  {NoStop}%
\bibitem [{\citenamefont {Goldenfeld}(1992)}]{Goldenfeld1992}%
  \BibitemOpen
  \bibfield  {author} {\bibinfo {author} {\bibfnamefont {N.}~\bibnamefont
  {Goldenfeld}},\ }\href@noop {} {\emph {\bibinfo {title} {Lectures on Phase
  Transitions and the Renormalization Group}}},\ \bibinfo {series} {Frontiers
  in Physics}, Vol.~\bibinfo {volume} {85}\ (\bibinfo  {publisher}
  {Addison--Wesley, The Advanced Book Program},\ \bibinfo {year}
  {1992})\BibitemShut {NoStop}%
\bibitem [{\citenamefont {Fisher}\ and\ \citenamefont
  {Wiodm}(1969)}]{Fisher1969}%
  \BibitemOpen
  \bibfield  {author} {\bibinfo {author} {\bibfnamefont {M.~E.}\ \bibnamefont
  {Fisher}}\ and\ \bibinfo {author} {\bibfnamefont {B.}~\bibnamefont {Wiodm}},\
  }\bibfield  {title} {\bibinfo {title} {Decay of correlations in linear
  systems},\ }\href@noop {} {\bibfield  {journal} {\bibinfo  {journal} {J.
  Chem. Phys.}\ }\textbf {\bibinfo {volume} {50}},\ \bibinfo {pages} {3756}
  (\bibinfo {year} {1969})}\BibitemShut {NoStop}%
\bibitem [{\citenamefont {Fisher}\ and\ \citenamefont
  {Widom}(2015)}]{Fisher2015}%
  \BibitemOpen
  \bibfield  {author} {\bibinfo {author} {\bibfnamefont {M.~E.}\ \bibnamefont
  {Fisher}}\ and\ \bibinfo {author} {\bibfnamefont {B.}~\bibnamefont {Widom}},\
  }\bibfield  {title} {\bibinfo {title} {Publisher's note: ``{D}ecay of
  correlations in linear systems'' [{J}.~{C}hem.~{P}hys.~50, 3756 (1969)]},\
  }\href@noop {} {\bibfield  {journal} {\bibinfo  {journal} {J. Chem. Phys.}\
  }\textbf {\bibinfo {volume} {143}},\ \bibinfo {pages} {3756} (\bibinfo {year}
  {2015})}\BibitemShut {NoStop}%
\bibitem [{\citenamefont {Evans}\ \emph {et~al.}(1994)\citenamefont {Evans},
  \citenamefont {Leote~de Carvalho}, \citenamefont {Henderson},\ and\
  \citenamefont {Hoyle}}]{Evans1994}%
  \BibitemOpen
  \bibfield  {author} {\bibinfo {author} {\bibfnamefont {R.}~\bibnamefont
  {Evans}}, \bibinfo {author} {\bibfnamefont {R.}~\bibnamefont {Leote~de
  Carvalho}}, \bibinfo {author} {\bibfnamefont {J.}~\bibnamefont {Henderson}},\
  and\ \bibinfo {author} {\bibfnamefont {D.}~\bibnamefont {Hoyle}},\ }\bibfield
   {title} {\bibinfo {title} {Asymptotic decay of correlations in liquids and
  their mixtures},\ }\href@noop {} {\bibfield  {journal} {\bibinfo  {journal}
  {J. Chem. Phys.}\ }\textbf {\bibinfo {volume} {100}},\ \bibinfo {pages} {591}
  (\bibinfo {year} {1994})}\BibitemShut {NoStop}%
\bibitem [{\citenamefont {Jane{\v{s}}}\ \emph {et~al.}(2019)\citenamefont
  {Jane{\v{s}}}, \citenamefont {Stumpf}, \citenamefont {Schmidt}, \citenamefont
  {Seifert},\ and\ \citenamefont {Smith}}]{Janes2019}%
  \BibitemOpen
  \bibfield  {author} {\bibinfo {author} {\bibfnamefont {J.~A.}\ \bibnamefont
  {Jane{\v{s}}}}, \bibinfo {author} {\bibfnamefont {H.}~\bibnamefont {Stumpf}},
  \bibinfo {author} {\bibfnamefont {D.}~\bibnamefont {Schmidt}}, \bibinfo
  {author} {\bibfnamefont {U.}~\bibnamefont {Seifert}},\ and\ \bibinfo {author}
  {\bibfnamefont {A.-S.}\ \bibnamefont {Smith}},\ }\bibfield  {title} {\bibinfo
  {title} {Statistical mechanics of an elastically pinned membrane: Static
  profile and correlations},\ }\href
  {https://doi.org/https://doi.org/10.1016/j.bpj.2018.12.003} {\bibfield
  {journal} {\bibinfo  {journal} {Biophys. J.}\ }\textbf {\bibinfo {volume}
  {116}},\ \bibinfo {pages} {283} (\bibinfo {year} {2019})}\BibitemShut
  {NoStop}%
\bibitem [{\citenamefont {Kadanoff}(1993)}]{Kadanoff1993}%
  \BibitemOpen
  \bibfield  {author} {\bibinfo {author} {\bibfnamefont {L.~P.}\ \bibnamefont
  {Kadanoff}},\ }\bibfield  {title} {\bibinfo {title} {Critical behavior.
  universality and scaling},\ }in\ \href@noop {} {\emph {\bibinfo {booktitle}
  {From Order To Chaos: Essays: Critical, Chaotic and Otherwise}}}\ (\bibinfo
  {publisher} {World Scientific},\ \bibinfo {year} {1993})\ pp.\ \bibinfo
  {pages} {222--239}\BibitemShut {NoStop}%
\bibitem [{\citenamefont {Lipowsky}(1990)}]{Lipowsky1990}%
  \BibitemOpen
  \bibfield  {author} {\bibinfo {author} {\bibfnamefont {R.}~\bibnamefont
  {Lipowsky}},\ }\bibfield  {title} {\bibinfo {title} {Shape fluctuations and
  critical phenomena},\ }in\ \href@noop {} {\emph {\bibinfo {booktitle}
  {Fundamental Problems in Statistical Mechanics VII: Proceedings of the
  Seventh International Summer School on Fundamental Problems in Statistical
  Mechanics, Altenburg, F.R. Germany, June 18-30, 1989}}},\ \bibinfo {editor}
  {edited by\ \bibinfo {editor} {\bibfnamefont {H.}~\bibnamefont {van
  Beijeren}}}\ (\bibinfo  {publisher} {North-Holland, Amsterdam},\ \bibinfo
  {year} {1990})\ pp.\ \bibinfo {pages} {139--170}\BibitemShut {NoStop}%
\end{thebibliography}
\end{document}